\documentclass[letterpaper,twocolumn,english,prc,aps,showpacs,superscriptaddress]{revtex4}

\usepackage{ae,aecompl}
\usepackage[T1]{fontenc}
\usepackage[latin9]{inputenc}
\usepackage{babel}
\usepackage{units}
\usepackage{amsmath}
\usepackage{amssymb}
\usepackage{graphicx}
\usepackage{subfigure}
\usepackage{tikz}
\usepackage{multirow}
\usepackage{ulem}

\makeatletter

\newcommand{\Mev}{\,MeV/A }
\newcommand{\mev}{\,MeV/A}
\newcommand{\Xesn}{$^{129}$\text{Xe}+$^{nat}$\text{Sn} }

\newcommand{\Snxe}{$^{129}$\text{Xe}+$^{122}$\text{Sn} }

\newcommand{\Sn}{$^{nat}$\text{Sn} }

\newcommand{\Xe}{$^{129}$\text{Xe} }

\newcommand{\Dv}{$\delta v$ }

\newcommand{\Dt}{$\delta t$ }

\newcommand{\anis}{W(0\textdegree)/W(90\textdegree)}
\newcommand{\ztot}{\text{Z$_{\text{tot}}$}}

\renewcommand{\bf}[1]{\begingroup\bfseries\mathversion{bold}#1\endgroup}

\DeclareTextFontCommand{\helvetica}{\fontfamily{phv}\selectfont}

\makeatother

\begin{document}

\title{Coulomb chronometry to probe the decay mechanism of hot nuclei}

\author{D. Gruyer}
\thanks{Present address : INFN -- Sezione di Firenze}
\email{diego.gruyer@fi.infn.it}
\affiliation{GANIL, CEA-DSM/CNRS-IN2P3, Bvd. Henri Becquerel, F-14076 Caen CEDEX, France}

\author{J.D. Frankland}
\affiliation{GANIL, CEA-DSM/CNRS-IN2P3, Bvd. Henri Becquerel, F-14076 Caen CEDEX, France}

\author{E. Bonnet}
\author{A. Chbihi}
\affiliation{GANIL, CEA-DSM/CNRS-IN2P3, Bvd. Henri Becquerel, F-14076 Caen CEDEX, France}

\author{G. Ademard}
\affiliation{Institut de Physique Nucl\'eaire, CNRS/IN2P3, Universit\'e Paris-Sud 11, F-91406 Orsay CEDEX, France}

\author{M. Boisjoli}
\affiliation{GANIL, CEA-DSM/CNRS-IN2P3, Bvd. Henri Becquerel, F-14076 Caen CEDEX, France}
\affiliation{D\'epartement de physique, de g\'enie physique et d'optique, Universit\'e Laval, Qu\'ebec, G1V 0A6 Canada}

\author{B. Borderie}
\affiliation{Institut de Physique Nucl\'eaire, CNRS/IN2P3, Universit\'e Paris-Sud 11, F-91406 Orsay CEDEX, France}

\author{R. Bougault}
\affiliation{LPC, CNRS/IN2P3, Ensicaen, Universit\'e de Caen, F-14050 Caen CEDEX, France}

\author{E. Galichet}
\affiliation{Institut de Physique Nucl\'eaire, CNRS/IN2P3, Universit\'e Paris-Sud 11, F-91406 Orsay CEDEX, France}
\affiliation{Conservatoire National des Arts et M\'etiers, F-75141 Paris Cedex 03, France}

\author{J. Gauthier}
\affiliation{D\'epartement de physique, de g\'enie physique et d'optique, Universit\'e Laval, Qu\'ebec, G1V 0A6 Canada}

\author{D. Guinet}
\author{P. Lautesse}
\affiliation{Institut de Physique Nucl\'eaire, Universit\'e Claude Bernard Lyon 1, CNRS/IN2P3, F-69622 Villeurbanne CEDEX, France}

\author{N. Le Neindre}
\affiliation{LPC, CNRS/IN2P3, Ensicaen, Universit\'e de Caen, F-14050 Caen CEDEX, France}

\author{E. Legou\'ee}
\affiliation{LPC, CNRS/IN2P3, Ensicaen, Universit\'e de Caen, F-14050 Caen CEDEX, France}

\author{I. Lombardo}
\affiliation{Dipartimento di Fisica, Universit\`a degli Studi di Napoli FEDERICO II, I-80126 Napoli, Italy}
\affiliation{Istituto Nazionale di Fisica Nucleare, Sezione di Napoli, Complesso Universitario di Monte S. Angelo, Via Cintia Edificio 6, I-80126 Napoli, Italy}

\author{O. Lopez}
\affiliation{LPC, CNRS/IN2P3, Ensicaen, Universit\'e de Caen, F-14050 Caen CEDEX, France}

\author{L. Manduci}
\affiliation{\'Ecole des Applications Militaires de l'\'Energie Atomique, B.P. 19, F-50115 Cherbourg, France}

\author{P. Marini}
\affiliation{GANIL, CEA-DSM/CNRS-IN2P3, Bvd. Henri Becquerel, F-14076 Caen CEDEX, France}

\author{K. Mazurek}
\affiliation{H. Niewodnicza\'{n}ski Institute of Nuclear Physics, PL-31342 Krak\'ow, Poland}
\author{P.N. Nadtochy}
\affiliation{Omsk State University, Mira prospekt 55-A, Omsk 644077, Russia}

\author{M. P\^arlog}
\affiliation{LPC, CNRS/IN2P3, Ensicaen, Universit\'e de Caen, F-14050 Caen CEDEX, France}
\affiliation{National Institute for Physics and Nuclear Engineering, RO-077125 Bucharest-M\u{a}gurele, Romania}

\author{M. F. Rivet}
\thanks{deceased}
\affiliation{Institut de Physique Nucl\'eaire, CNRS/IN2P3, Universit\'e Paris-Sud 11, F-91406 Orsay CEDEX, France}

\author{R. Roy}
\affiliation{D\'epartement de physique, de g\'enie physique et d'optique, Universit\'e Laval, Qu\'ebec, G1V 0A6 Canada}

\author{E. Rosato}
\thanks{deceased}
\author{G. Spadaccini}
\affiliation{Dipartimento di Fisica, Universit\`a degli Studi di Napoli FEDERICO II, I-80126 Napoli, Italy}
\affiliation{Istituto Nazionale di Fisica Nucleare, Sezione di Napoli, Complesso Universitario di Monte S. Angelo, Via Cintia Edificio 6, I-80126 Napoli, Italy}

\author{G. Verde}
\affiliation{Institut de Physique Nucl\'eaire, CNRS/IN2P3, Universit\'e Paris-Sud 11, F-91406 Orsay CEDEX, France}
\affiliation{Istituto Nazionale di Fisica Nucleare, Sezione di Catania, 64 Via Santa Sofia, I-95123, Catania, Italy}

\author{E. Vient}
\affiliation{LPC, CNRS/IN2P3, Ensicaen, Universit\'e de Caen, F-14050 Caen CEDEX, France}

\author{M. Vigilante}
\affiliation{Dipartimento di Fisica, Universit\`a degli Studi di Napoli FEDERICO II, I-80126 Napoli, Italy}
\affiliation{Istituto Nazionale di Fisica Nucleare, Sezione di Napoli, Complesso Universitario di Monte S. Angelo, Via Cintia Edificio 6, I-80126 Napoli, Italy}
\author{J.P. Wieleczko}
\affiliation{GANIL, CEA-DSM/CNRS-IN2P3, Bvd. Henri Becquerel, F-14076 Caen CEDEX, France}

\collaboration{INDRA collaboration}\noaffiliation

\begin{abstract}
In \Xesn central collisions from 8 to 25\mev,
the three-fragment exit channel occurs with a significant cross section.
We show that these fragments arise from two successive binary splittings of a heavy composite system. 
The sequence of fragment production is determined.
Strong Coulomb proximity effects are observed in the three-fragment final state. 
A comparison with Coulomb trajectory calculations shows that the time scale between the consecutive break-ups
decreases with increasing bombarding energy, becoming quasi-simultaneous above
excitation energy $E^*=4.0\pm0.5$\mev.
This transition from sequential to simultaneous break-up was interpreted as the signature of the onset of multifragmentation for the 
three-fragment exit channel in this system.
\end{abstract}


\maketitle

\section{Introduction}

In central heavy  ion collisions at bombarding energies around $10-20$\mev, namely well above the Coulomb barrier but below the Fermi energy regime,
different types of reaction mechanism leading to the production of one, two, three, or more heavy fragments in the exit channel are possible, namely fusion-fission, quasifission, and deeply inelastic collisions \cite{Huizenga1982Distinctive,Moretto1983Phenomenology,Toke1985Quasifission}.
Only by detecting all reaction products in coincidence and achieving a full
kinematical reconstruction event by event can we hope to better understand the underlying reaction and decay mechanisms.
Such exclusive experimental data are relatively scarce for multibody exit channels in this
energy range \cite{Glassel1983Observation,Charity1991Results,Casini1993Fission,PhysRevC.81.024605}
leaving room for ambiguities in the interpretation of the reaction mechanism.
New theoretical efforts are made to cover this energy range including time dependent microscopic approaches \cite{PhysRevLett.103.042701,Sekizawa2013Timedependent}, 
transport models \cite{PhysRevC.78.044614} and molecular dynamics calculations \cite{Li2013Dynamical}, which require comparison with new exclusive measurements in order
to advance.

\begin{figure}[h]
\includegraphics[width=0.99\linewidth]{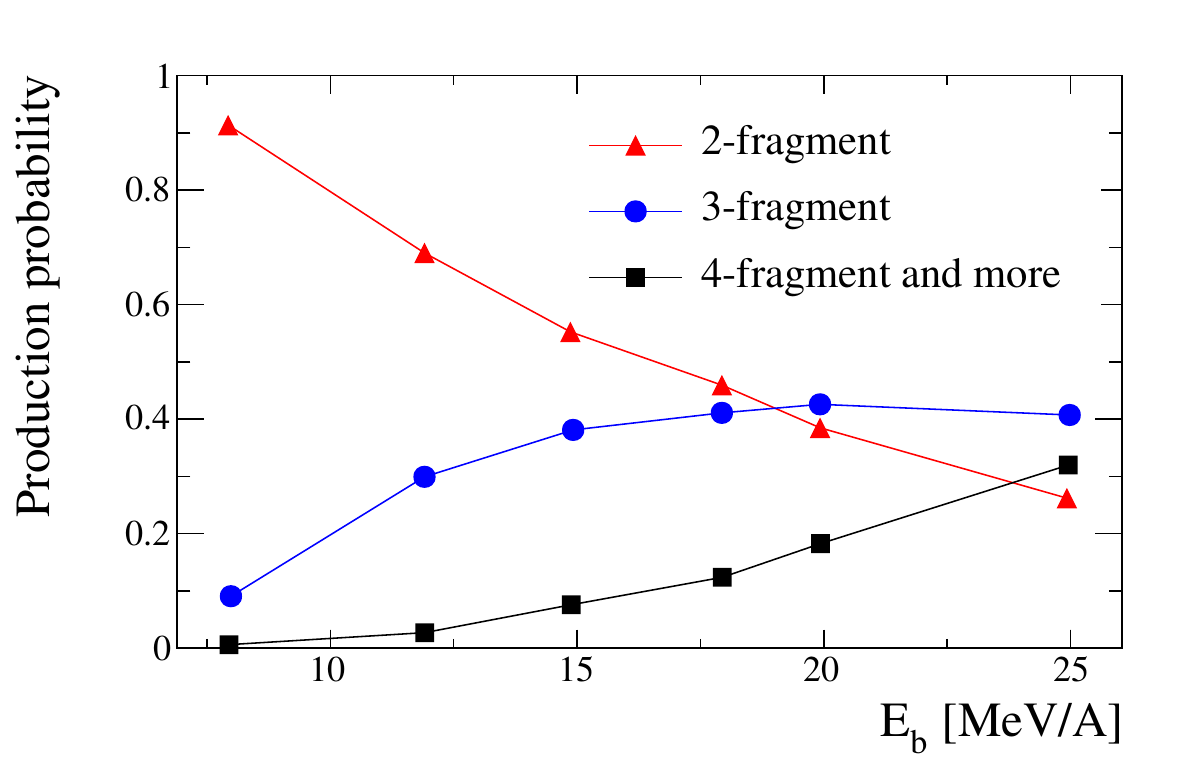}
\caption{(color online). Evolution of different exit channel production probabilities as a function of the beam energy for \Xesn central collisions.}
\label{fig:sec}
\end{figure}

Recent exclusive data on \Xesn central collisions measured with the INDRA $4\pi$ charged particle
multidetector \cite{Pouthas1995INDRA} show that 
at 8\Mev bombarding energy almost all events contain two heavy fragments in the exit channel with a total charge close to that of the incident nuclei (including evaporated light
charged particles) \cite{Chbihi012099}.
Above 12\Mev bombarding energy (see Fig.\ref{fig:sec}), the three-fragment exit channel becomes significant, overcoming the two-fragment production rate above 18\mev.
The question we want to address in this paper is the underlying mechanisms
responsible for these three-fragment events: are they the result of deeply-inelastic reactions (followed by fission of one of the two partners, or with the third 
fragment resulting from a neck formed between projectile and target), or do they result from the decay of a composite system (not necessarily fully-equilibrated) ? Is the break-up 
a sequential continuation of low-energy fission processes to higher available energies, or is it a precursor of the simultaneous nuclear disassembly (multifragmentation) 
observed at higher energies for this same system \cite{Hudan2003Characteristics,Piantelli2008FreezeOut}?

To answer these questions, a dynamical characterization of the decay mechanism is needed, based on a full kinematical reconstruction of the multibody exit channel.
In particular, we will show that the determination of the order in which fragments are produced and the estimation of the involved time-scales allow to disentangle sequential fission and simultaneous three-fragment break-up.
This information is of great importance in view of constraining reaction models with predictive power in this energy regime.

\begin{figure*}[t]
     {\subfigure[ 12\,MeV/A\label{fig:ztotcflot:12}]{\includegraphics[width=0.24\linewidth]{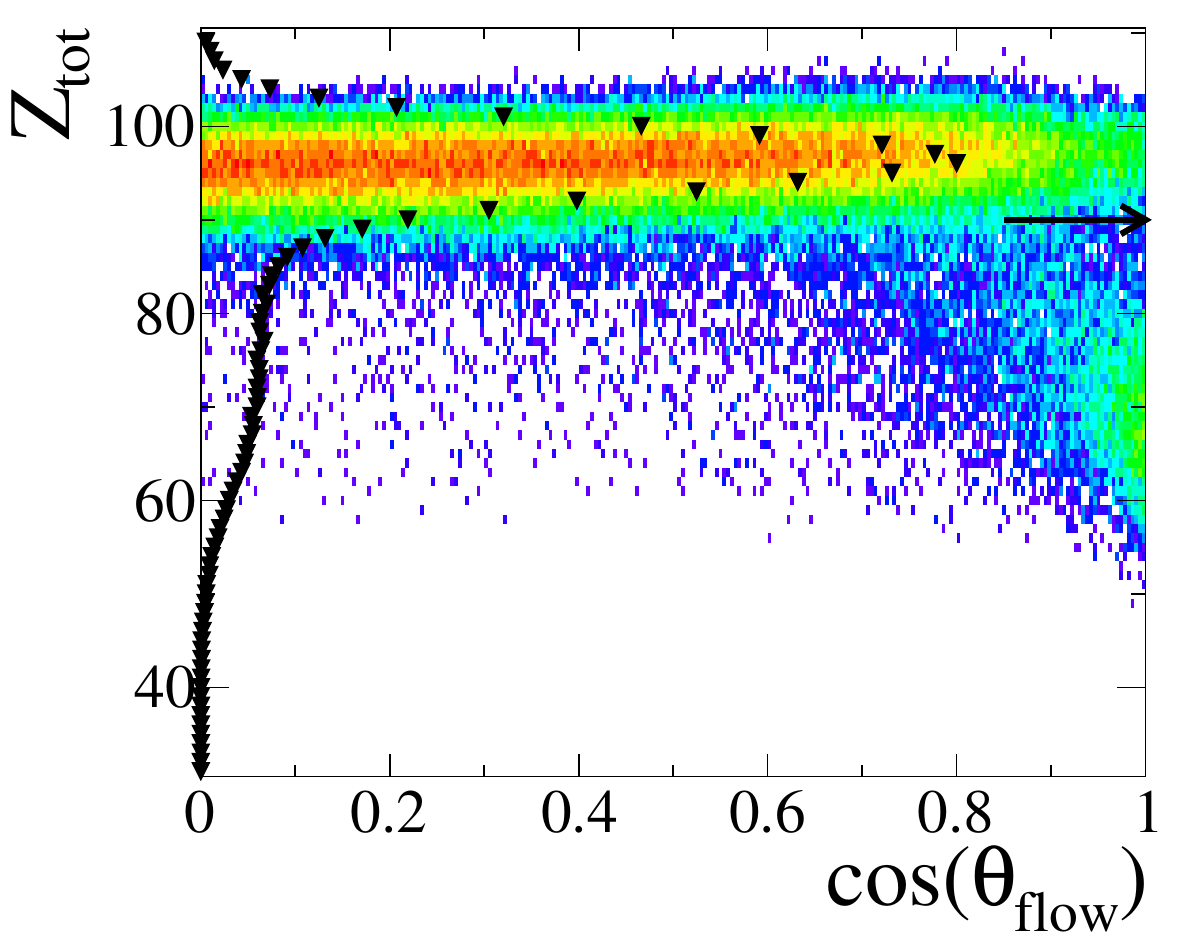}}}
     {\subfigure[ 15\,MeV/A\label{fig:ztotcflot:15}]{\includegraphics[width=0.24\linewidth]{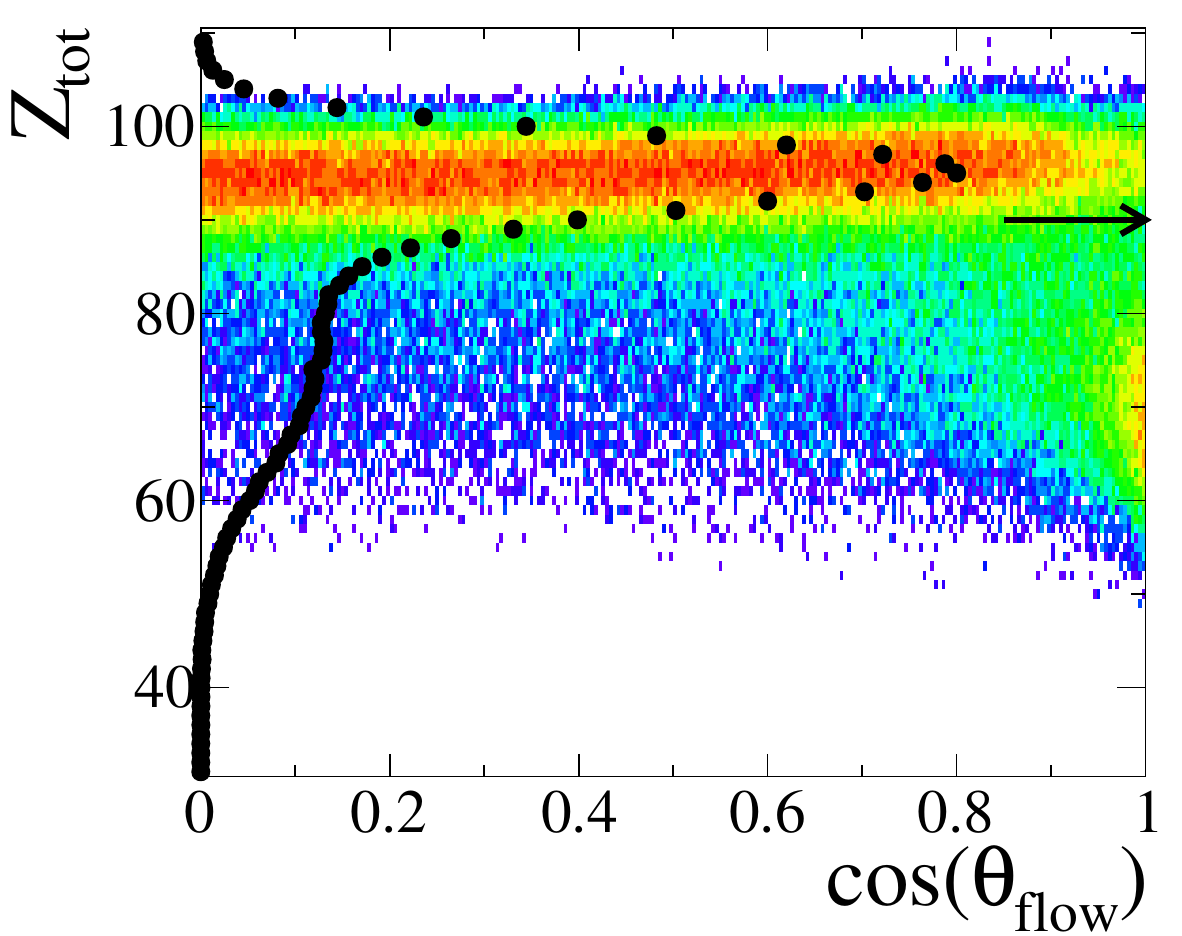}}}
     {\subfigure[ 18\,MeV/A\label{fig:ztotcflot:18}]{\includegraphics[width=0.24\linewidth]{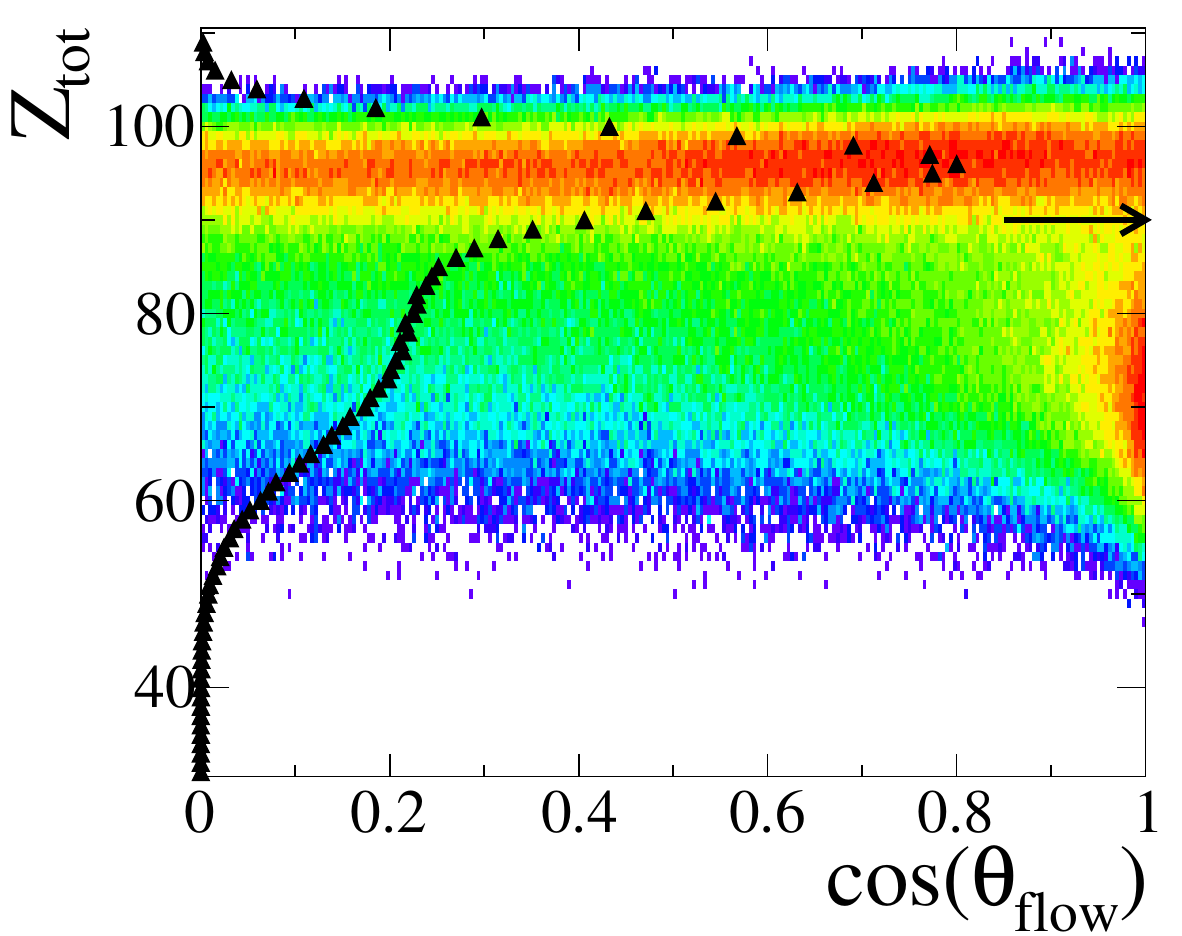}}}
     {\subfigure[ 20\,MeV/A\label{fig:ztotcflot:20}]{\includegraphics[width=0.24\linewidth]{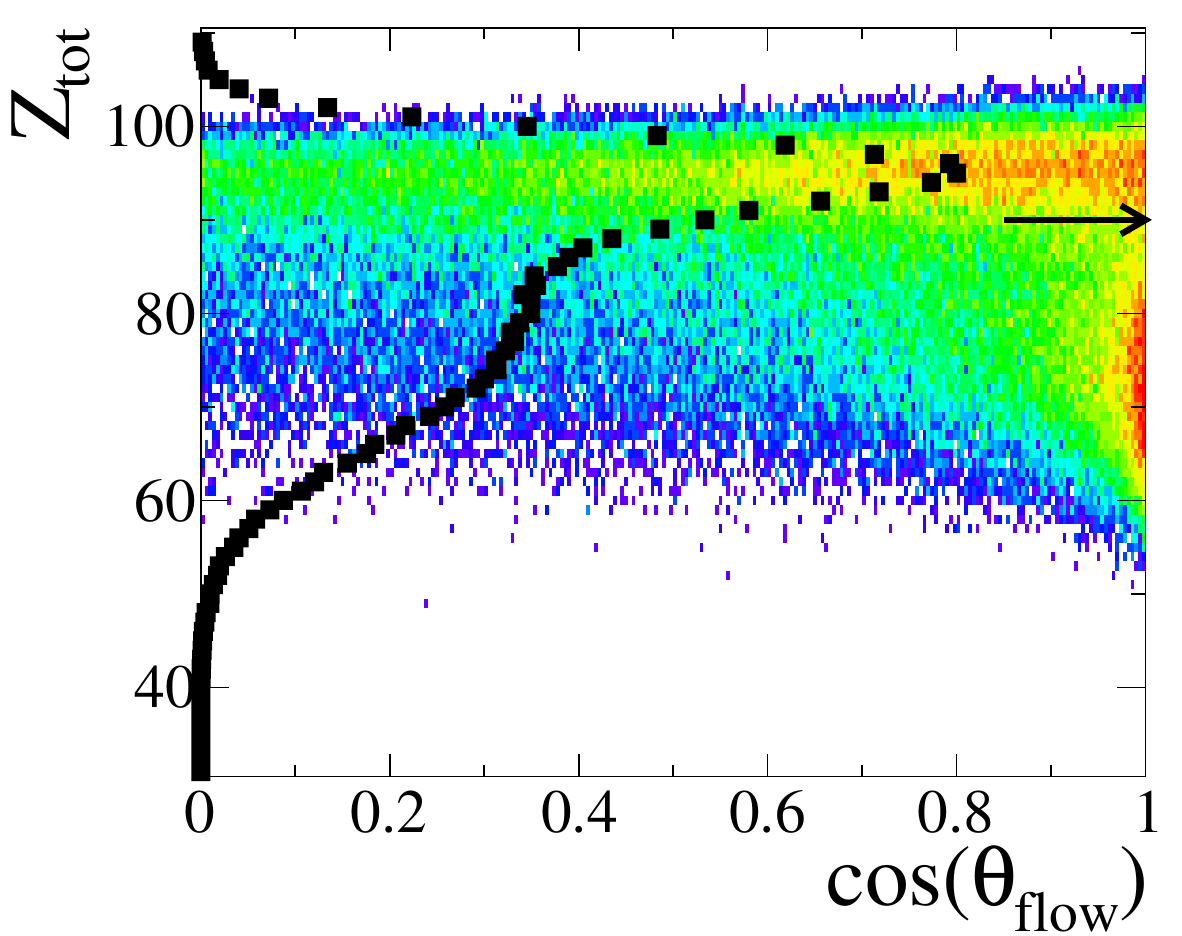}}}
     {\subfigure[ 12\,MeV/A\label{fig:vt:12}]       {\includegraphics[width=0.24\linewidth]{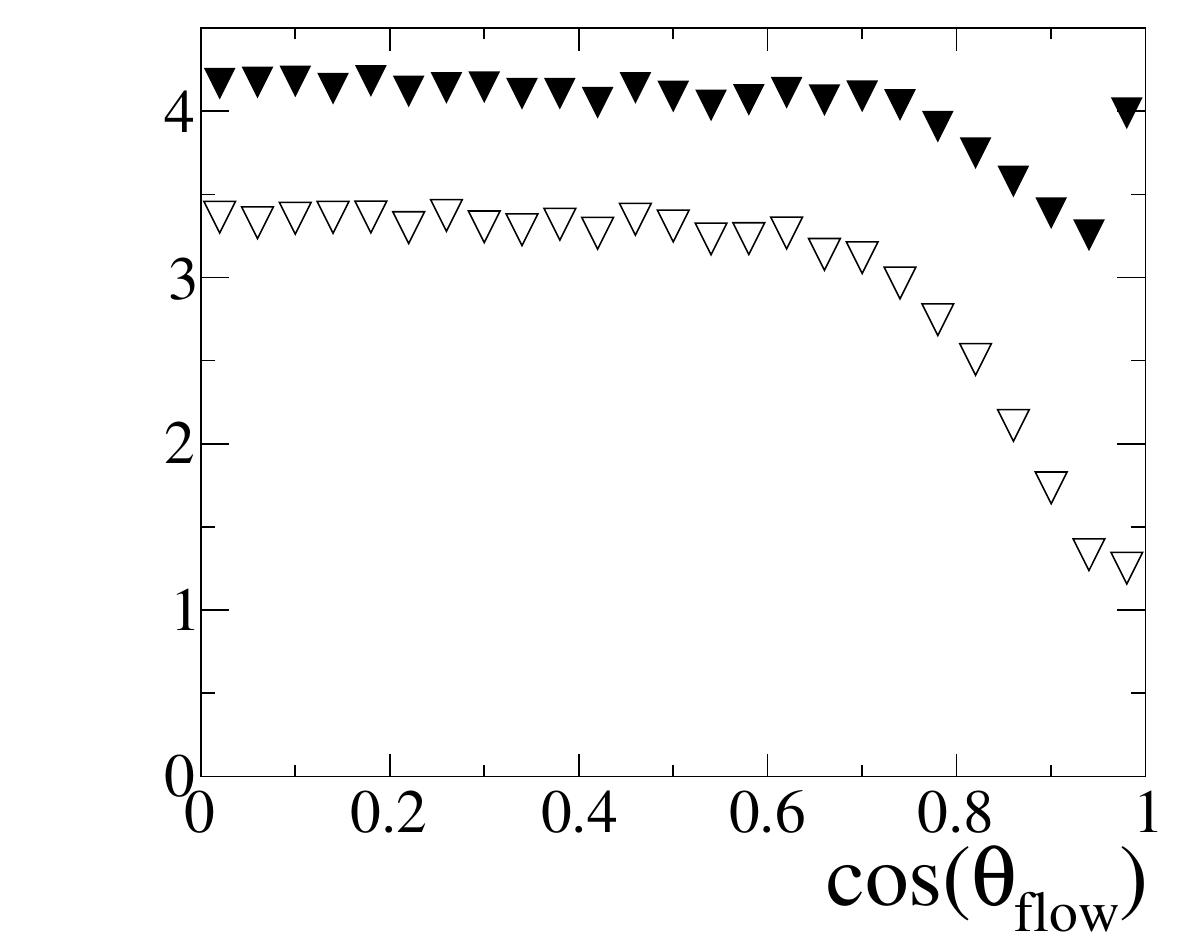}}}
     {\subfigure[ 15\,MeV/A\label{fig:vt:15}]       {\includegraphics[width=0.24\linewidth]{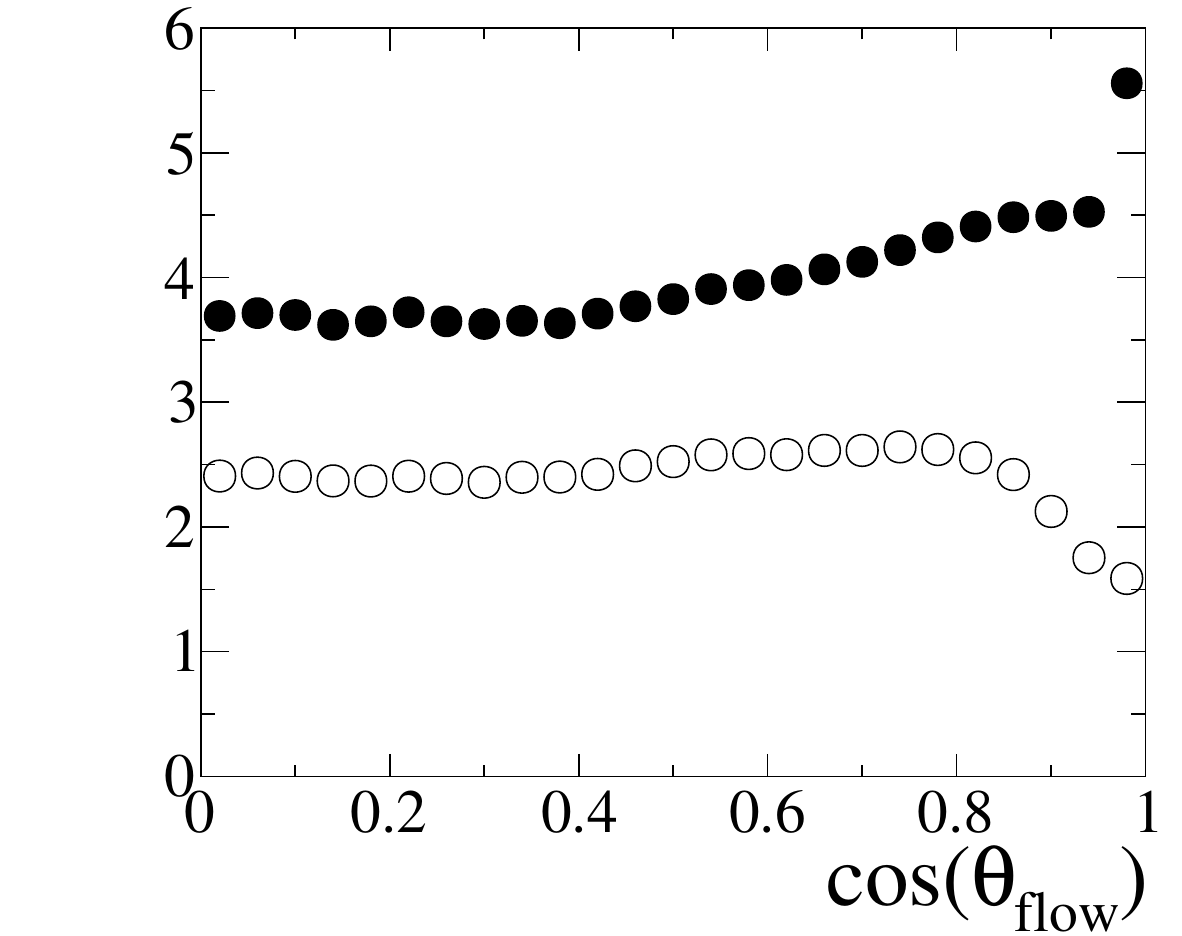}}}
     {\subfigure[ 18\,MeV/A\label{fig:vt:18}]       {\includegraphics[width=0.24\linewidth]{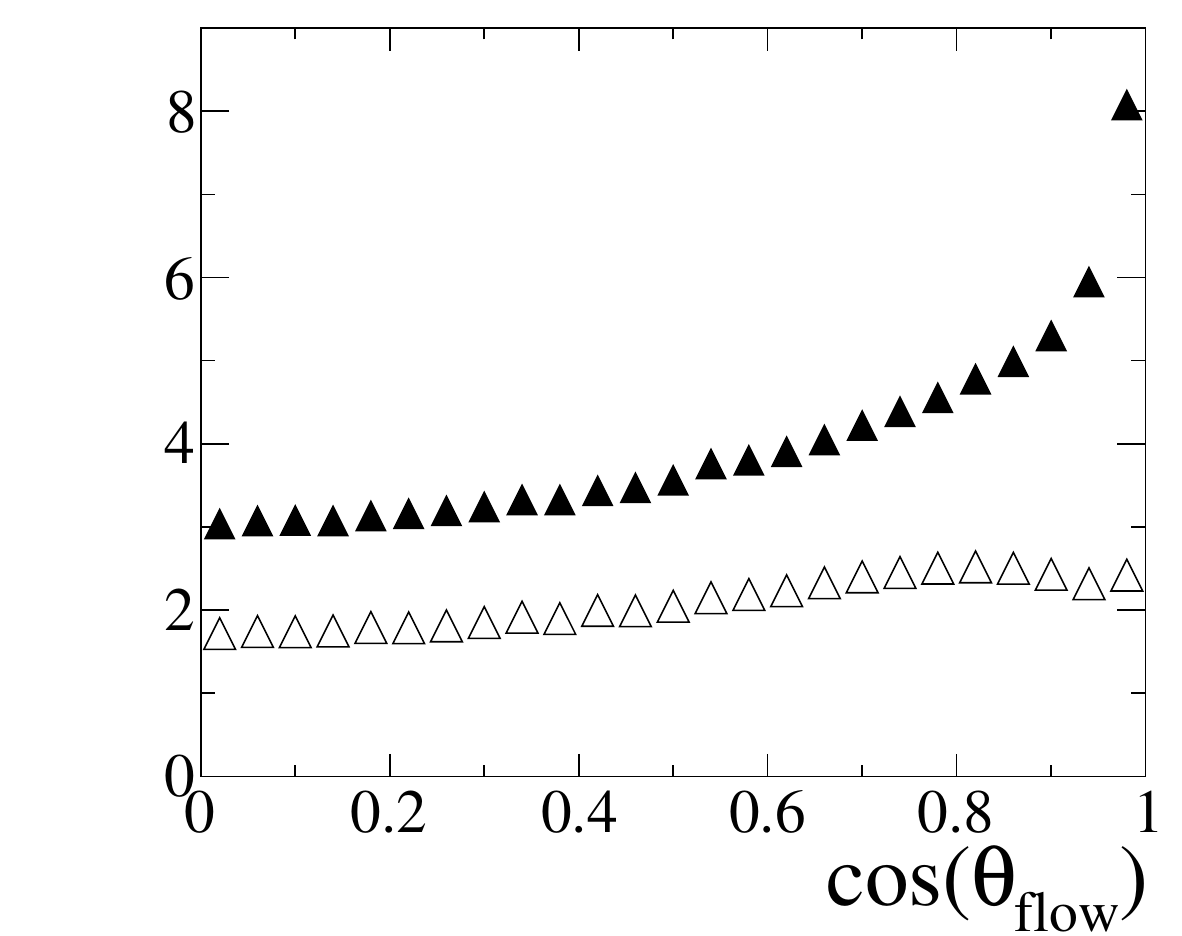}}}
     {\subfigure[ 20\,MeV/A\label{fig:vt:20}]       {\includegraphics[width=0.24\linewidth]{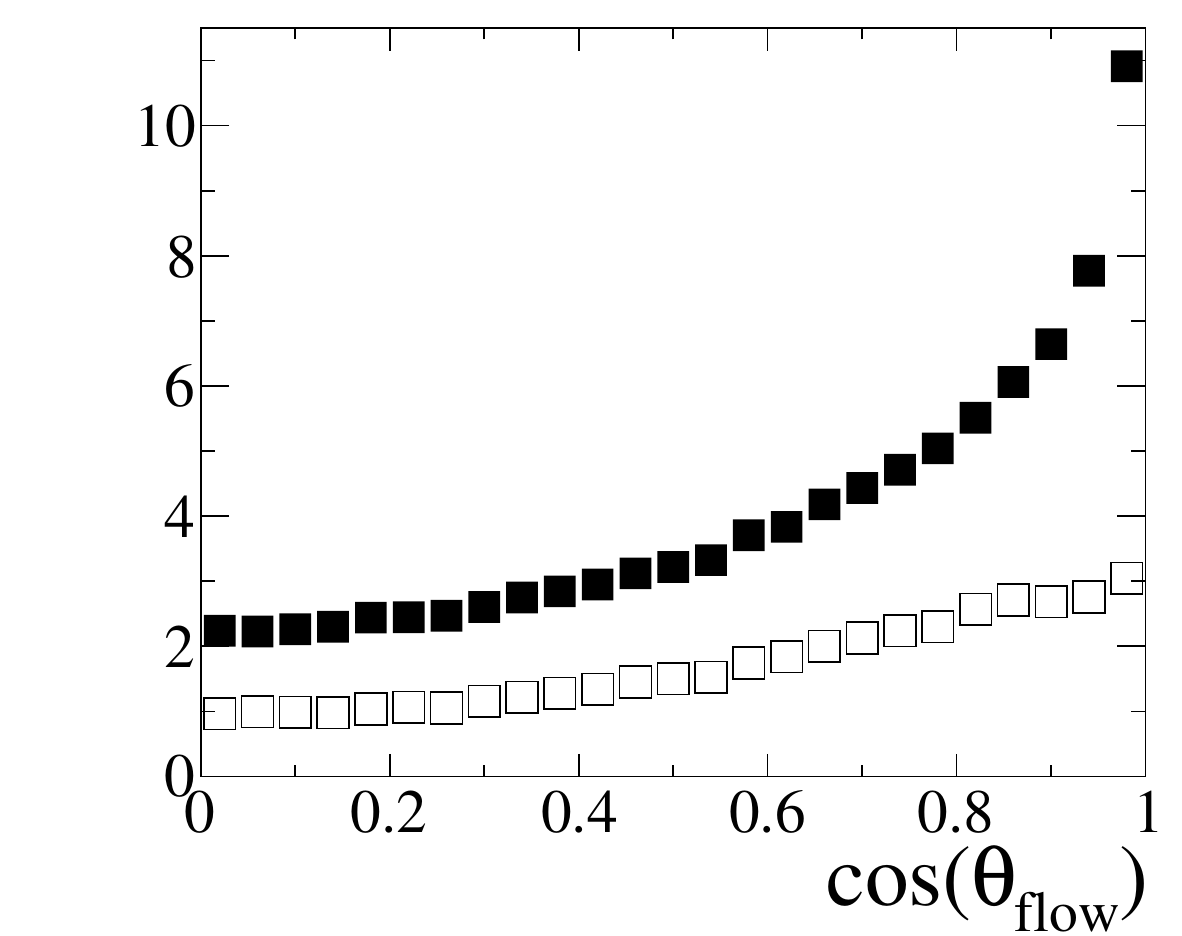}}}
\caption{(color on-line). 
(a$-$d) Experimental correlations between the cosinus of the flow angle, $\cos{(\theta_{\text{flow}})}$, (see text) and the total detected charge (Z$_{\text{tot}}$) event by event, for events with three heavy fragments ($Z>10$) in the exit channel. Round symbols show the distribution of Z$_{\text{tot}}$.
(e$-$h) Distribution of the cosinus of the flow angle ($\cos{(\theta_{\text{flow}})}$) for three-fragment events: (full symbols) before and (open symbols) 
after the selection in total detected charge (Z$_{\text{tot}}>90$) indicated by the arrow in each of figures (a$-$d).}
\label{fig:sel}
\end{figure*}

Several methods have been proposed for time-scale measurement in peripheral heavy-ion collisions,
which are dominated by deep-inelastic reactions, neck formation and decay, and so-called dynamical fission of projectile- or target-like nuclei \cite{Casini1993Fission,Stefanini1995Analysis,DeFilippo2005Time}.
Such methods were recently used to probe the isospin equilibration between projectile and target nuclei 
\cite{McIntosh2010Shortlived,Hudan2012Tracking,PhysRevC.87.061601,DeFilippo2012Correlations}.
However, they are not applicable to cases where an intermediate composite system is formed, hence losing the
distinction between ``projectile-like" or ``target-like" fragments,
as it may occur in central collisions.

In the case of central collisions, two-fragment correlation functions have been used to extract emission time scales in multifragmentation events, 
typically observed at intermediate energies~\cite{Durand1998Physics,Beaulieu2000Signals,Verde2006Correlations,Tabacaru2006371,Borderie2008551,PhysRevC.58.270}. 
The extracted emission properties are affected by space-time ambiguities.
Moreover, distortions of the 
correlation function shape induced by momentum and energy conservation laws~\cite{PhysRevC.78.064903}, collective motion and reaction plane 
orientation effects~\cite{Kampfer1993Velocity,Verde2007Correlation}, while small or negligible in the case of light particle correlation studies, 
may become important and difficult to deal with in the case of massive fragment-fragment correlations~\cite{Verde2006Correlations,Verde2007Correlation}. 

In this paper, we propose a new Coulomb ''chronometer`` suitable for three-fragment exit channels in central collisions.
The proposed method is similar to that used for the study of three-fragment coincidences in \Snxe collisions at 12.5\Mev bombarding energy 
\cite{Harrach1982Direct}, with the crucial addition of the knowledge of the fragment emission sequence, and the use of a 4$\pi$-multidetector, 
which reduces biasing the detected exit channels.
We have used this chronometer to study the underlying prodution mechanism of 3-fragment exit channels and to extract the evolution of fragment 
emission time-scales in \Xesn central collisions from 8 to 25\Mev bombarding energy.

\section{Experimental details}

\subsection{Experimental setup\label{sc:experiment} }

Collisions of \Xesn at 8, 12, 15, 18, 20, and 25 \Mev were measured using the INDRA $4\pi$ charged
product array \cite{Pouthas1995INDRA} at the GANIL accelerator facility.
The \Xe beam at 25 \Mev was directly delivered by the coupling of the two main cyclotrons, CSS1 and CSS2.
However, this combination does not allow to obtain incident energies between 8 and 20\mev. 
Therefore the \Xe beam was first accelerated by the coupled
cyclotrons to 27\Mev with a $40^+$ charge state and then decelerated to the required beam energies of 20, 18, 15, 12 and 8\Mev using a carbon degrader foil placed in the beam line whose orientation was modified to give different effective thicknesses. 
The charge state and purity of the \Xe beam after the degrader were ensured using the alpha-spectrometer of GANIL,
whose $B\rho$ setting was optimized for each incident energy. For the two lowest energies,
8 and 12\mev, more than one charge state were transmitted, inducing uncertainties on these incident energies: 
$\delta$E = 0.5 (0.2)\Mev at 8 (12)\Mev beam energy.

The \Xe beam then impinged on a self-supported 350\,$\mu$g/cm$^2$-thick \Sn 
target placed inside the INDRA detector array \cite{Pouthas1995INDRA}. 
This charged product multidetector, composed of 336 detection cells arranged 
in 17 rings centered on the beam axis, covers 90\% of the solid angle.
The first ring (2\textdegree~to 3\textdegree) is made of 12
telescopes composed of 300\,$\mu$m silicon wafer (Si) and CsI(Tl) scintillator (14 cm thick). Rings 2 to 9
(3\textdegree~to 45\textdegree) are composed of 12 or 24 three-member detection telescopes : a 5\,cm thick ionization chamber (IC) 
with 2.5\,$\mu$m Mylar windows operated with 20-50\,mbar of $C_3F_8$ gas; a 300\,$\mu$m or
150\,$\mu$m silicon wafer; and a CsI(Tl) scintillator (14 to 10\,cm thick) coupled to a photomultiplier
tube. Rings 10 to 17 (45\textdegree~to 176\textdegree) are composed of 24, 16 or 8 two-member telescopes: an 
ionization chamber and a CsI(Tl) scintillator of 8, 6 or 5\,cm thickness.
Events were recorded with an on-line trigger requiring at least 2 independent telescopes
hit in coincidence.

In the offline analysis, charged reaction products were identified from $\Delta E-E$ correlations between successive IC-Si, 
Si-CsI(Tl) or IC-CsI(Tl) detectors. In the IC-Si telescopes, where most of the heavy reaction products are stopped at these energies,
extrapolation of experimental $\Delta E-E$ maps using range-energy tables \cite{NorthcliffeSchilling,HubertBimbotGauvin} was used, 
achieving charge identification with unit resolution up to $Z\sim 20$ and with a resolution lower than 5 charge units for $Z\sim 80$. 
In addition, energetic light ions ($Z<5$) punching through to the CsI(Tl) scintillators were isotopically identified by pulse-shape 
discrimination (PSD) of the fast and slow components of the light output. At forward angles (<45\textdegree), coherency checks 
between Si-CsI and CsI-PSD identification allowed to discriminate neutrons which undergo reactions with the nuclei of the CsI scintillator.

Resulting charge identification
thresholds are around 0.5\Mev for the lightest fragments ($Z\sim10$) and 1.5$-$2\Mev for the
heaviest ($Z\geq50$) (see Fig.1 of \cite{Frankland2001SingleSource}). This means that slow-moving ($\lesssim2$\,cm/ns) heavy ions such 
as target-like fragments from the least dissipative binary inelastic reactions have a very low probability of correct identification, 
being stopped in, or only just punching through, the IC. Nonetheless, a minimum atomic number can be estimated for such products based on 
$\Delta E_{\text{IC}}$, which allows to exclude events where such fragments are present from the analysis.

\subsection{Event selection \label{sc:selection}}
In this analysis, we considered only kinematically complete (well-detected) events with 
three identified heavy fragments ($Z > 10$) in the exit channel.
To select such a set of events, we consider the total charge detected in each event (\ztot) and the angle $\theta_{\text{flow}}$ which characterises the global orientation of each event with respect to the beam axis \cite{Frankland2001SingleSource}. Here, the kinetic energy tensor \cite{Cugnon1983Global} used to determine $\theta_{\text{flow}}$
was built using the three detected fragments in each event.

Event-by-event correlations between these two global variables
are presented in Fig.\ref{fig:sel}(a$-$d) for different beam energies. It should be reminded that in order to build these correlations, we require the detection and identification of three heavy fragments ($Z>10$) in coincidence, therefore not all reactions are represented: most notably, slow target-like fragments have a very low probability of correct identification (see Sec.\ref{sc:experiment}), therefore the least dissipative reactions are under-represented with respect to collisions in which a significant momentum transfer occurs. This bias is more noticeable the lower the beam energy, as then only the most dissipative binary collisions can impart sufficient momentum to target-like fragments for them to be identified. Conversely, the detection efficiency is highest and less dependent on beam energy for collisions with a full momentum transfer, \textit{i.e.} fully-damped binary collisions or fusion. 

The $\ztot-\theta_{\text{flow}}$ correlations in Fig.\ref{fig:sel}(a$-$d) show that three fragment events are dominated by two contributions with different kinematical properties, 
whatever the beam energy.
The first contribution, for which $\ztot\sim60-80$, has a strongly forward-peaked $\theta_{\text{flow}}$ distribution, with most events being oriented in the beam direction.
It can be seen (symbols in Fig.\ref{fig:sel}(a$-$d)) that the relative proportion of this contribution increases with the beam energy. 
These two observations are consistent with what is expected for most reactions proceeding by a dissipative binary collision in the first step, 
for which target-like fragments have a small probability to be identified (see previous paragraph).

The second contribution corresponds to a strong peak
in the $\ztot$ distribution around $\ztot\sim90-100$ (we recall that the total charge of projectile and target is $104$). This contribution populates all $\theta_{\text{flow}}$ angles, and it should be noted that the total detected charge is independent of the orientation of events in this case.
Fig.\ref{fig:sel}(e$-$h) shows the flow angle distributions for
all 3-fragment events (full symbols) or only the contribution with $\ztot>90$ (open symbols).
The effect of this selection is to effectively suppress the forward-peaked anisotropy in the full distributions which is associated with the low-$\ztot$ contribution.
The remaining events 
have a quasi-isotropic distribution of $\theta_{\text{flow}}$. The reduced yield at the most forward angles for 12 and 15 \Mev (Fig.\ref{fig:sel}(e$-$f)) can be ascribed to lower detection efficiency (no IC-Si telescopes at laboratory angles $<3^o$).
The total kinetic energy of the fragments for these events is independent of the flow angle, which indicates that they have the same degree of dissipation. 
Therefore the events selected with $\ztot>90$ which will be used in the following are compatible with reactions where the first step is either a fully-damped 
deeply-inelastic collision, quasifission, or fusion-fission. The associated measured cross-section for the selected events, calculated using the integrated 
beam-current and corrected for acquisition dead time, is almost independent of bombarding energy and represents $\sim50$mb. We will return to the exact nature 
of these reactions later.

\begin{center}
\begin{figure}[ht]
     \includegraphics[width=0.9\linewidth]{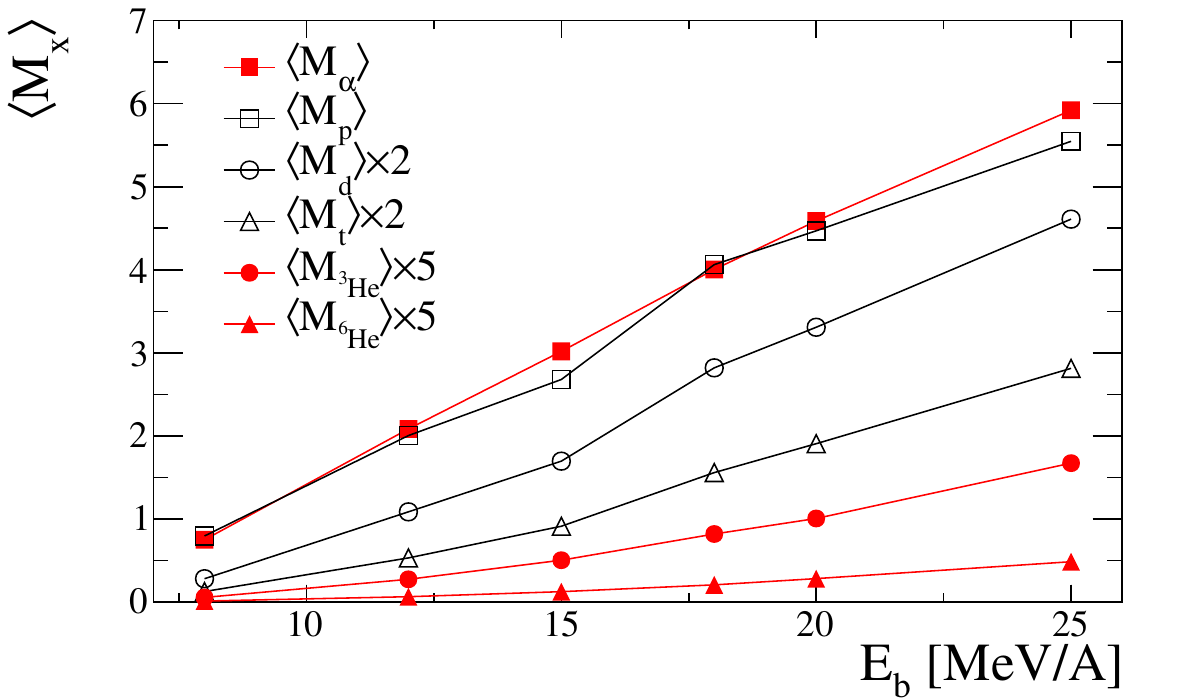}
\caption{Evolution of the light charged particle multiplicities as a function of the beam energy for \Xesn central collisions.}
\label{fig:mult}
\end{figure}
\end{center}

The evolution of the average multiplicity of light charged particle detected in coincidence with the three heavy fragment
as a function of the beam energy is diplayed on Fig. \ref{fig:mult}. Whatever the nature of the emitted particle,
the average multiplicity increases quasi-linearly with increasing bombarding energy.

\section{From sequential to simultaneous break-up}
\subsection{Qualitative evolution}
To begin the analysis of the three-fragment exit channels, we will show in a qualitative way the evolution of the decay process from two sequential splittings towards simultaneous fragmentation.
If two successive independent splittings occur, three possible sequences of splittings have to be considered.
For instance, in one sequence, the first splitting leads to a fragment of charge $Z_1$ and another fragment which,
later, undergoes fission leading to $Z_2$ and $Z_3$.
Let us call this sequence \textit{1}. The sequences \textit{2} and \textit{3} are readily deduced by circular permutation of the indices.

\begin{figure*}[!t]
     {\subfigure[ 12\,MeV/A\label{fig:dlzP:12}]{\includegraphics[width=0.24\linewidth]{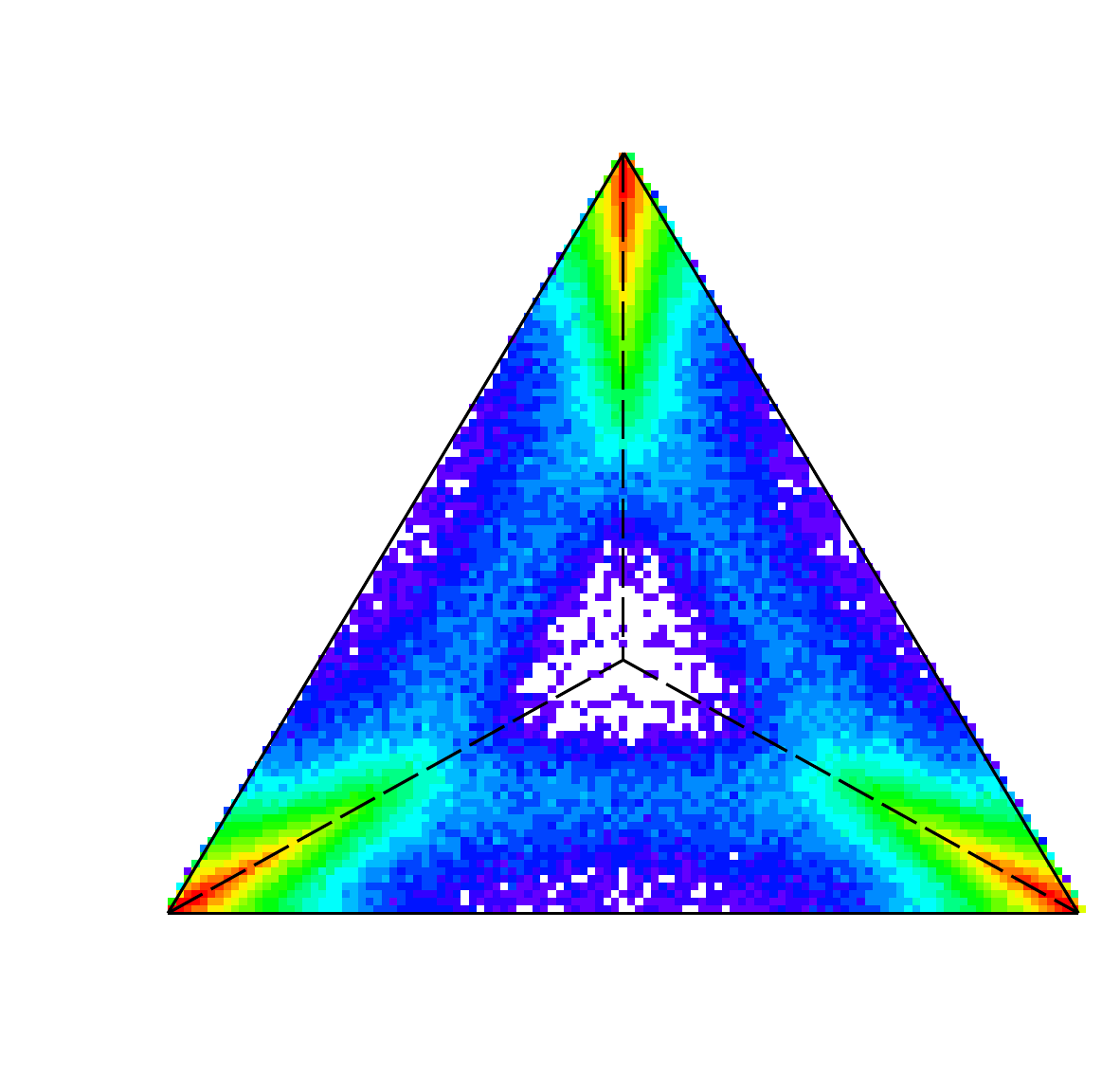}}}
     {\subfigure[ 15\,MeV/A\label{fig:dlzP:15}]{\includegraphics[width=0.24\linewidth]{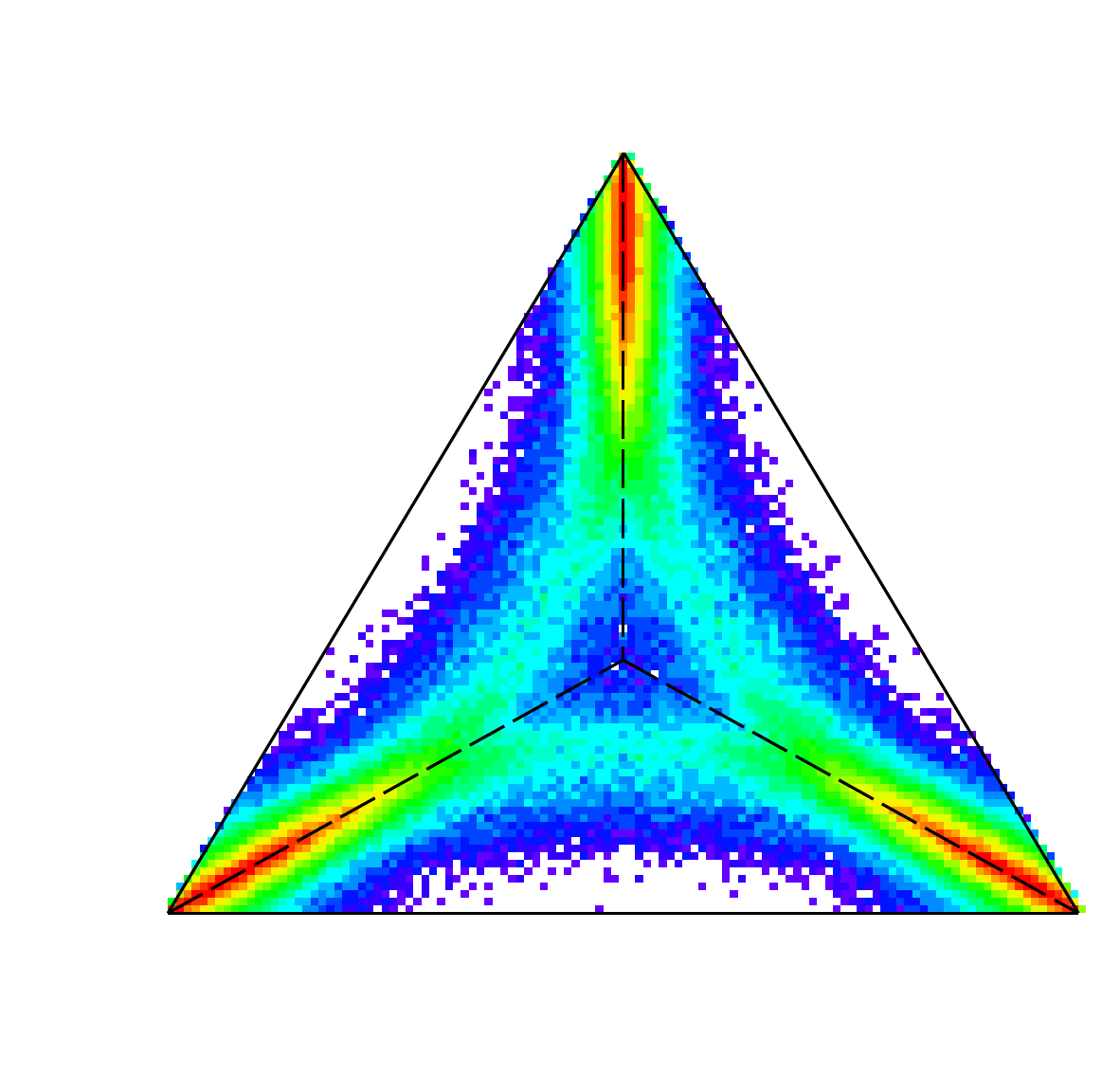}}}
     {\subfigure[ 20\,MeV/A\label{fig:dlzP:20}]{\includegraphics[width=0.24\linewidth]{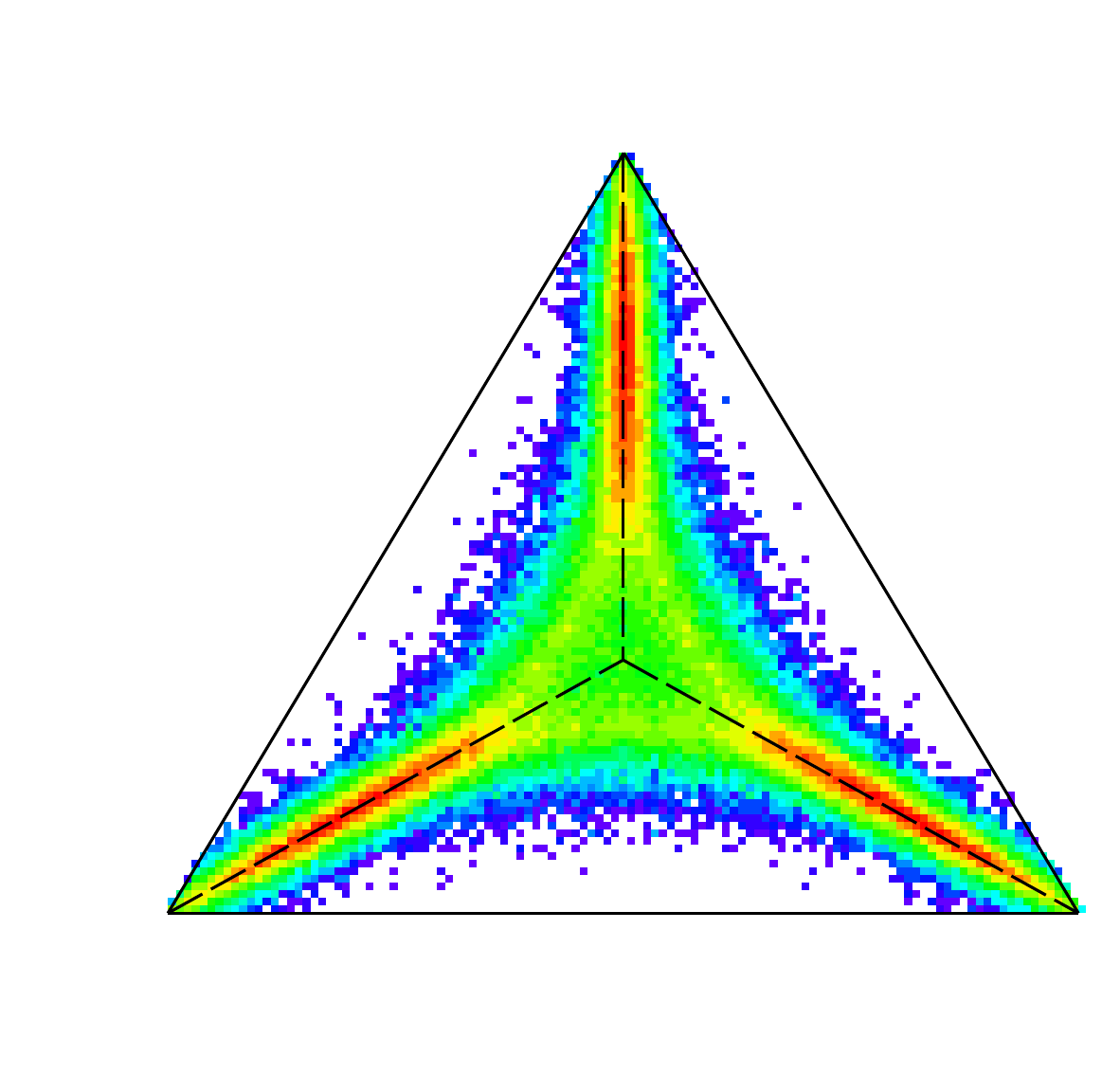}}}
     {\subfigure[ 25\,MeV/A\label{fig:dlzP:25}]{\includegraphics[width=0.24\linewidth]{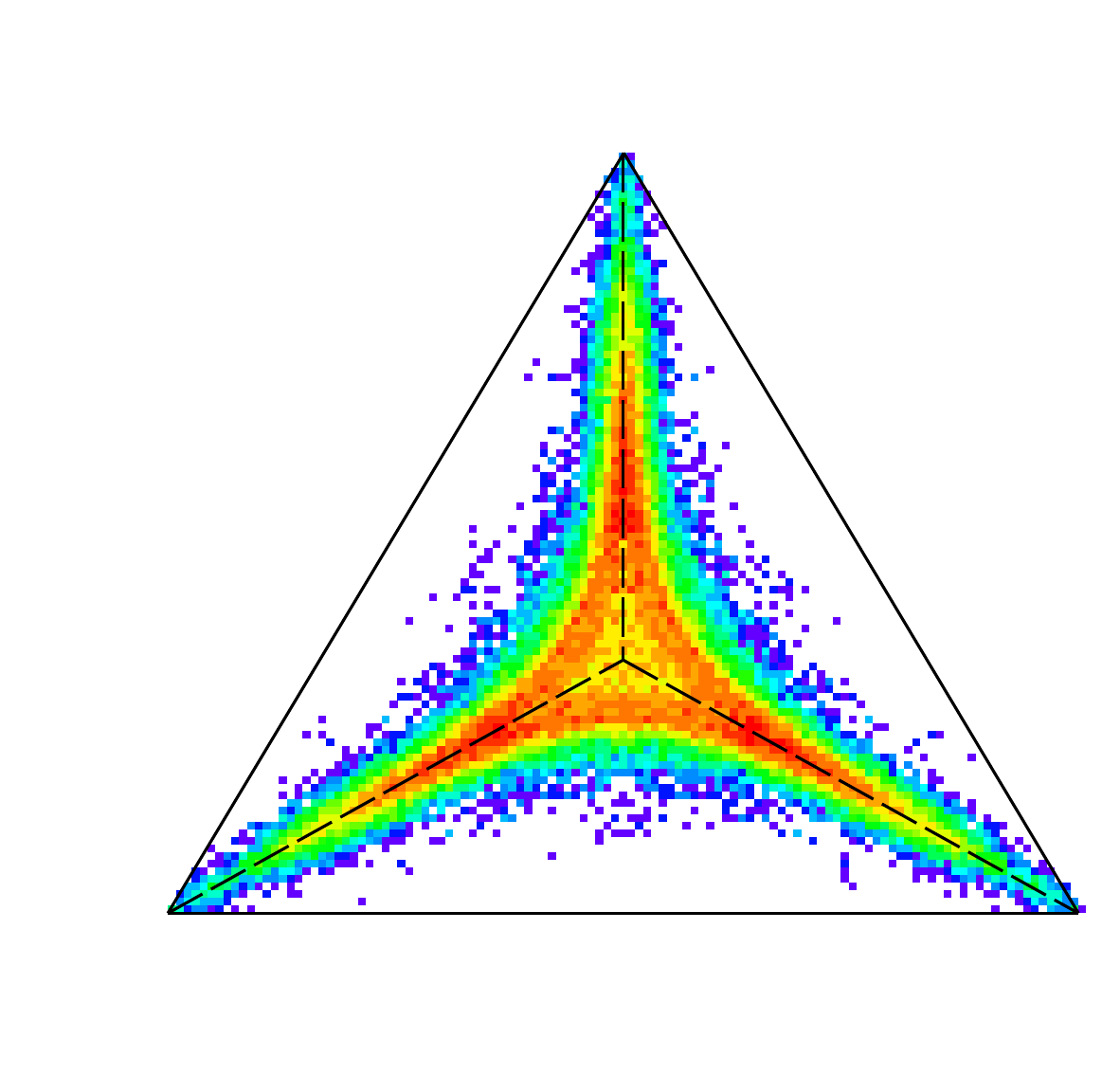}}}\hfill
\caption{(color online). Dalitz plot of $P_i$ (see text) for \Xesn central collisions at different beam energies.}
\label{fig:dlzP}
\end{figure*}

Bizard et al. \cite{Bizard1992413} proposed a method to show qualitatively the nature of the process.
To test the compatibility of an event with the sequence of splittings \textit{i}, we compare the experimental
relative velocities between fragments with those expected for fission.
For each event we build the following quantities:
\begin{align}
 P_{i} = (v_{i(jk)}^{exp} - v_{i(jk)}^{viola})^2 + (v_{jk}^{exp} - v_{jk}^{viola})^2 
\label{eq:Bizard}
\end{align}
where $i$ = \textit{1}, \textit{2}, \textit{3} the index of the fragment produced in the first splitting; 
$v_{\alpha\beta}^{exp}$ is the experimental relative velocity between fragments $\alpha$ and $\beta$ ; 
and $v_{\alpha\beta}^{viola}$ is the expected relative velocity for fission, taken from the Viola systematic \cite{PhysRevC.31.1550} 
extended to include asymmetric fission \cite{Hinde1987318}. 
The first (second) term in Eq.(\ref{eq:Bizard}) refers to the first (second) splitting. 
The three values of $P_i$ are calculated for each event and represented in Dalitz plots (Fig.\ref{fig:dlzP}). 
\bf{In this diagram, the distance of each point from the three sides of the triangle are $a_1$, $a_2$ and $a_3$; with
$a_i = P_i/(P_1+P_2+P_3)$. Therefore, the population of the Dalitz plot reflects the relative values of $P_1$, $P_2$, and $P_3$.}

At 12\Mev bombarding energy (Fig.\ref{fig:dlzP:12}), events populate mainly three branches parallel to the edges of 
the Dalitz plot, which correspond to the three sequences of sequential break-up ($P_i \ll P_j, P_k$). 
Simultaneous break-up events would be located close to the centre of this plot ($P_i \sim P_j \sim P_k$), where few events are observed.
\bf{The strong accumulations of events on the corners ($P_i \sim P_j \ll P_k$) correspond to the intersection of two sequential branches.
For these  particular kinematic configurations, two sequences cannot be disentangled.}
Consequently, Fig.\ref{fig:dlzP:12} shows that, for this energy, three-fragment events arise mainly from two sequential splittings.

It should be noted that at all bombarding energies, all three splitting sequences are nearly equally populated, 
showing that there is no biasing of the exit channels due to the experimental apparatus. 
This is very different to the previous study of 12.5\Mev \Snxe reactions by Gl\"assel \textit{et al.} where 
only one sequence was well-detected (see Fig.11 of \cite{Glassel1983Observation}), due to the two PPAC detectors 
used in that study being positioned for optimal detection of fission fragments of projectile-like fragments 
following deep-inelastic collisions \cite{Harrach1982Direct}. The advantage of $4\pi$ detection with high granularity 
means that we are able to study all possible exit channels without such bias.

With increasing beam energy (Fig.\ref{fig:dlzP}(b$-$d)), the three branches are still present but become closer and closer to the centre of the Dalitz plot. 
This indicates that fragment production becomes more and more simultaneous with increasing beam energy, and the deexcitation process 
evolves continuously from two sequential splittings towards simultaneous fragmentation.

In the following sections we will quantify this effect by measuring the time $\delta t$ between the two splittings.
First we must determine, event by event, in which order fragments have been produced.

\subsection{Sequence of splittings}
To establish the sequence of splitting event by event, we start from the hypothesis that fragments are produced sequentially, 
which was shown in the previous paragraph to be reasonable at least at the lowest beam energies (Fig.\ref{fig:dlzP}).
As mentioned above, three sequences of splittings have to be considered. 
In each possible sequence, one pair of fragments is the result of the second splitting, and should therefore have a relative velocity 
close to that expected for fission \cite{PhysRevC.31.1550, Hinde1987318}. 
Therefore, to identify the sequence of splittings event by event, we need only to find the pair with the most fission-like relative velocity
and we trivially deduce that the remaining fragment resulted from the first step.
This procedure amounts to computing, for each event, the three following quantities:
\begin{align}
 p_{i} = (v_{jk}^{exp} - v_{jk}^{viola})^2, 
\label{eq:Moi}
\end{align}
which corresponds to the second term of Eq.\eqref{eq:Bizard}.
The lower the value of $p_i$, the larger the probability of the considered event to have been generated by the sequence of splittings \textit{i}.
In each event, the smallest value of $p_{i}$ determines the sequence $i$ of splittings. 
This procedure has been tested on simulated three-fragment break-up events (see Appendix \ref{an:eff}) and was found to be, at worst, 66\% efficient in the most pessimistic scenario.
This efficiency is increased to 83\% when limiting to the angular range used
to extract the inter-splitting time (see Sec.\ref{sc:inter-splitting-time}).

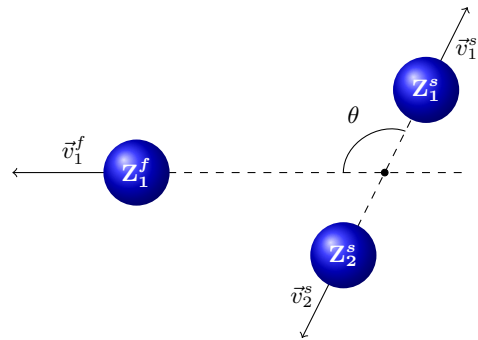
\begin{figure}[h]
\center
\begin{tikzpicture}[scale=1.1]
\path [-,thin,opacity=1,color=black,dashed,above] (2+1,4) edge node {} (6.+1,4);
\path [-,thin,opacity=1,color=black,dashed,above] (4.5+1,3) edge node {} (5.5+1,5);
\draw [color=black,opacity=1, fill=black] (5+1,4) circle (0.04);
\path [->,thin,opacity=1,color=black,above] (2+1,4)   edge node {$\vec{v}^{f}_{1}$} (0.5+1,4);
\path [->,thin,opacity=1,color=black,right] (5.5+1,5) edge node {$\vec{v}^{s}_{1}$} (6+1,6);
\path [->,thin,opacity=1,color=black,left] (4.5+1,3) edge node {$\vec{v}^{s}_{2}$} (4+1,2);
\shade[ball color = blue,opacity=1] (2+1,4) circle (0.4);
\shade[ball color = blue,opacity=1] (5.5+1,5) circle (0.4);
\shade[ball color = blue,opacity=1] (4.5+1,3) circle (0.4);
\path [-,thin,opacity=1,color=black] (4.5+1,4) edge [out=90, in=160]  (5.25+1,4.5);
\node [color=black,left] at (4.8+1,4.7) {{$\theta$}};
\node [color=white] at (2+1,4)    {\bf{Z$^f_1$}};
\node [color=white] at (5.5+1,5.) {\bf{Z$^s_1$}};
\node [color=white] at (4.5+1,3)  {\bf{Z$^s_2$}};
\end{tikzpicture}
\caption{Definition of the relevant kinematic observables for the three-fragment exit channel, in the rest frame of the intermediate system Z$^f_{2}$.}
\label{fig:obs}
\end{figure}

Once the sequence of splittings is known event by event, fragments can be sorted according to their order of production and
the intermediate system can be reconstructed. Let us now call Z$^f_{1}$ and Z$^f_{2}$, the two nuclei 
coming from the first splitting. The fragment Z$^f_{2}$ breaks in Z$^s_{1}$ and Z$^s_{2}$ during the second step (see Fig.\ref{fig:obs}).

\begin{table}[h!]
\center\begin{tabular}{r c c c c c c}
\hline
	 & $\langle \text{Z}_{tot}\rangle$ & $\langle \text{Z}_{src}\rangle$& $\langle \text{Z}^f_{1}\rangle$ & $\langle \text{Z}^f_{2}\rangle$& $\langle \text{Z}^s_{i}\rangle$  & $\sigma(\text{Z}^s_{i})$ \\
\hline
\hline

8\mev & 97.6 & 95.1 & 28.6 & 66.5 & 33.2 & 10.8\\
12\mev & 96.3 & 89.2 & 25.6 & 63.7 & 31.8 & 11.7\\
15\mev & 95.2 & 85.0 & 24.6 & 60.4 & 30.2 & 11.6\\
18\mev & 95.4 & 80.9 & 24.4 & 56.5 & 28.2 & 10.9\\
20\mev & 94.8 & 78.1 & 24.5 & 53.6 & 26.8 & 10.2\\
25\mev & 94.1 & 72.3 & 24.9 & 47.4 & 23.7 & 8.6\\
\hline
\end{tabular}
\caption{Mean charges of the two splittings and standard deviation of the charge distribution of the second splitting for \Xesn central collisions. 
$\langle \text{Z}_{src}\rangle = \langle \text{Z}^f_{1} + \text{Z}^f_{2}\rangle$ (see text) and the exponent $f$ ($s$) stands for the first (second) splitting.
\label{tab:charges}}
\end{table}

\begin{center}
\begin{figure}[h]
\includegraphics[width=0.9\linewidth]{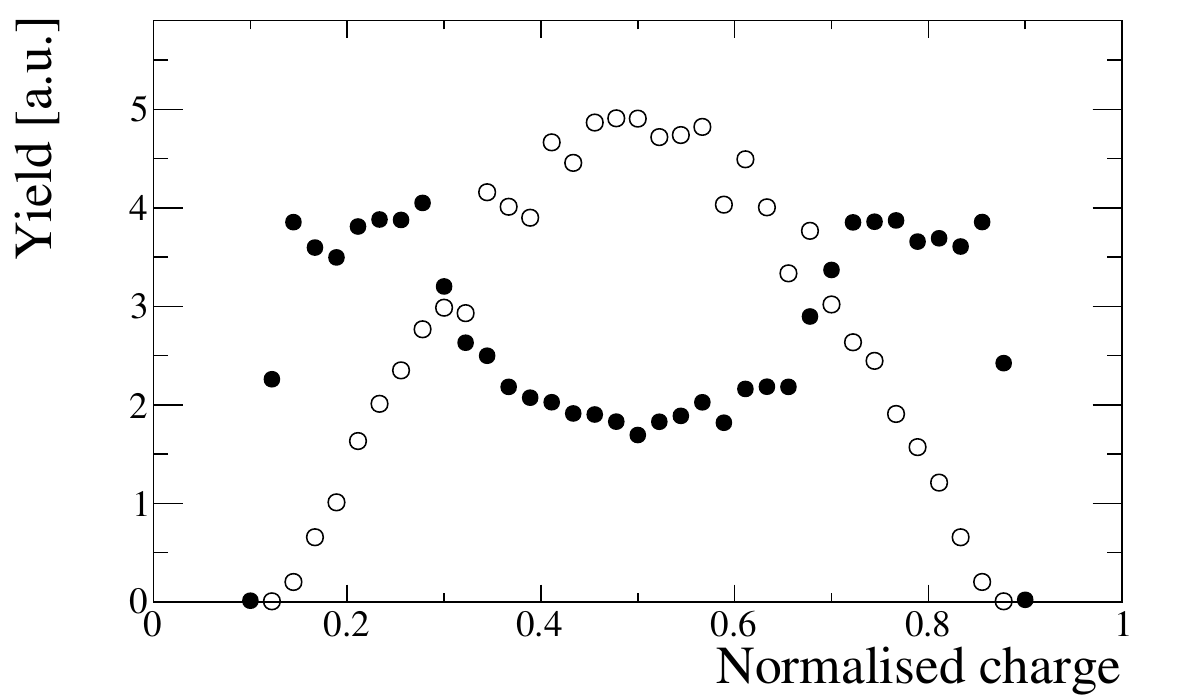}
\caption{Exemple of normalised charge distribution of the reconstructed first (full symbols) 
and second (empty symbols) splitting obtained for \Xesn reaction at 12\,\mev.\label{fig:zf12}}

\includegraphics[width=0.9\linewidth]{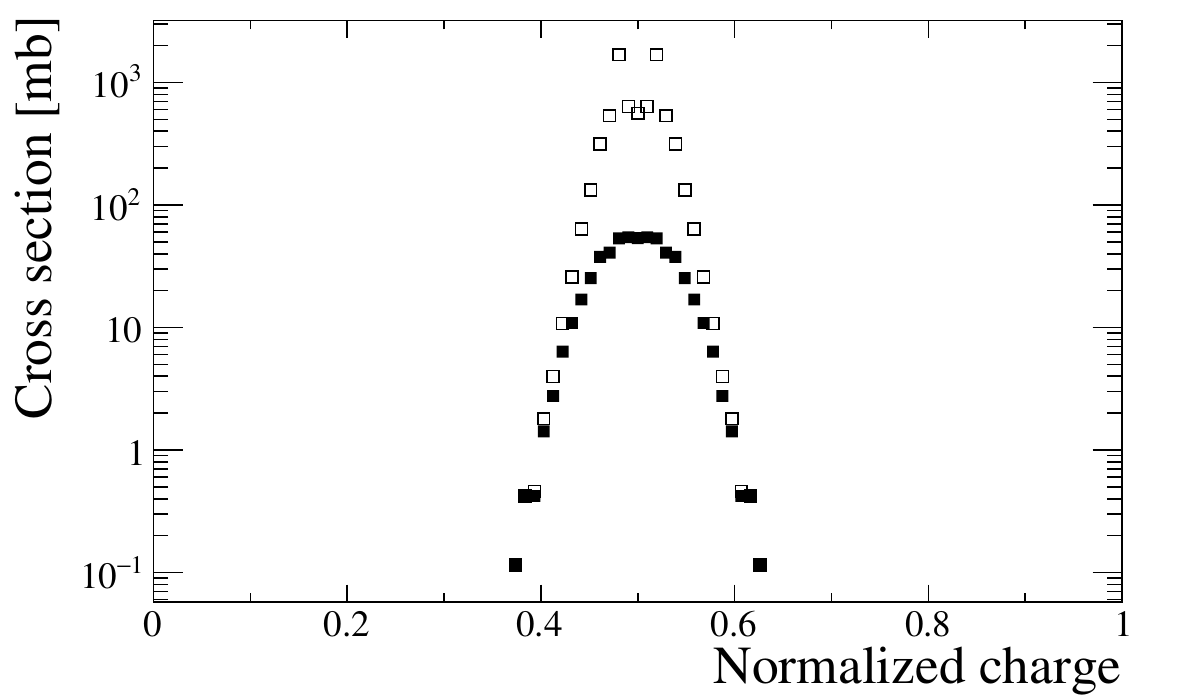}
\caption{Charge distribution for \Xesn reaction at 12\,\Mev simulated with Deep Inelastic Transfers (DIT) \cite{TassanGot1991Deep} for: (empty symbols) all deep inelastic events, 
(full symbols) the most dissipative deep inelastic events. (see Appendix \ref{an:dit} for details)}
\label{fig:zfdit}
\end{figure}
\end{center}

The mean atomic numbers of the fragments produced in each splitting are given in Tab.\ref{tab:charges}.
It can be seen that the first splitting is strongly asymmetric: indeed, the reconstructed fragment charge distribution for these two initial fragments, 
Z$^f_{1}$ and Z$^f_{2}$, presents two well-separated bumps (see Fig. \ref{fig:zf12}). 
It is then the larger of the two, $\text{Z}^f_{2}$, which subsequently undergoes a second splitting, giving a symmetric charge distribution peaked 
at $\text{Z}^f_{2}/2$ (see Fig. \ref{fig:zf12}). For the 12\Mev bombarding energy, this is in contradiction with the findings of \cite{Glassel1982Direct} where an asymmetric second 
fission was reported. Indeed, the authors of that work found a dependency of the mass asymmetry of the second step on the fission orientation: 
in the present work, the (a)symmetry of both splittings is independent of their relative orientation.

The mean total charge of the three fragments, $\langle \text{Z}_{src}\rangle$ in 
Tab.\ref{tab:charges}, decreases from 95 at 8\Mev bombarding energy to 69 at 25\mev. 
As the total detected charge for all events is fixed by the selection $\ztot>90$ (see Sec. \ref{sc:selection}), 
this decrease reflects the increasing multiplicity of emitted light charged particles with increasing bombarding energy (see Fig. \ref{fig:mult}), 
due to both preequilibrium emission \cite{Rosch1989Preequilibrium} and evaporation from the excited fragments or from any of 
the intermediate compound systems \cite{Hudan2003Characteristics}. It should be noted that as bombarding energy increases, 
the difference in the mean charge of the three final fragments becomes smaller, and at 25\Mev all three fragments have a mean atomic number $Z\sim 23$.

\bf{
Although the charge/mass asymmetry of the first splitting (Fig. \ref{fig:zf12}) seems at first to be counter-intuitive, it can be explained
considering the probability of sequential fission. Indeed, a symmetric first splitting will have little probability of
sequential fission as the fission barriers of both fragments will be large. On the other hand, if the first splitting is asymmetric,
the heaviest fragment will have a smaller fission barrier resulting in a larger sequential fission probability. 
Therefore the selection of three body events preferentially select out asymmetric initial splitting,
whatever the underlying reaction mechanism (fusion/fission, deep inelastic, quasi-fission). 

The charge distribution of the first splitting is also very broad. This is not consistent with a binary reaction scenario as a first step. 
Figure \ref{fig:zfdit} shows the charge distribution of binary events
simulated with the Deep Inelastic Transfers (DIT) model \citep{TassanGot1991Deep} for the \Xesn at 12\Mev reaction (see Appendix \ref{an:dit} for details). 
The obtained charge distribution, even for the most dissipative events,
is too narrow to give rise to the reconstructed charge distribution of the first splitting (full symbols on Fig. \ref{fig:zf12}): 
the most asymmetric splitting obtained in the calculation ($Z_{\mathrm{TLF}}=39$, $Z_{\mathrm{PLF}}=65$) has an associated cross-section of 114 $\mu$b
while the mean charge partition experimentally observed is ($Z_{1}^f=26$, $Z_{2}^f=64$) (see Tab.\ref{tab:charges}). }
Therefore in the following we will assume that the first step of the reactions leading to three-fragment exit channels 
is the formation of composite systems with $Z\sim 80\-- 100$ , which subsequently undergo fission (first splitting). 

\begin{figure*}[!t]
     {\subfigure[ 12\,MeV/A\label{fig:thDist:12}]{\includegraphics[width=0.24\linewidth]{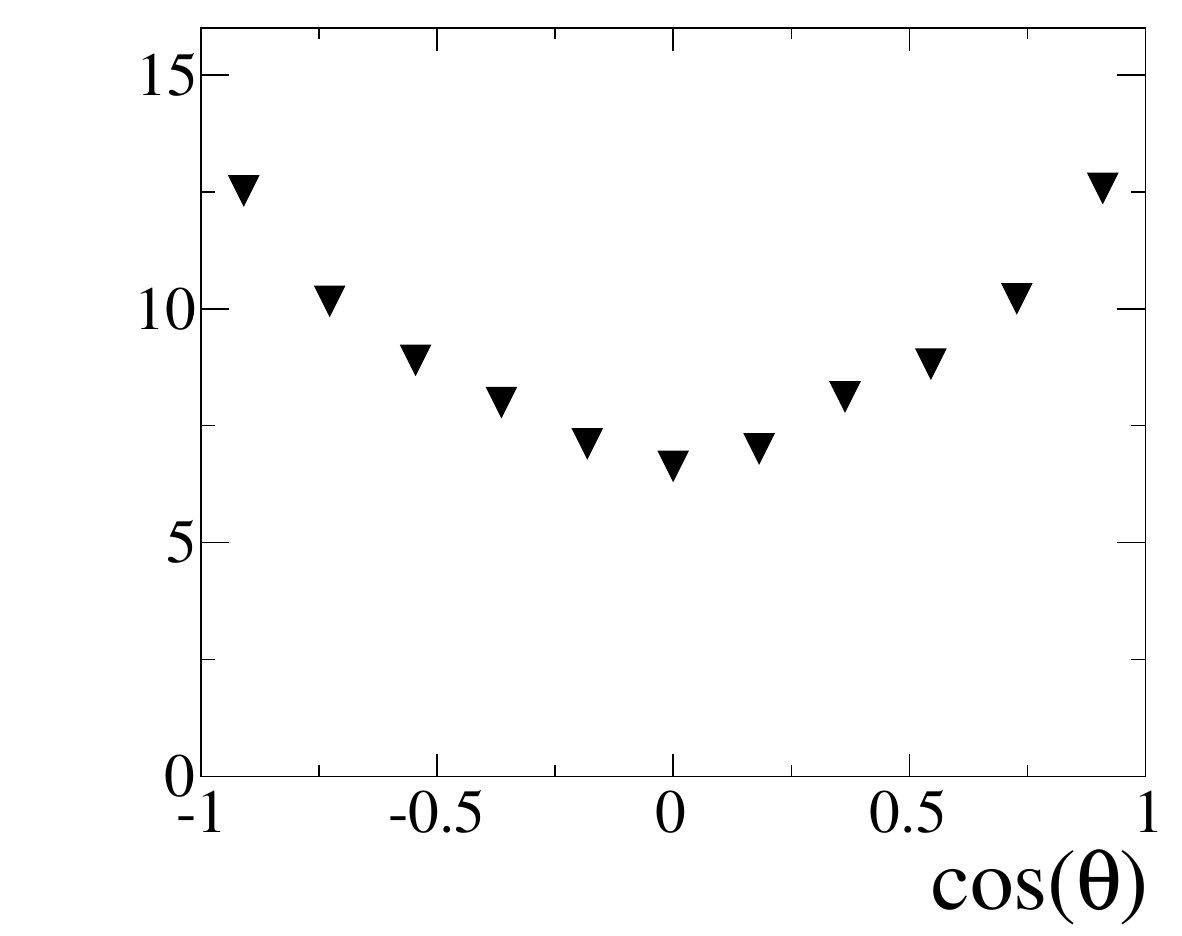}}}
     {\subfigure[ 15\,MeV/A\label{fig:thDist:15}]{\includegraphics[width=0.24\linewidth]{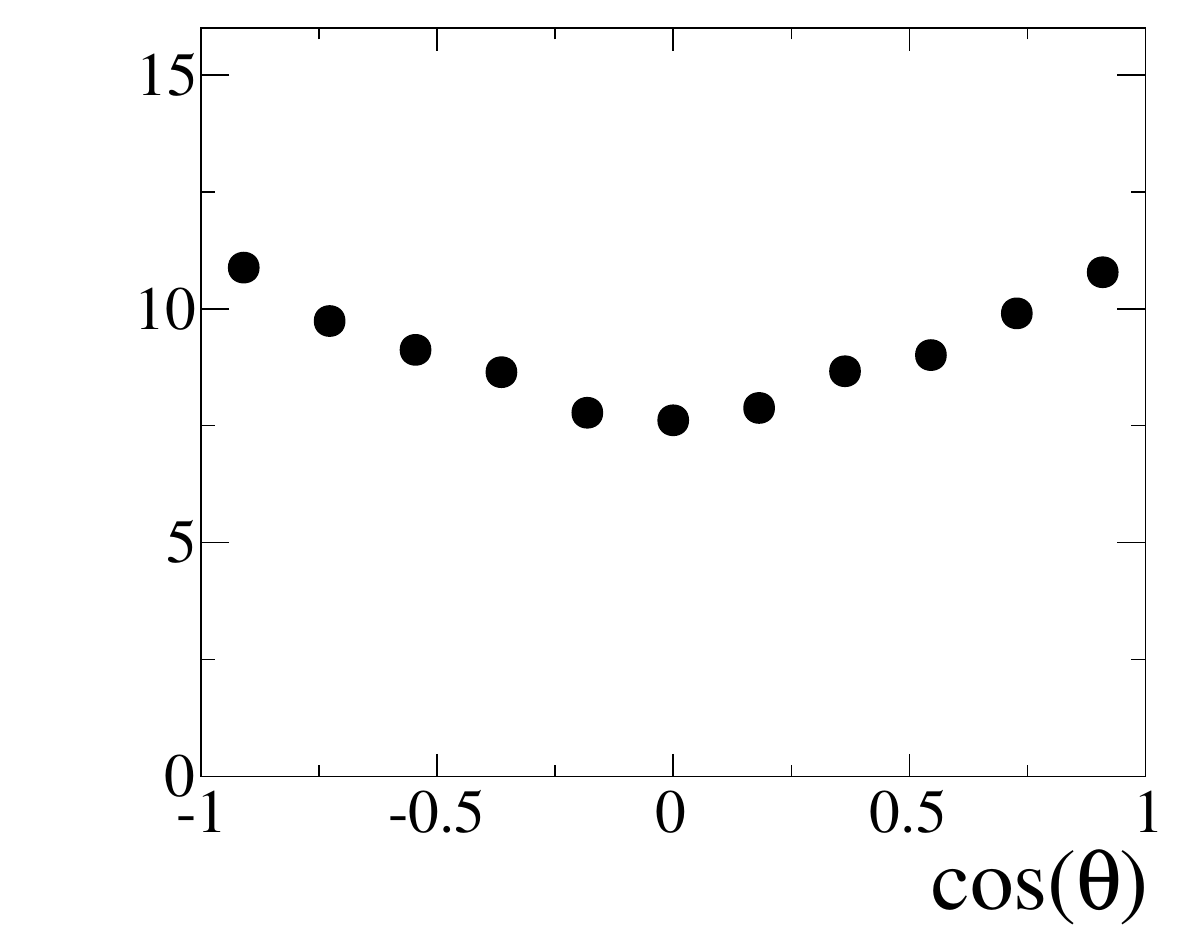}}}
     {\subfigure[ 18\,MeV/A\label{fig:thDist:18}]{\includegraphics[width=0.24\linewidth]{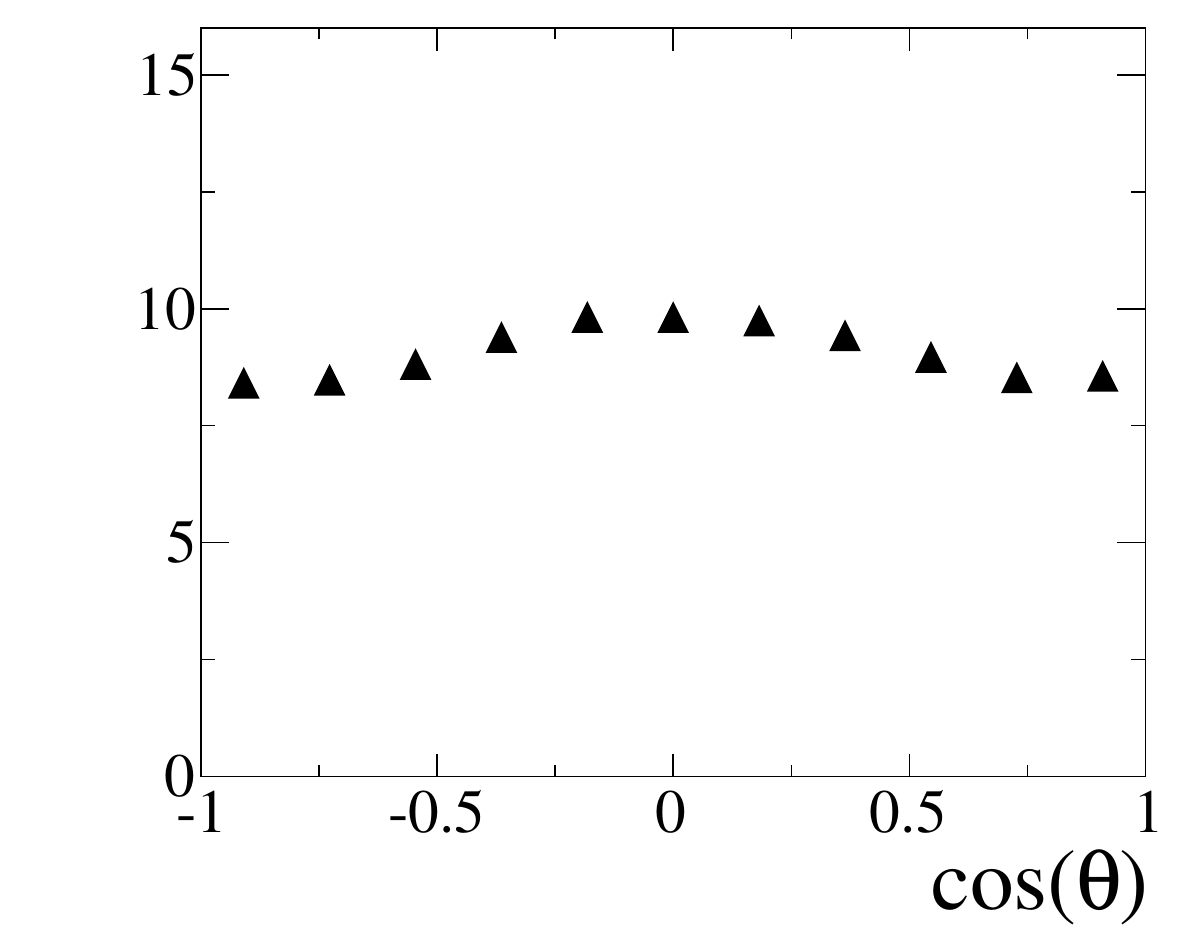}}}
     {\subfigure[ 20\,MeV/A\label{fig:thDist:20}]{\includegraphics[width=0.24\linewidth]{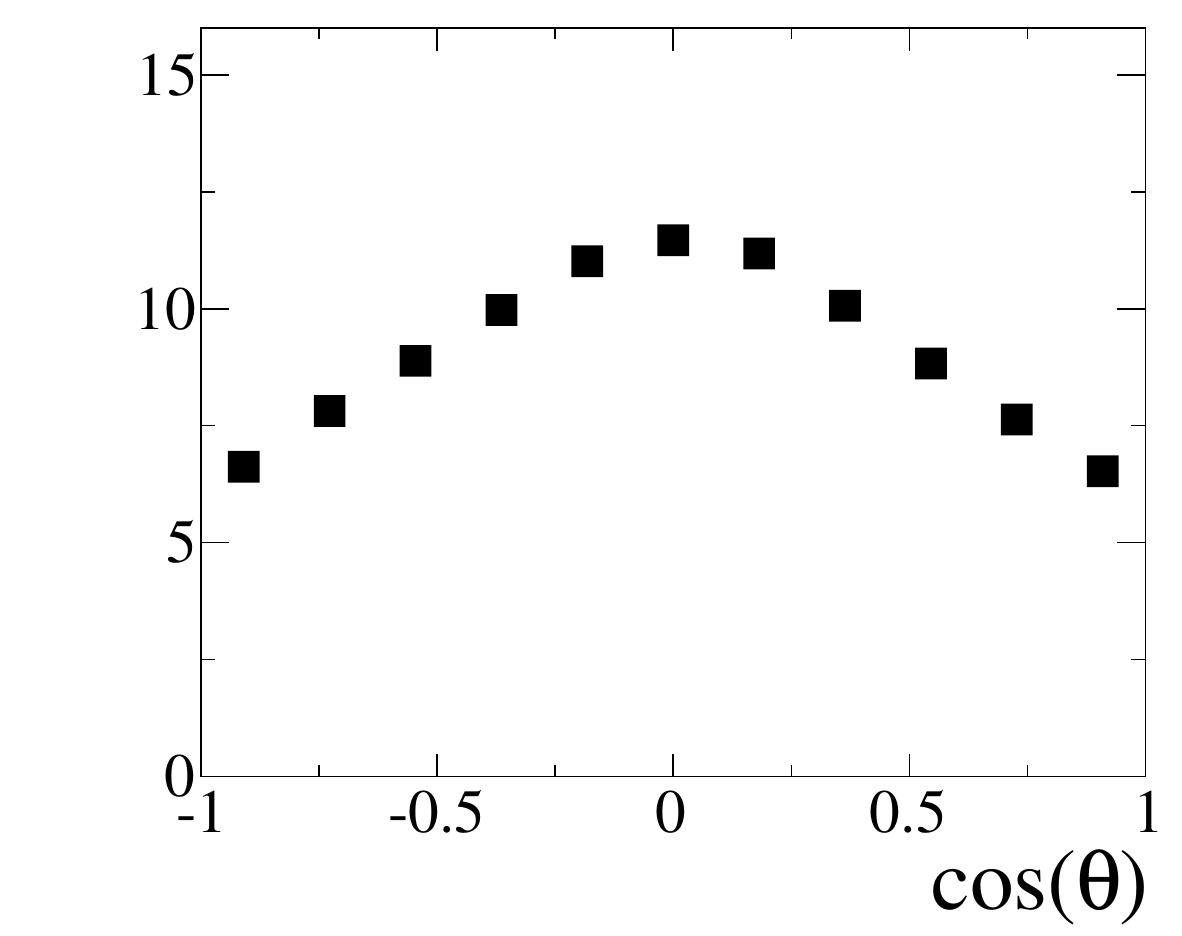}}}
     {\subfigure[ 12\,MeV/A\label{fig:vt:12}]    {\includegraphics[width=0.24\linewidth]{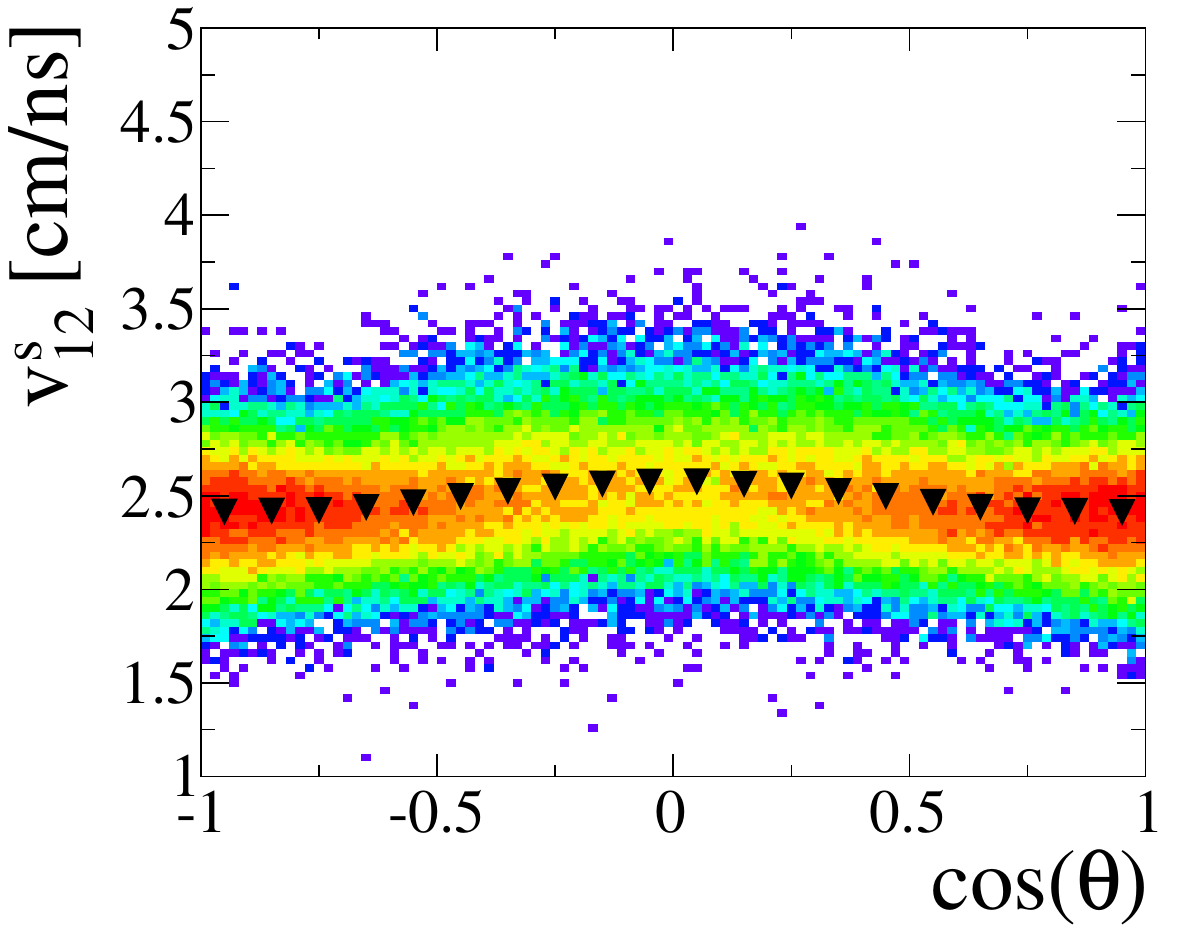}}}
     {\subfigure[ 15\,MeV/A\label{fig:vt:15}]    {\includegraphics[width=0.24\linewidth]{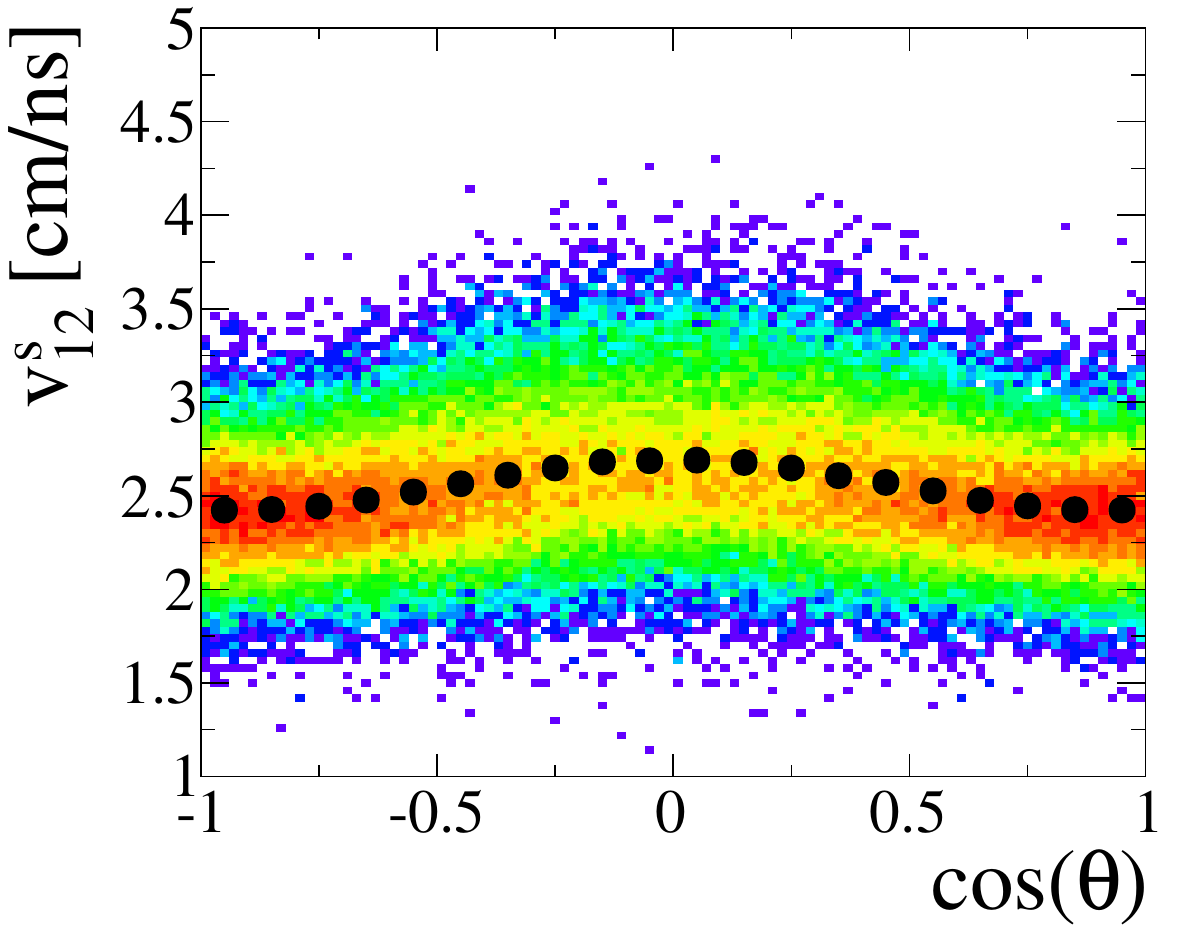}}}
     {\subfigure[ 18\,MeV/A\label{fig:vt:18}]    {\includegraphics[width=0.24\linewidth]{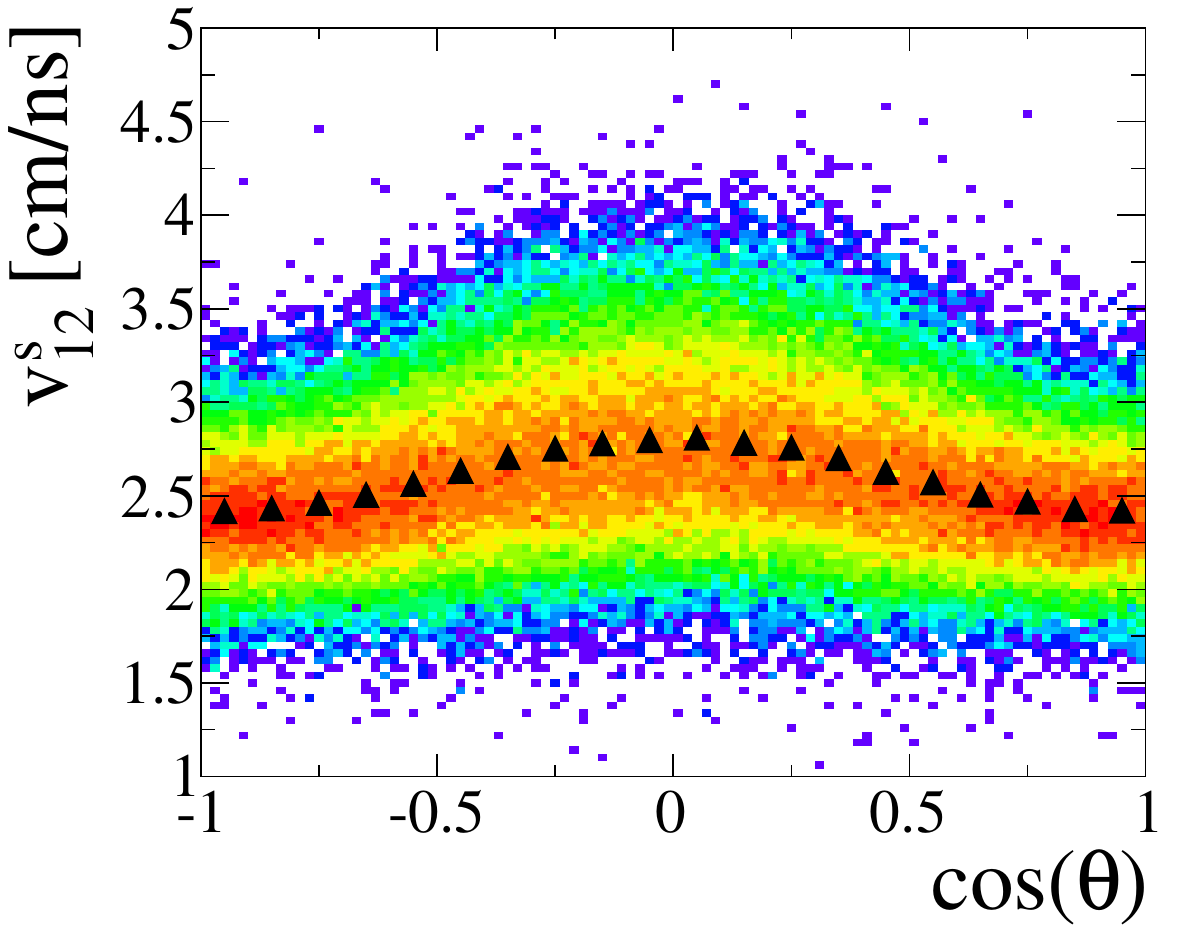}}}
     {\subfigure[ 20\,MeV/A\label{fig:vt:20}]    {\includegraphics[width=0.24\linewidth]{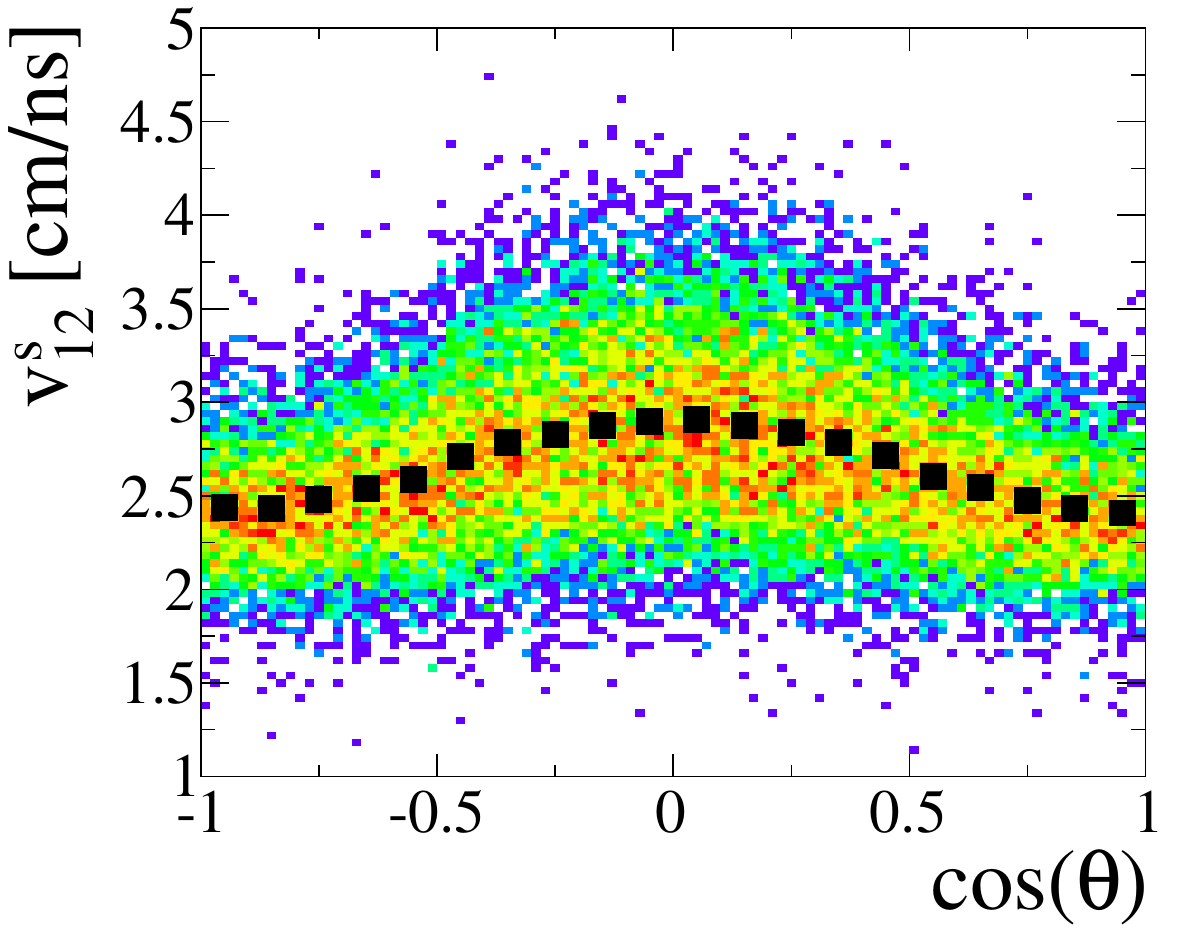}}}
\caption{(color on-line). 
(a$-$d) Distribution of the inter-splitting angle $\theta$.
(e$-$h) Correlation between the inter-splitting angle $\theta$ and the relative velocity of the second splitting $v^{s}_{12}$.
Vertical error bars are smaller than the size of the points
}
\label{fig:vthist}
\end{figure*}

\subsection{Angular distribution}
We now characterize the two splittings by their relative orientation.
Fig.\ref{fig:vthist}(a$-$d) show the distribution of the angle $\theta$ between the two separation axes (see Fig.\ref{fig:obs}) for different beam energies.
At the lowest beam energies, the angular distribution 
presents a ``U'' shape (Fig.\ref{fig:vthist}(a)) which is characteristic of fission of an equilibrated system \cite{Vigdor1980Anisotropy} with angular momentum. 
With increasing beam energy, the angular distribution flattens (Fig.\ref{fig:vthist}(b$-$c)) and then
develops a maximum centered on $\theta\sim90$\textdegree (Fig.\ref{fig:vthist}(d)), leading to anisotropy values $\text{\anis}<1$. 
The latter behaviour is unexpected for an isolated fissioning system and suggests the presence of large final state interactions,
where the Coulomb field of the first emitted fragment focuses the other two more perpendicularly to the first separation axis.
It is clear that the presence of such an anisotropy
requires the second splitting to take place at a distance from the first emitted fragment of the same order of magnitude as the distance between the 
centers of the fissioning fragments at scission. 
It is these Coulomb proximity effects which we will now use in order to deduce the time interval between the two splittings.

\subsection{Inter-splitting time \label{sc:inter-splitting-time}}
To estimate the mean inter-splitting time ($\delta t$), we used the correlation between the inter-splitting angle $\theta$ 
and the relative velocity of the second splitting: $v^{s}_{12} = \parallel \vec{v}^{s}_1 - \vec{v}^{s}_2 \parallel$ (see Fig.\ref{fig:obs}).
In fact, for long inter-splitting times the second splitting occurs far from the first emitted fragment.
The relative velocity $v^{s}_{12}$ is then only determined by the mutual repulsion between
Z$^s_{1}$ and Z$^s_{2}$ and should not depend on the relative orientation of the two splittings. 
However, for short inter-splitting time the second splitting occurs close to the first emitted fragment.
The relative velocity $v^{s}_{12}$ is modified by the Coulomb field of Z$^f_{1}$ and depends on the relative 
orientation of the two splittings. In this case, $v^{s}_{12}$ should present a maximum for $\theta=90$\textdegree.
We used this Coulomb proximity effect as a chronometer to measure the inter-splitting time $\delta t$.

\begin{center}
\begin{figure}[ht]
     \includegraphics[width=0.99\linewidth]{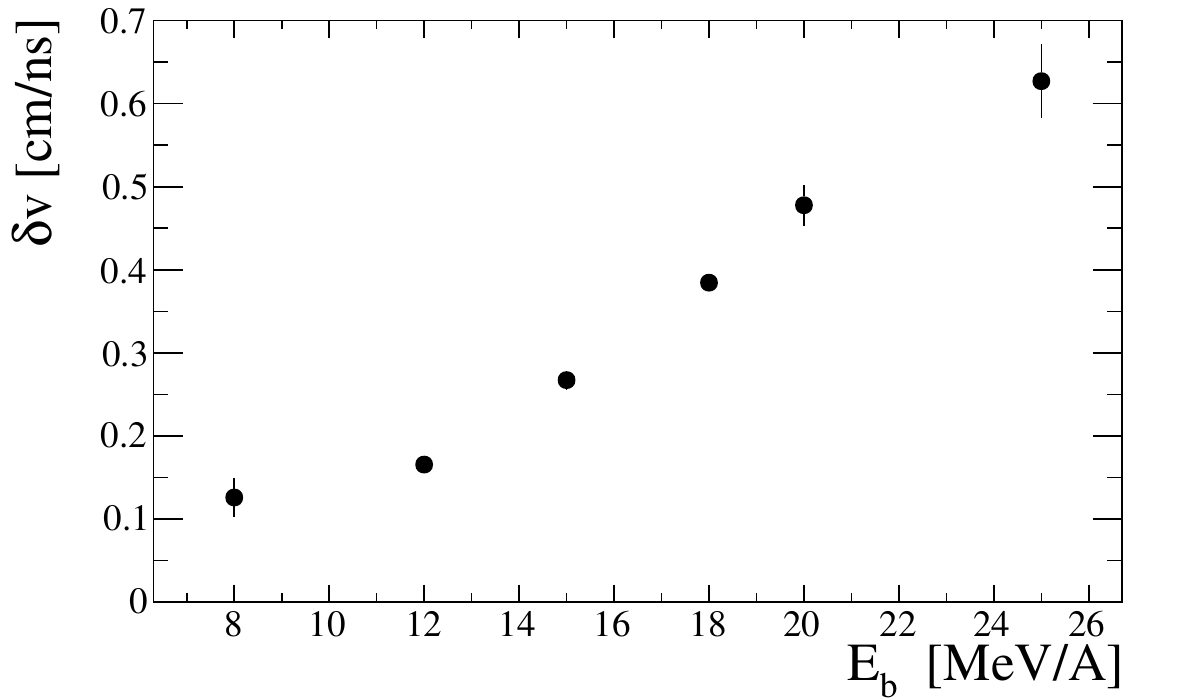}
\caption{Evolution of the Coulomb distortion parameter $\delta v$ as a function of the beam energy for \Xesn central collisions.}
\label{fig:vt}
\end{figure}
\end{center}

Experimental correlations between $v^{s}_{12}$ and $\theta$ are presented in Fig.\ref{fig:vthist}(e$-$h), for different beam energies.
These correlations present a maximum at $\theta\sim90$\textdegree,~which is more pronounced as the beam energy increases.
We quantify this effect by the Coulomb distortion parameter $\delta v = v^{s}_{12}(90\text{\textdegree}) - v^{s}_{12}(0\text{\textdegree})$.
In practice, $\delta v$ is computed in the ranges $| \cos(\theta) |<0.05$ ($\theta\sim90\text{\textdegree}$) and $| \cos(\theta)|>0.9$ ($\theta\sim0\text{\textdegree}$).
In this angular range, the sequence identification procedure presents an efficiency of 83\% (see Appendix \ref{an:eff}).
$\delta v$ increases with the beam energy (Fig.\ref{fig:vt}), indicating that the second splitting occured closer and closer to the first emitted fragment.

\setcounter{subfigure}{0}

\begin{figure}[ht]
\begin{center}
\includegraphics[width=0.99\linewidth]{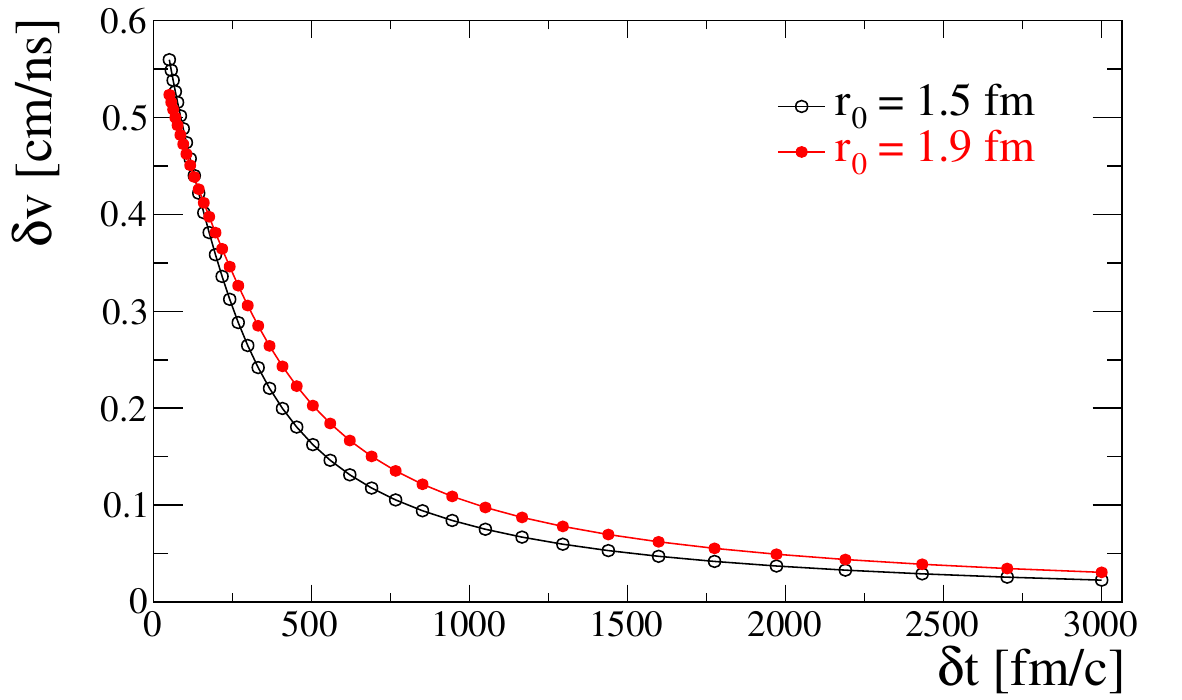}
\caption{Evolution of the Coulomb distortion parameter $\delta v$ as a function of the inter-splitting time $\delta t$ obtained from the Coulomb trajectory calculation.
The charges used correspond to that given in Tab.\ref{tab:charges} for 15\Mev beam energy.}
\label{fig:dvdt}
\end{center}
\end{figure}

To translate $\delta v$ in terms of inter-splitting time $\delta t$, we performed Coulomb trajectory calculations for point charges,
which simulate sequential break-ups using mean charges given in Tab.\ref{tab:charges}.
The initial conditions of the calculation were
chosen in order to reproduce the systematics of asymmetric fission \cite{Hinde1987318}: for each step the two fissioning fragments were separated by
a distance $d_{ij} = r_0(A_i^{1/3}+A_j^{1/3})$ with $r_0=1.9$ fm. 
$\delta v$ is then computed by varying $\delta t$ to get the calibration function presented in Fig.\ref{fig:dvdt}.

\begin{figure}[ht]
\includegraphics[width=0.99\linewidth]{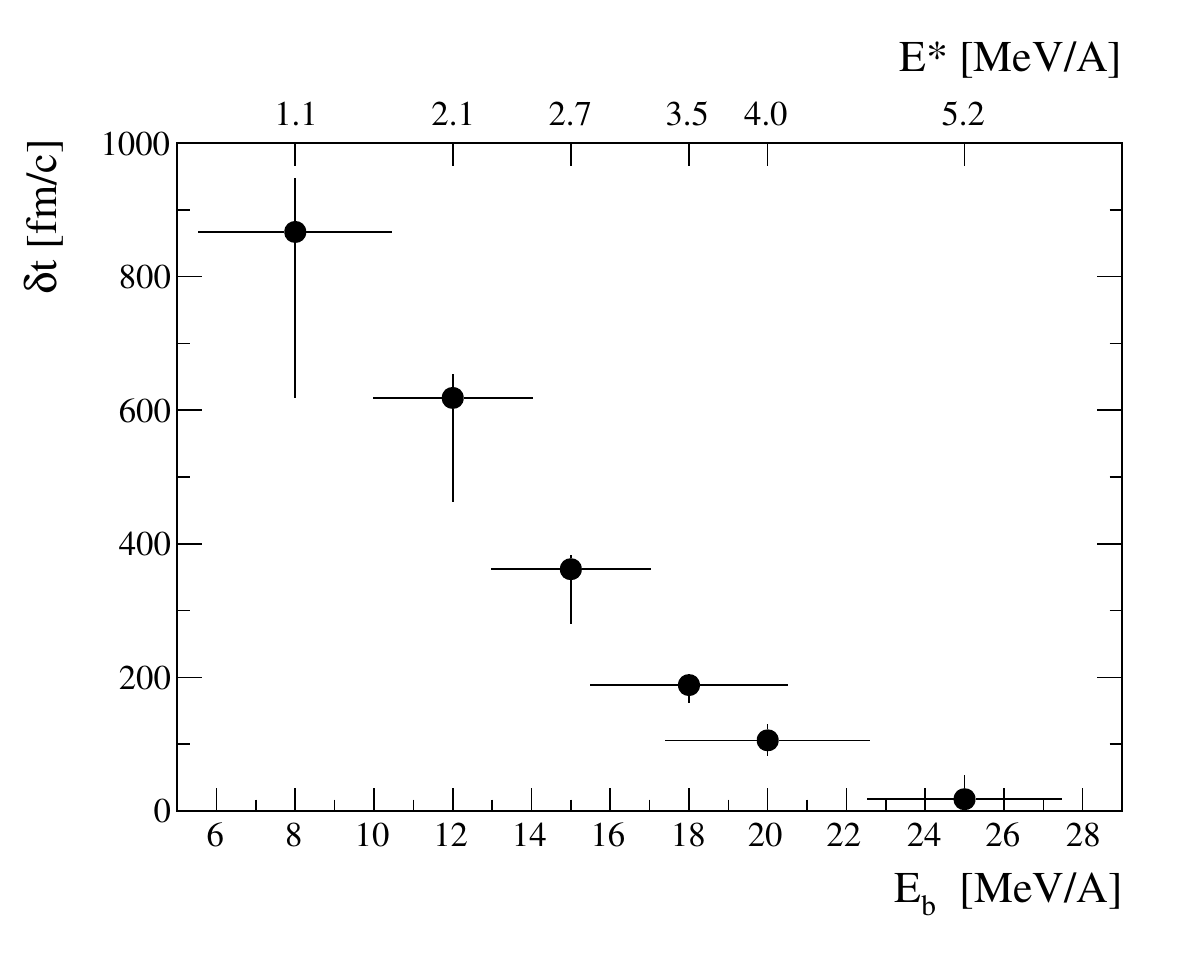}
\caption{Evolution of the mean inter-splitting time $\delta t$ as a function of the beam energy (lower scale) and the estimated excitation energy of 
the initial composite systems (upper scale) produced in \Xesn central collisions. Horizontal error bars refer to the upper scale.
\label{fig:dte}
}
\end{figure}

Finally, we obtained the evolution of the inter-splitting time as a function of the beam energy (Fig.\ref{fig:dte}). 
The vertical error bars in Fig.\ref{fig:dte} reflect the statistical uncertainties on the measurement of $\delta v$ (Fig.\ref{fig:vt}) 
and take into account variations of the initial conditions in the
trajectory calculations: $r_0 = 1.9-1.5$ fm (see Fig.\ref{fig:dvdt}). We verified that the experimental apparatus does not introduce significant systematic errors on the average values. 

A clear decrease of the inter-splitting time with increasing beam energy is observed in Fig.\ref{fig:dte}.
At 8 and 12\mev, the inter-splitting time \Dt is greater than 500\,fm/c ($1.7\times 10^{-21}$ s.). It shows that, for the lower beam energies, 
fragments arise from two successive splittings well separated in time, validating our starting hypothesis.
As the beam energy increases from 12\Mev to 20\mev, \Dt decreases monotonically from 600\,fm/c to about 100\,fm/c. 
At 25\mev, \Dt becomes compatible with 0 ($\text{\Dt} = 20 \pm 20$\,fm/c).
It reflects in fact here the sensitivity limit of the method.
Indeed, our trajectory calculations show that below $\delta t \sim 100$\,fm/c the two nuclei resulting from the first splitting 
do not have sufficient time to move apart beyond the range of the nuclear forces 
before the second splitting occurs. For such a short time, fragment emissions cannot be treated independently, 
and it is no longer meaningful to speak of a sequential process.
This inter-splitting time is reached around 20\mev.
It should be recalled that, concurrently with this decrease in the break-up time-scale, the mean charges of the three final fragments become more and more similar (see Tab.\ref{tab:charges}), culminating in the quasi-simultaneous production of three equal-sized fragments. In this case, one is justified in speaking of the onset of a multi-fragment break-up process which appears as a natural evolution of the sequential fission decay processes observed at lower energies. 

\section{Discussion \label{sc:discussion}}
Our results show that the three-fragment exit channel in central \Xesn collisions is compatible with
successive binary splittings of composite heavy systems with estimated atomic numbers $Z\sim 80\-- 100$. 
The mean lifetime of the second fission-like step becomes
shorter and shorter with increasing bombarding energy,
leading to a decay which is indistinguishable from simultaneous multi-fragment break-up
above 20\mev.

For each beam energy, the excitation energy of the initial composite system
has been estimated using a standard calorimetric procedure \cite{Cussol1993Chargedparticle,Marie199715,Bonnet20091}, including
the light charged particles detected in coincidence.
The mean values are given in the upper scale of Fig.\ref{fig:dte}.
At the lowest beam energies, where sequential fission is the dominant decay mode,
this energy has to be seen as an upper limit for the excitation energy
of the nucleus undergoing the second splitting, assuming that the excitation energy of the initial composite system is partitioned between the partners of the first scission;
however, at the highest energies this estimate corresponds directly to the excitation
energy of the system undergoing simultaneous three-body decay, and it gives the threshold energy for the onset of this process at $E^{*}\sim4$\mev.

The inter-splitting times reported in Fig.\ref{fig:dte} are in good agreement with fragment emission times extracted for
excited gold nuclei formed in $\pi^{-}+Au$ reactions \cite{Beaulieu2000Signals}
over the whole excitation energy range, although the mechanism forming the initial excited system is very 
different in these two reactions. Break-up times for similar-sized
nuclei formed in heavy-ion induced reactions \cite{Bougault1989Timescale, Louvel1994Rapid, Durand1995Nuclear, Bauge1993Observation}
show the same trend, but time scales for excitation energies
below 5\Mev are systematically larger than those of \cite{Beaulieu2000Signals},
and measurements from different reactions give widely varying results.
This discrepancy can be due to angular momentum or compression-expansion effects
which are negligible in hadron induced reaction \cite{Beaulieu2000Signals} 
but depend on the entrance channel in heavy-ion collisions \cite{Bonnet20091}.
This issue could be fixed with a systematic study of fragment emission times over a broad range of excitation energy and system size, and also
by extending the presented method to exit channels with four- and more fragments.

Compared to previous studies of three-fragment events for the \Snxe system at 12.5\Mev, we find an inter-splitting time $\delta t=2\times 10^{-21}$ s, 
which is of the same order of magnitude as in \cite{Harrach1982Direct} (factor of 2 greater), but the characteristics of each of the two sequential splittings 
are found to be very different in our analysis. In \cite{Glassel1982Direct} the authors concluded that the dominant mechanism was a deep-inelastic collision 
followed by an asymmetric and strongly aligned break-up of one of the two outgoing fragments, as has since been observed to dominate the reaction cross-section 
for heavy ion collisions at bombarding energies up to and around the Fermi energy \cite{Bocage2000Dynamical,Colin2003Dynamical,Filippo2005Dynamical}. 
On the contrary, we observe fully-relaxed and globally isotropic events with a small associated ($\sim50$mb) cross-section, for which the highly asymmetric 
first scission is incompatible with deep-inelastic or quasifission reactions, after which the heavier of the two primary fission fragments rapidly undergoes a second, 
symmetric, fission whose characteristic angular distribution only deviates from the statistical expectation due to Coulomb proximity effects.

In light of the preceding discussion, it seems clear to us that there is, in fact, no real contradiction between our analysis and that of 
\cite{Harrach1982Direct,Glassel1982Direct,Glassel1983Observation}: the three-fragment events in the two studies do not correspond to the 
same class of reactions. The experimental set-up of Gl\"assel \textit{et al.} was ``optimized for three-body coincidences arising from
the sequential fission of deep-inelastic collision fragments emitted into forward CM angles''
\cite{Glassel1983Observation}. On the other hand, the use of a $4\pi$ multidetector such as INDRA imposes no such \textit{a priori} bias on the studied reactions, and brings additional selectivity allowing to study low cross-section phenomena, typical of central collisions, which were previously unattainable.

\section{Conclusion}
In summary, we proposed a new chronometer which profit from Coulomb proximity effects observed in the three-fragment final state.
This is made possible thanks to highly exclusive measurements performed with INDRA. 
The originality of the method relies on the unambiguous determination of the sequence of splitting. 
This method is applied to probe the decay mechanism responsible for 
the three-fragment exit channel observed in \Xesn central collisions at bombarding energies from 8 to 25\mev.
We showed that these fragments arise from successive binary splittings occurring on shorter and shorter time scales.
The involved time scale becomes compatible with simultaneous three-fragment break-up  above $E^*=4.0\pm0.5$\mev,
which can be interpreted as the signature of the onset of multifragmentation.

\begin{acknowledgments}
The authors would like to thank Dominique Durand for useful discussions and 
the staff of the GANIL Accelerator
facility for their continued support during the experiments. D. G.
gratefully acknowledges the financial support of the Commissariat
\`a l'\'energie Atomique et aux \'energies alternatives and the Conseil R\'egional de Basse-Normandie.
The work was partially sponsored by the French-Polish agreements IN2P3-COPIN
(Project No.~09-136) and the Russian Foundation for Basic Research,
Research Project No.~13-02-00168 (Russia). 
\end{acknowledgments}

\appendix

\section{Efficiency of the sequence identification procedure}
\label{an:eff}

To test the validity of the proposed procedure of sequence identification, we simulated 
300 three-fragment break-ups using the experimentally-measured fragment charges, for each event measured at 12\Mev beam energy. 
The sequence of splitting, as well as the relative orientation of the two splittings $\theta$ (see Fig.\ref{fig:obs}) were set randomly. 
For each splitting, the two fissioning fragments were separated by a distance $d_{ij} = r_0(A_i^{1/3}+A_j^{1/3})$  with $r_0=1.4$\,fm. 
This value of $r_0$ is voluntarily much smaller than that expected for fission ($r_0\sim1.9$\,fm \cite{PhysRevC.31.1550,Hinde1987318}) 
in order to test the method in a non-ideal case. We used a typical inter-splitting time of $300$\,fm/c (see Fig.\ref{fig:dte}).
Simulated events were then filtered using a simulation of the INDRA detector response. 
Finally, the experimental procedure of sequence identification was applied to these simulated events.

The correlation between the true and the extracted sequence of splittings is presented in Tab.\ref{tab:seq}(a).
It can be seen that our method is rather efficient, even in a far-from-ideal case: the well-identified events (in bold) represent approximately
66\% of the total number of simulated events. The remaining 34\% correspond to particular relative orientations of the two splittings
where the method does not allow to distinguish accurately two sequences. 
These ambiguities are mainly located around $\theta\sim50\pm10$\textdegree~and $\theta\sim130\pm10$\textdegree.

The identification efficiency can be increased up to 83\% (Tab.\ref{tab:seq}(b)) by considering only events with $| \cos(\theta)|>0.9$ or $| \cos(\theta) |<0.05$ ($\theta\sim90$\textdegree or $0$\textdegree),
which corresponds to the angular range where the Coulomb distortion parameter \Dv is computed.

\begin{table}[!ht]
\begin{center}{
\renewcommand{\arraystretch}{2.2}
\hfill
\subtable[All events\label{tab:seqa}]{
\begin{tabular}{l | c | c | c | c}
\cline{2-4}
 \textit{3}~~~ &6.5	       &7.3	       &\bf{21.8}      &~~~~~\\
\cline{2-4}
 \textit{2}~~~ &5.1	       &\bf{21.9}      &6.8	       &~~~~~\\
\cline{2-4}
 \textit{1}~~~ &\bf{21.7}      &4.0	       &4.6	       &~~~~~\\
\cline{2-4}
 \multicolumn{1}{c}{}&  \multicolumn{1}{c}{\textit{1}}& \multicolumn{1}{c}{\textit{2}} & \multicolumn{1}{c}{\textit{3}} &\\
\end{tabular}}
\hfill
\subtable[$\theta\sim90$\textdegree or $0$\textdegree\label{tab:seqb}]{
\begin{tabular}{ l | c | c | c | c}
\cline{2-4}
 \textit{3}~~~ &3.7	       &6.0	       &\bf{28.6}      &~~~~~\\
\cline{2-4}
 \textit{2}~~~ &0.5	       &\bf{26.3}      &4.9	       &~~~~~\\
\cline{2-4}
 \textit{1}~~~ &\bf{28.5}      &0.3	       &1.3	       &~~~~~\\
\cline{2-4}
 \multicolumn{1}{c}{}&  \multicolumn{1}{c}{\textit{1}}& \multicolumn{1}{c}{\textit{2}} & \multicolumn{1}{c}{\textit{3}}& \\
\end{tabular}}
\hfill}
\caption{Correlation between the true (x axis) and the identified (y axis) sequence of splittings for: (a) all simulated events, (b) events with $\theta\sim90$\textdegree, $0$\textdegree
~used to extract the inter-splitting time (see text).\label{tab:seq}}
\end{center}
\end{table}

\section{DIT simulations for \Xesn collisions at 12\mev}
\label{an:dit}


\begin{table}[ht]
\begin{tabular}{rcccc}
\hline 
 & $l_{\mathrm{max}}$ ($\hbar$) & $\sigma_{\mathrm{R}}$ [mb] & $\sigma_{\mathrm{fus}}$ [mb] & $\overline{l}_{\mathrm{fus}}$ [$\hbar$]\tabularnewline
\hline 
\hline 
Systematics \citep{Wilcke1980Reaction} &  517 & 3821 & 92 & -\\
DIT \citep{TassanGot1991Deep} &  519 & 3831 & 92$\pm2$ & 78\\
\hline
\end{tabular}
\caption{
Reaction parameters for the system \Xesn at
bombarding energy $12$\mev: (first row) calculated according to
systematics given in \citep{Wilcke1980Reaction}; (second row) results
of the DIT calculations.\label{tab:par}}
\end{table}

Calculations using the Deep Inelastic Transfers (DIT) model of \citep{TassanGot1991Deep}
have been performed for collisions of the heavy quasi-symmetric system
\Xesn at bombarding energy 12\mev.

\begin{figure}[ht]
\includegraphics[width=0.99\columnwidth]{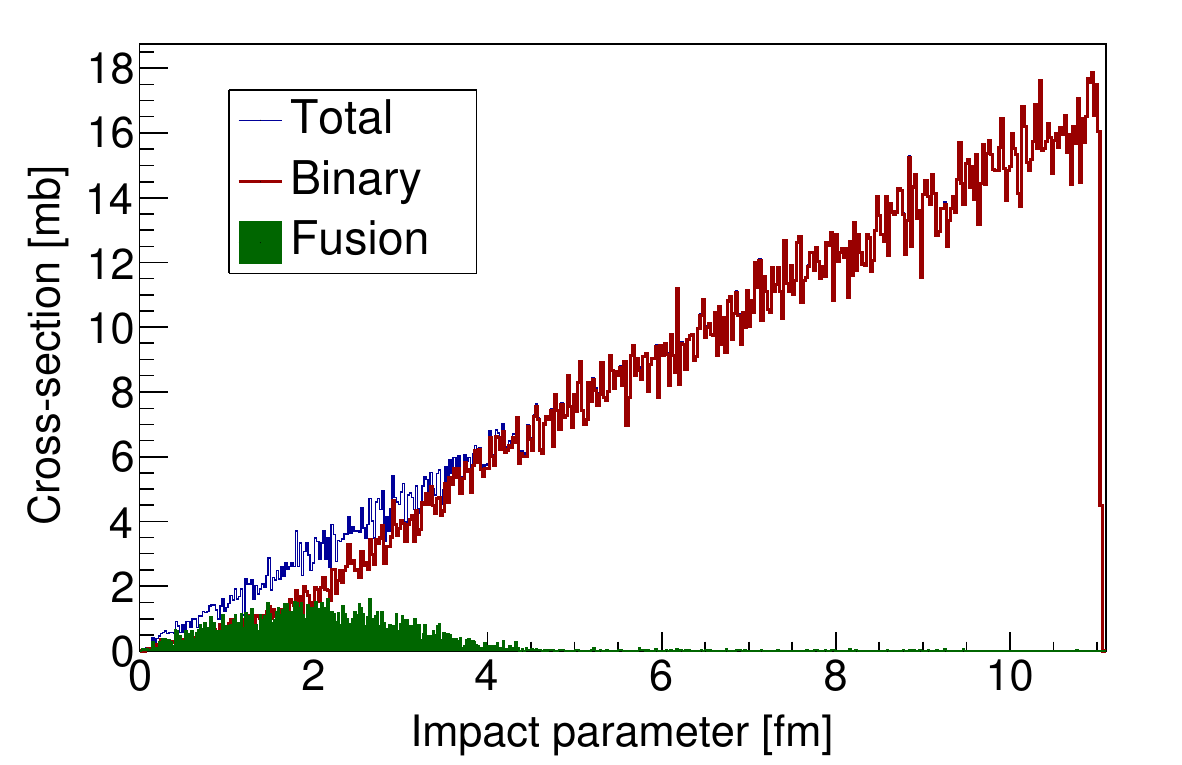}
\caption{Impact parameter distribution of fusion and deep-inelastic (binary) events.\label{fig:bdit}}
\end{figure}

In this study, $10^{5}$ events were generated with DIT corresponding
to a total reaction cross-section of $\sigma_{\mathrm{R}}\sim3.8$
barns (see Table \ref{tab:par}). The calculated
reaction and fusion cross-sections are very close to those given by
the systematics of \citep{Wilcke1980Reaction}. Figure \ref{fig:bdit}
shows the impact parameter distributions calculated for binary exit
channels and fusion events. Fusion occurs over quite a wide range
of (small) impact parameters ($b<4$ fm). The corresponding spin distribution
of the compound nuclei has a mean value of 78$\hbar$.

\begin{figure}[ht]
\includegraphics[width=0.99\columnwidth]{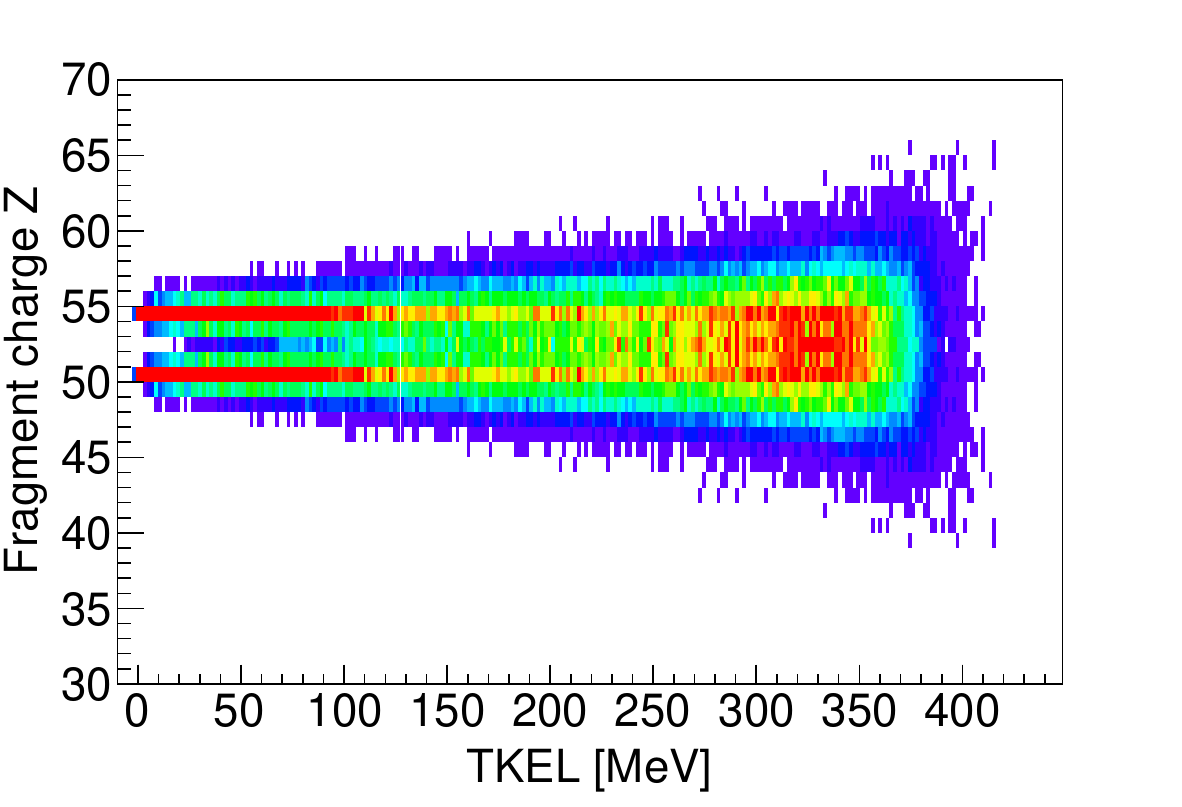}
\caption{Correlation between the fragment charge (Z) and the total kinetic energy loss (TKEL) for binary exit channels.\label{fig:ztkel} }
\end{figure}

Figure \ref{fig:ztkel} shows the distribution of projectile-/target-like
fragment atomic number $Z$ as a function of total kinetic energy
loss (TKEL). As expected, this distribution broadens with increasing
dissipation but remains centered around the mean atomic number of
$(Z_{\mathrm{P}}+Z_{\mathrm{T}})/2=52$. Figure \ref{fig:zfdit} shows the charge
distribution of PLF/TLF fragments for all binary events, and for a selection of the most dissipative
reactions (TKEL$\geqslant 350$\mev), corresponding to a total cross-section of 278 mb. The distribution
is symmetric, and there is no significant cross-section for highly-asymmetric exit channels.

\begin{table}[ht]
\begin{tabular}{rccccc}
\hline 
$Z_{\mathrm{PLF}}$ & 61 & 62 & 63 & 64 & 65\tabularnewline
$Z_{\mathrm{TLF}}$ & 43 & 42 & 41 & 40 & 39\tabularnewline
\hline 
\hline 
$\sigma$ {[}mb{]} & 2.72 & 1.38 & 0.38 & 0.42 & 0.11\tabularnewline
$\Delta\sigma$ {[}mb{]} & 0.32 & 0.23 & 0.12 & 0.13 & 0.07\tabularnewline
\hline 
\end{tabular}
\caption{Calculated cross-sections for the most asymmetric PLF-TLF splittings
for very dissipative events (TKEL$\geq350$ MeV) and their estimated
(statistical) uncertainty.\label{tab:sig}}
\end{table}

Table \ref{tab:sig} details the calculated
cross-section for each of the most asymmetric PLF-TLF splits observed
for these most dissipative reactions. The most asymmetric splitting
observed, ($Z_{\mathrm{PLF}}=65$, $Z_{\mathrm{TLF}}=39$), has an
associated cross-section of 114 $\mu$b. This corresponds to 3 simulated
collisions. We have therefore calculated a (statistical) uncertainty
for this and the other calculated cross-sections using the simple
Poissonian $\sqrt{N}$ prescription, which allows to show that increasing
the total number of simulated collisions would not significantly increase
the deduced cross-section for asymmetric splittings.

\bibliography{paper}

\begin{thebibliography}{53}
\expandafter\ifx\csname natexlab\endcsname\relax\def\natexlab#1{#1}\fi
\expandafter\ifx\csname bibnamefont\endcsname\relax
  \def\bibnamefont#1{#1}\fi
\expandafter\ifx\csname bibfnamefont\endcsname\relax
  \def\bibfnamefont#1{#1}\fi
\expandafter\ifx\csname citenamefont\endcsname\relax
  \def\citenamefont#1{#1}\fi
\expandafter\ifx\csname url\endcsname\relax
  \def\url#1{\texttt{#1}}\fi
\expandafter\ifx\csname urlprefix\endcsname\relax\def\urlprefix{URL }\fi
\providecommand{\bibinfo}[2]{#2}
\providecommand{\eprint}[2][]{\url{#2}}

\bibitem[{\citenamefont{Huizenga et~al.}(1982)\citenamefont{Huizenga,
  Schr\"{o}der, Birkelund, and Wilcke}}]{Huizenga1982Distinctive}
\bibinfo{author}{\bibfnamefont{J.~R.} \bibnamefont{Huizenga}},
  \bibinfo{author}{\bibfnamefont{W.~U.} \bibnamefont{Schr\"{o}der}},
  \bibinfo{author}{\bibfnamefont{J.~R.} \bibnamefont{Birkelund}},
  \bibnamefont{and} \bibinfo{author}{\bibfnamefont{W.~W.}
  \bibnamefont{Wilcke}}, \bibinfo{journal}{Nuclear Physics A}
  \textbf{\bibinfo{volume}{387}}, \bibinfo{pages}{257} (\bibinfo{year}{1982}).

\bibitem[{\citenamefont{Moretto}(1983)}]{Moretto1983Phenomenology}
\bibinfo{author}{\bibfnamefont{L.~G.} \bibnamefont{Moretto}},
  \bibinfo{journal}{Nuclear Physics A} \textbf{\bibinfo{volume}{409}},
  \bibinfo{pages}{115} (\bibinfo{year}{1983}).

\bibitem[{\citenamefont{Toke et~al.}(1985)\citenamefont{Toke, Bock, Dai, Gobbi,
  Gralla, Hildenbrand, Kuzminski, M\"{u}ller, Olmi, Stelzer
  et~al.}}]{Toke1985Quasifission}
\bibinfo{author}{\bibfnamefont{J.}~\bibnamefont{Toke}},
  \bibinfo{author}{\bibfnamefont{R.}~\bibnamefont{Bock}},
  \bibinfo{author}{\bibfnamefont{G.~X.} \bibnamefont{Dai}},
  \bibinfo{author}{\bibfnamefont{A.}~\bibnamefont{Gobbi}},
  \bibinfo{author}{\bibfnamefont{S.}~\bibnamefont{Gralla}},
  \bibinfo{author}{\bibfnamefont{K.~D.} \bibnamefont{Hildenbrand}},
  \bibinfo{author}{\bibfnamefont{J.}~\bibnamefont{Kuzminski}},
  \bibinfo{author}{\bibfnamefont{W.~F.~J.} \bibnamefont{M\"{u}ller}},
  \bibinfo{author}{\bibfnamefont{A.}~\bibnamefont{Olmi}},
  \bibinfo{author}{\bibfnamefont{H.}~\bibnamefont{Stelzer}},
  \bibnamefont{et~al.}, \bibinfo{journal}{Nuclear Physics A}
  \textbf{\bibinfo{volume}{440}}, \bibinfo{pages}{327} (\bibinfo{year}{1985}).

\bibitem[{\citenamefont{Gl\"{a}ssel et~al.}(1983)\citenamefont{Gl\"{a}ssel,
  Harrach, Specht, and Grodzins}}]{Glassel1983Observation}
\bibinfo{author}{\bibfnamefont{P.}~\bibnamefont{Gl\"{a}ssel}},
  \bibinfo{author}{\bibfnamefont{D.}~\bibnamefont{Harrach}},
  \bibinfo{author}{\bibfnamefont{H.~J.} \bibnamefont{Specht}},
  \bibnamefont{and} \bibinfo{author}{\bibfnamefont{L.}~\bibnamefont{Grodzins}},
  \bibinfo{journal}{Zeitschrift f\"{u}r Physik A Atoms and Nuclei}
  \textbf{\bibinfo{volume}{310}}, \bibinfo{pages}{189} (\bibinfo{year}{1983}).

\bibitem[{\citenamefont{Charity et~al.}(1991)\citenamefont{Charity, Freifelder,
  Gobbi, Herrmann, Hildenbrand, Rami, Stelzer, Wessels, Casini, Maurenzig
  et~al.}}]{Charity1991Results}
\bibinfo{author}{\bibfnamefont{R.~J.} \bibnamefont{Charity}},
  \bibinfo{author}{\bibfnamefont{R.}~\bibnamefont{Freifelder}},
  \bibinfo{author}{\bibfnamefont{A.}~\bibnamefont{Gobbi}},
  \bibinfo{author}{\bibfnamefont{N.}~\bibnamefont{Herrmann}},
  \bibinfo{author}{\bibfnamefont{K.~D.} \bibnamefont{Hildenbrand}},
  \bibinfo{author}{\bibfnamefont{F.}~\bibnamefont{Rami}},
  \bibinfo{author}{\bibfnamefont{H.}~\bibnamefont{Stelzer}},
  \bibinfo{author}{\bibfnamefont{J.~P.} \bibnamefont{Wessels}},
  \bibinfo{author}{\bibfnamefont{G.}~\bibnamefont{Casini}},
  \bibinfo{author}{\bibfnamefont{P.~R.} \bibnamefont{Maurenzig}},
  \bibnamefont{et~al.}, \bibinfo{journal}{Zeitschrift f\"{u}r Physik A Hadrons
  and Nuclei} \textbf{\bibinfo{volume}{341}}, \bibinfo{pages}{53}
  (\bibinfo{year}{1991}).

\bibitem[{\citenamefont{Casini et~al.}(1993)\citenamefont{Casini, Bizzeti,
  Maurenzig, Olmi, Stefanini, Wessels, Charity, Freifelder, Gobbi, Herrmann
  et~al.}}]{Casini1993Fission}
\bibinfo{author}{\bibfnamefont{G.}~\bibnamefont{Casini}},
  \bibinfo{author}{\bibfnamefont{P.~G.} \bibnamefont{Bizzeti}},
  \bibinfo{author}{\bibfnamefont{P.~R.} \bibnamefont{Maurenzig}},
  \bibinfo{author}{\bibfnamefont{A.}~\bibnamefont{Olmi}},
  \bibinfo{author}{\bibfnamefont{A.~A.} \bibnamefont{Stefanini}},
  \bibinfo{author}{\bibfnamefont{J.~P.} \bibnamefont{Wessels}},
  \bibinfo{author}{\bibfnamefont{R.~J.} \bibnamefont{Charity}},
  \bibinfo{author}{\bibfnamefont{R.}~\bibnamefont{Freifelder}},
  \bibinfo{author}{\bibfnamefont{A.}~\bibnamefont{Gobbi}},
  \bibinfo{author}{\bibfnamefont{N.}~\bibnamefont{Herrmann}},
  \bibnamefont{et~al.}, \bibinfo{journal}{Physical Review Letters}
  \textbf{\bibinfo{volume}{71}}, \bibinfo{pages}{2567} (\bibinfo{year}{1993}).

\bibitem[{\citenamefont{Wilczy\ifmmode~\acute{n}\else \'{n}\fi{}ski
  et~al.}(2010)\citenamefont{Wilczy\ifmmode~\acute{n}\else \'{n}\fi{}ski,
  Skwira-Chalot, Siwek-Wilczy\ifmmode~\acute{n}\else \'{n}\fi{}ska, Pagano,
  Amorini, Anzalone, Auditore, Baran, Brzychczyk, Cardella
  et~al.}}]{PhysRevC.81.024605}
\bibinfo{author}{\bibfnamefont{J.}~\bibnamefont{Wilczy\ifmmode~\acute{n}\else
  \'{n}\fi{}ski}},
  \bibinfo{author}{\bibfnamefont{I.}~\bibnamefont{Skwira-Chalot}},
  \bibinfo{author}{\bibfnamefont{K.}~\bibnamefont{Siwek-Wilczy\ifmmode~\acute{n}\else
  \'{n}\fi{}ska}}, \bibinfo{author}{\bibfnamefont{A.}~\bibnamefont{Pagano}},
  \bibinfo{author}{\bibfnamefont{F.}~\bibnamefont{Amorini}},
  \bibinfo{author}{\bibfnamefont{A.}~\bibnamefont{Anzalone}},
  \bibinfo{author}{\bibfnamefont{L.}~\bibnamefont{Auditore}},
  \bibinfo{author}{\bibfnamefont{V.}~\bibnamefont{Baran}},
  \bibinfo{author}{\bibfnamefont{J.}~\bibnamefont{Brzychczyk}},
  \bibinfo{author}{\bibfnamefont{G.}~\bibnamefont{Cardella}},
  \bibnamefont{et~al.}, \bibinfo{journal}{Physical Review C}
  \textbf{\bibinfo{volume}{81}}, \bibinfo{pages}{024605}
  (\bibinfo{year}{2010}).

\bibitem[{\citenamefont{Golabek and Simenel}(2009)}]{PhysRevLett.103.042701}
\bibinfo{author}{\bibfnamefont{C.}~\bibnamefont{Golabek}} \bibnamefont{and}
  \bibinfo{author}{\bibfnamefont{C.}~\bibnamefont{Simenel}},
  \bibinfo{journal}{Physical Review Letters} \textbf{\bibinfo{volume}{103}},
  \bibinfo{pages}{042701} (\bibinfo{year}{2009}).

\bibitem[{\citenamefont{Sekizawa and Yabana}(2013)}]{Sekizawa2013Timedependent}
\bibinfo{author}{\bibfnamefont{K.}~\bibnamefont{Sekizawa}} \bibnamefont{and}
  \bibinfo{author}{\bibfnamefont{K.}~\bibnamefont{Yabana}},
  \bibinfo{journal}{Physical Review C} \textbf{\bibinfo{volume}{88}},
  \bibinfo{pages}{014614} (\bibinfo{year}{2013}).

\bibitem[{\citenamefont{Ryabov et~al.}(2008)\citenamefont{Ryabov, Karpov,
  Nadtochy, and Adeev}}]{PhysRevC.78.044614}
\bibinfo{author}{\bibfnamefont{E.~G.} \bibnamefont{Ryabov}},
  \bibinfo{author}{\bibfnamefont{A.~V.} \bibnamefont{Karpov}},
  \bibinfo{author}{\bibfnamefont{P.~N.} \bibnamefont{Nadtochy}},
  \bibnamefont{and} \bibinfo{author}{\bibfnamefont{G.~D.} \bibnamefont{Adeev}},
  \bibinfo{journal}{Physical Review C} \textbf{\bibinfo{volume}{78}},
  \bibinfo{pages}{044614} (\bibinfo{year}{2008}).

\bibitem[{\citenamefont{Li et~al.}(2013)\citenamefont{Li, Yan, Jiang, and
  Wang}}]{Li2013Dynamical}
\bibinfo{author}{\bibfnamefont{Y.}~\bibnamefont{Li}},
  \bibinfo{author}{\bibfnamefont{S.}~\bibnamefont{Yan}},
  \bibinfo{author}{\bibfnamefont{X.}~\bibnamefont{Jiang}}, \bibnamefont{and}
  \bibinfo{author}{\bibfnamefont{L.}~\bibnamefont{Wang}},
  \bibinfo{journal}{Nuclear Physics A} \textbf{\bibinfo{volume}{902}},
  \bibinfo{pages}{1} (\bibinfo{year}{2013}).

\bibitem[{\citenamefont{Pouthas et~al.}(1995)\citenamefont{Pouthas, Borderie,
  Dayras, Plagnol, Rivet, Saint-Laurent, Steckmeyer, Auger, Bacri, Barbey
  et~al.}}]{Pouthas1995INDRA}
\bibinfo{author}{\bibfnamefont{J.}~\bibnamefont{Pouthas}},
  \bibinfo{author}{\bibfnamefont{B.}~\bibnamefont{Borderie}},
  \bibinfo{author}{\bibfnamefont{R.}~\bibnamefont{Dayras}},
  \bibinfo{author}{\bibfnamefont{E.}~\bibnamefont{Plagnol}},
  \bibinfo{author}{\bibfnamefont{M.-F.} \bibnamefont{Rivet}},
  \bibinfo{author}{\bibfnamefont{F.}~\bibnamefont{Saint-Laurent}},
  \bibinfo{author}{\bibfnamefont{J.~C.} \bibnamefont{Steckmeyer}},
  \bibinfo{author}{\bibfnamefont{G.}~\bibnamefont{Auger}},
  \bibinfo{author}{\bibfnamefont{C.~O.} \bibnamefont{Bacri}},
  \bibinfo{author}{\bibfnamefont{S.}~\bibnamefont{Barbey}},
  \bibnamefont{et~al.}, \bibinfo{journal}{Nuclear Instruments and Methods in
  Physics Research Section A: Accelerators, Spectrometers, Detectors and
  Associated Equipment} \textbf{\bibinfo{volume}{357}}, \bibinfo{pages}{418}
  (\bibinfo{year}{1995}).

\bibitem[{\citenamefont{Chbihi et~al.}(2013)\citenamefont{Chbihi, Manduci,
  Moisan, Bonnet, Frankland, Roy, and Verde}}]{Chbihi012099}
\bibinfo{author}{\bibfnamefont{A.}~\bibnamefont{Chbihi}},
  \bibinfo{author}{\bibfnamefont{L.}~\bibnamefont{Manduci}},
  \bibinfo{author}{\bibfnamefont{J.}~\bibnamefont{Moisan}},
  \bibinfo{author}{\bibfnamefont{E.}~\bibnamefont{Bonnet}},
  \bibinfo{author}{\bibfnamefont{J.~D.} \bibnamefont{Frankland}},
  \bibinfo{author}{\bibfnamefont{R.}~\bibnamefont{Roy}}, \bibnamefont{and}
  \bibinfo{author}{\bibfnamefont{G.}~\bibnamefont{Verde}},
  \bibinfo{journal}{Journal of Physics: Conference Series}
  \textbf{\bibinfo{volume}{420}}, \bibinfo{pages}{012099}
  (\bibinfo{year}{2013}).

\bibitem[{\citenamefont{Hudan et~al.}(2003)\citenamefont{Hudan, Chbihi,
  Frankland, Mignon, Wieleczko, Auger, Bellaize, Borderie, Botvina, Bougault
  et~al.}}]{Hudan2003Characteristics}
\bibinfo{author}{\bibfnamefont{S.}~\bibnamefont{Hudan}},
  \bibinfo{author}{\bibfnamefont{A.}~\bibnamefont{Chbihi}},
  \bibinfo{author}{\bibfnamefont{J.~D.} \bibnamefont{Frankland}},
  \bibinfo{author}{\bibfnamefont{A.}~\bibnamefont{Mignon}},
  \bibinfo{author}{\bibfnamefont{J.~P.} \bibnamefont{Wieleczko}},
  \bibinfo{author}{\bibfnamefont{G.}~\bibnamefont{Auger}},
  \bibinfo{author}{\bibfnamefont{N.}~\bibnamefont{Bellaize}},
  \bibinfo{author}{\bibfnamefont{B.}~\bibnamefont{Borderie}},
  \bibinfo{author}{\bibfnamefont{A.}~\bibnamefont{Botvina}},
  \bibinfo{author}{\bibfnamefont{R.}~\bibnamefont{Bougault}},
  \bibnamefont{et~al.}, \bibinfo{journal}{Physical Review C}
  \textbf{\bibinfo{volume}{67}}, \bibinfo{pages}{064613}
  (\bibinfo{year}{2003}).

\bibitem[{\citenamefont{Piantelli et~al.}(2008)\citenamefont{Piantelli,
  Borderie, Bonnet, Le~Neindre, Raduta, Rivet, Bougault, Chbihi, Dayras,
  Frankland et~al.}}]{Piantelli2008FreezeOut}
\bibinfo{author}{\bibfnamefont{S.}~\bibnamefont{Piantelli}},
  \bibinfo{author}{\bibfnamefont{B.}~\bibnamefont{Borderie}},
  \bibinfo{author}{\bibfnamefont{E.}~\bibnamefont{Bonnet}},
  \bibinfo{author}{\bibfnamefont{N.}~\bibnamefont{Le~Neindre}},
  \bibinfo{author}{\bibfnamefont{A.}~\bibnamefont{Raduta}},
  \bibinfo{author}{\bibfnamefont{M.~F.} \bibnamefont{Rivet}},
  \bibinfo{author}{\bibfnamefont{R.}~\bibnamefont{Bougault}},
  \bibinfo{author}{\bibfnamefont{A.}~\bibnamefont{Chbihi}},
  \bibinfo{author}{\bibfnamefont{R.}~\bibnamefont{Dayras}},
  \bibinfo{author}{\bibfnamefont{J.~D.} \bibnamefont{Frankland}},
  \bibnamefont{et~al.}, \bibinfo{journal}{Nuclear Physics A}
  \textbf{\bibinfo{volume}{809}}, \bibinfo{pages}{111} (\bibinfo{year}{2008}).

\bibitem[{\citenamefont{Stefanini et~al.}(1995)\citenamefont{Stefanini, Casini,
  Maurenzig, Olmi, Charity, Freifelder, Gobbi, Herrmann, Hildenbrand, Petrovici
  et~al.}}]{Stefanini1995Analysis}
\bibinfo{author}{\bibfnamefont{A.~A.} \bibnamefont{Stefanini}},
  \bibinfo{author}{\bibfnamefont{G.}~\bibnamefont{Casini}},
  \bibinfo{author}{\bibfnamefont{P.~R.} \bibnamefont{Maurenzig}},
  \bibinfo{author}{\bibfnamefont{A.}~\bibnamefont{Olmi}},
  \bibinfo{author}{\bibfnamefont{R.~J.} \bibnamefont{Charity}},
  \bibinfo{author}{\bibfnamefont{R.}~\bibnamefont{Freifelder}},
  \bibinfo{author}{\bibfnamefont{A.}~\bibnamefont{Gobbi}},
  \bibinfo{author}{\bibfnamefont{N.}~\bibnamefont{Herrmann}},
  \bibinfo{author}{\bibfnamefont{K.~D.} \bibnamefont{Hildenbrand}},
  \bibinfo{author}{\bibfnamefont{M.}~\bibnamefont{Petrovici}},
  \bibnamefont{et~al.}, \bibinfo{journal}{Zeitschrift f\"{u}r Physik A Hadrons
  and Nuclei} \textbf{\bibinfo{volume}{351}}, \bibinfo{pages}{167}
  (\bibinfo{year}{1995}).

\bibitem[{\citenamefont{De~Filippo
  et~al.}(2005{\natexlab{a}})\citenamefont{De~Filippo, Pagano, Wilczy\'{n}ski,
  Amorini, Anzalone, Auditore, Baran, Berceanu, Blicharska, Brzychczyk
  et~al.}}]{DeFilippo2005Time}
\bibinfo{author}{\bibfnamefont{E.}~\bibnamefont{De~Filippo}},
  \bibinfo{author}{\bibfnamefont{A.}~\bibnamefont{Pagano}},
  \bibinfo{author}{\bibfnamefont{J.}~\bibnamefont{Wilczy\'{n}ski}},
  \bibinfo{author}{\bibfnamefont{F.}~\bibnamefont{Amorini}},
  \bibinfo{author}{\bibfnamefont{A.}~\bibnamefont{Anzalone}},
  \bibinfo{author}{\bibfnamefont{L.}~\bibnamefont{Auditore}},
  \bibinfo{author}{\bibfnamefont{V.}~\bibnamefont{Baran}},
  \bibinfo{author}{\bibfnamefont{I.}~\bibnamefont{Berceanu}},
  \bibinfo{author}{\bibfnamefont{J.}~\bibnamefont{Blicharska}},
  \bibinfo{author}{\bibfnamefont{J.}~\bibnamefont{Brzychczyk}},
  \bibnamefont{et~al.}, \bibinfo{journal}{Physical Review C}
  \textbf{\bibinfo{volume}{71}}, \bibinfo{pages}{044602}
  (\bibinfo{year}{2005}{\natexlab{a}}).

\bibitem[{\citenamefont{McIntosh et~al.}(2010)\citenamefont{McIntosh, Hudan,
  Black, Mercier, Metelko, Yanez, de~Souza, Chbihi, Famiano, Fr\'{e}geau
  et~al.}}]{McIntosh2010Shortlived}
\bibinfo{author}{\bibfnamefont{A.~B.} \bibnamefont{McIntosh}},
  \bibinfo{author}{\bibfnamefont{S.}~\bibnamefont{Hudan}},
  \bibinfo{author}{\bibfnamefont{J.}~\bibnamefont{Black}},
  \bibinfo{author}{\bibfnamefont{D.}~\bibnamefont{Mercier}},
  \bibinfo{author}{\bibfnamefont{C.~J.} \bibnamefont{Metelko}},
  \bibinfo{author}{\bibfnamefont{R.}~\bibnamefont{Yanez}},
  \bibinfo{author}{\bibfnamefont{R.~T.} \bibnamefont{de~Souza}},
  \bibinfo{author}{\bibfnamefont{A.}~\bibnamefont{Chbihi}},
  \bibinfo{author}{\bibfnamefont{M.}~\bibnamefont{Famiano}},
  \bibinfo{author}{\bibfnamefont{M.~O.} \bibnamefont{Fr\'{e}geau}},
  \bibnamefont{et~al.}, \bibinfo{journal}{Physical Review C}
  \textbf{\bibinfo{volume}{81}}, \bibinfo{pages}{034603}
  (\bibinfo{year}{2010}).

\bibitem[{\citenamefont{Hudan et~al.}(2012)\citenamefont{Hudan, McIntosh,
  de~Souza, Bianchin, Black, Chbihi, Famiano, Fr\'{e}geau, Gauthier, Mercier
  et~al.}}]{Hudan2012Tracking}
\bibinfo{author}{\bibfnamefont{S.}~\bibnamefont{Hudan}},
  \bibinfo{author}{\bibfnamefont{A.~B.} \bibnamefont{McIntosh}},
  \bibinfo{author}{\bibfnamefont{R.~T.} \bibnamefont{de~Souza}},
  \bibinfo{author}{\bibfnamefont{S.}~\bibnamefont{Bianchin}},
  \bibinfo{author}{\bibfnamefont{J.}~\bibnamefont{Black}},
  \bibinfo{author}{\bibfnamefont{A.}~\bibnamefont{Chbihi}},
  \bibinfo{author}{\bibfnamefont{M.}~\bibnamefont{Famiano}},
  \bibinfo{author}{\bibfnamefont{M.~O.} \bibnamefont{Fr\'{e}geau}},
  \bibinfo{author}{\bibfnamefont{J.}~\bibnamefont{Gauthier}},
  \bibinfo{author}{\bibfnamefont{D.}~\bibnamefont{Mercier}},
  \bibnamefont{et~al.}, \bibinfo{journal}{Physical Review C}
  \textbf{\bibinfo{volume}{86}}, \bibinfo{pages}{021603}
  (\bibinfo{year}{2012}).

\bibitem[{\citenamefont{Brown et~al.}(2013)\citenamefont{Brown, Hudan, deSouza,
  Gauthier, Roy, Shetty, Souliotis, and Yennello}}]{PhysRevC.87.061601}
\bibinfo{author}{\bibfnamefont{K.}~\bibnamefont{Brown}},
  \bibinfo{author}{\bibfnamefont{S.}~\bibnamefont{Hudan}},
  \bibinfo{author}{\bibfnamefont{R.~T.} \bibnamefont{deSouza}},
  \bibinfo{author}{\bibfnamefont{J.}~\bibnamefont{Gauthier}},
  \bibinfo{author}{\bibfnamefont{R.}~\bibnamefont{Roy}},
  \bibinfo{author}{\bibfnamefont{D.~V.} \bibnamefont{Shetty}},
  \bibinfo{author}{\bibfnamefont{G.~A.} \bibnamefont{Souliotis}},
  \bibnamefont{and} \bibinfo{author}{\bibfnamefont{S.~J.}
  \bibnamefont{Yennello}}, \bibinfo{journal}{Physical Review C}
  \textbf{\bibinfo{volume}{87}}, \bibinfo{pages}{061601}
  (\bibinfo{year}{2013}).

\bibitem[{\citenamefont{De~Filippo et~al.}(2012)\citenamefont{De~Filippo,
  Pagano, Russotto, Amorini, Anzalone, Auditore, Baran, Berceanu, Borderie,
  Bougault et~al.}}]{DeFilippo2012Correlations}
\bibinfo{author}{\bibfnamefont{E.}~\bibnamefont{De~Filippo}},
  \bibinfo{author}{\bibfnamefont{A.}~\bibnamefont{Pagano}},
  \bibinfo{author}{\bibfnamefont{P.}~\bibnamefont{Russotto}},
  \bibinfo{author}{\bibfnamefont{F.}~\bibnamefont{Amorini}},
  \bibinfo{author}{\bibfnamefont{A.}~\bibnamefont{Anzalone}},
  \bibinfo{author}{\bibfnamefont{L.}~\bibnamefont{Auditore}},
  \bibinfo{author}{\bibfnamefont{V.}~\bibnamefont{Baran}},
  \bibinfo{author}{\bibfnamefont{I.}~\bibnamefont{Berceanu}},
  \bibinfo{author}{\bibfnamefont{B.}~\bibnamefont{Borderie}},
  \bibinfo{author}{\bibfnamefont{R.}~\bibnamefont{Bougault}},
  \bibnamefont{et~al.}, \bibinfo{journal}{Physical Review C}
  \textbf{\bibinfo{volume}{86}}, \bibinfo{pages}{014610}
  (\bibinfo{year}{2012}).

\bibitem[{\citenamefont{Durand}(1998)}]{Durand1998Physics}
\bibinfo{author}{\bibfnamefont{D.}~\bibnamefont{Durand}},
  \bibinfo{journal}{Nuclear Physics A} \textbf{\bibinfo{volume}{630}},
  \bibinfo{pages}{52} (\bibinfo{year}{1998}).

\bibitem[{\citenamefont{Beaulieu et~al.}(2000)\citenamefont{Beaulieu, Lefort,
  Kwiatkowski, de~Souza, Hsi, Pienkowski, Back, Bracken, Breuer, Cornell
  et~al.}}]{Beaulieu2000Signals}
\bibinfo{author}{\bibfnamefont{L.}~\bibnamefont{Beaulieu}},
  \bibinfo{author}{\bibfnamefont{T.}~\bibnamefont{Lefort}},
  \bibinfo{author}{\bibfnamefont{K.}~\bibnamefont{Kwiatkowski}},
  \bibinfo{author}{\bibfnamefont{R.~T.} \bibnamefont{de~Souza}},
  \bibinfo{author}{\bibnamefont{Hsi}},
  \bibinfo{author}{\bibfnamefont{L.}~\bibnamefont{Pienkowski}},
  \bibinfo{author}{\bibfnamefont{B.}~\bibnamefont{Back}},
  \bibinfo{author}{\bibfnamefont{D.~S.} \bibnamefont{Bracken}},
  \bibinfo{author}{\bibfnamefont{H.}~\bibnamefont{Breuer}},
  \bibinfo{author}{\bibfnamefont{E.}~\bibnamefont{Cornell}},
  \bibnamefont{et~al.}, \bibinfo{journal}{Physical Review Letters}
  \textbf{\bibinfo{volume}{84}}, \bibinfo{pages}{5971} (\bibinfo{year}{2000}).

\bibitem[{\citenamefont{Verde et~al.}(2006)\citenamefont{Verde, Chbihi, Ghetti,
  and Helgesson}}]{Verde2006Correlations}
\bibinfo{author}{\bibfnamefont{G.}~\bibnamefont{Verde}},
  \bibinfo{author}{\bibfnamefont{A.}~\bibnamefont{Chbihi}},
  \bibinfo{author}{\bibfnamefont{R.}~\bibnamefont{Ghetti}}, \bibnamefont{and}
  \bibinfo{author}{\bibfnamefont{J.}~\bibnamefont{Helgesson}},
  \bibinfo{journal}{The European Physical Journal A - Hadrons and Nuclei}
  \textbf{\bibinfo{volume}{30}}, \bibinfo{pages}{81} (\bibinfo{year}{2006}).

\bibitem[{\citenamefont{T\u{a}b\u{a}caru
  et~al.}(2006)\citenamefont{T\u{a}b\u{a}caru, Rivet, Borderie, P\^arlog,
  Bouriquet, Chbihi, Frankland, Wieleczko, Bonnet, Bougault
  et~al.}}]{Tabacaru2006371}
\bibinfo{author}{\bibfnamefont{G.}~\bibnamefont{T\u{a}b\u{a}caru}},
  \bibinfo{author}{\bibfnamefont{M.-F.} \bibnamefont{Rivet}},
  \bibinfo{author}{\bibfnamefont{B.}~\bibnamefont{Borderie}},
  \bibinfo{author}{\bibfnamefont{M.}~\bibnamefont{P\^arlog}},
  \bibinfo{author}{\bibfnamefont{B.}~\bibnamefont{Bouriquet}},
  \bibinfo{author}{\bibfnamefont{A.}~\bibnamefont{Chbihi}},
  \bibinfo{author}{\bibfnamefont{J.}~\bibnamefont{Frankland}},
  \bibinfo{author}{\bibfnamefont{J.}~\bibnamefont{Wieleczko}},
  \bibinfo{author}{\bibfnamefont{E.}~\bibnamefont{Bonnet}},
  \bibinfo{author}{\bibfnamefont{R.}~\bibnamefont{Bougault}},
  \bibnamefont{et~al.}, \bibinfo{journal}{Nuclear Physics A}
  \textbf{\bibinfo{volume}{764}}, \bibinfo{pages}{371 } (\bibinfo{year}{2006}).

\bibitem[{\citenamefont{Borderie and Rivet}(2008)}]{Borderie2008551}
\bibinfo{author}{\bibfnamefont{B.}~\bibnamefont{Borderie}} \bibnamefont{and}
  \bibinfo{author}{\bibfnamefont{M.-F.} \bibnamefont{Rivet}},
  \bibinfo{journal}{Progress in Particle and Nuclear Physics}
  \textbf{\bibinfo{volume}{61}}, \bibinfo{pages}{551 } (\bibinfo{year}{2008}).

\bibitem[{\citenamefont{Popescu et~al.}(1998)\citenamefont{Popescu, Glasmacher,
  Dinius, Gaff, Gelbke, Handzy, Huang, Kunde, Lynch, Martin
  et~al.}}]{PhysRevC.58.270}
\bibinfo{author}{\bibfnamefont{R.}~\bibnamefont{Popescu}},
  \bibinfo{author}{\bibfnamefont{T.}~\bibnamefont{Glasmacher}},
  \bibinfo{author}{\bibfnamefont{J.~D.} \bibnamefont{Dinius}},
  \bibinfo{author}{\bibfnamefont{S.~J.} \bibnamefont{Gaff}},
  \bibinfo{author}{\bibfnamefont{C.~K.} \bibnamefont{Gelbke}},
  \bibinfo{author}{\bibfnamefont{D.~O.} \bibnamefont{Handzy}},
  \bibinfo{author}{\bibfnamefont{M.~J.} \bibnamefont{Huang}},
  \bibinfo{author}{\bibfnamefont{G.~J.} \bibnamefont{Kunde}},
  \bibinfo{author}{\bibfnamefont{W.~G.} \bibnamefont{Lynch}},
  \bibinfo{author}{\bibfnamefont{L.}~\bibnamefont{Martin}},
  \bibnamefont{et~al.}, \bibinfo{journal}{Phys. Rev. C}
  \textbf{\bibinfo{volume}{58}}, \bibinfo{pages}{270} (\bibinfo{year}{1998}),
  \urlprefix\url{http://link.aps.org/doi/10.1103/PhysRevC.58.270}.

\bibitem[{\citenamefont{Chajecki and Lisa}(2008)}]{PhysRevC.78.064903}
\bibinfo{author}{\bibfnamefont{Z.}~\bibnamefont{Chajecki}} \bibnamefont{and}
  \bibinfo{author}{\bibfnamefont{M.}~\bibnamefont{Lisa}},
  \bibinfo{journal}{Physical Review C} \textbf{\bibinfo{volume}{78}},
  \bibinfo{pages}{064903} (\bibinfo{year}{2008}).

\bibitem[{\citenamefont{K\"{a}mpfer et~al.}(1993)\citenamefont{K\"{a}mpfer,
  Kotte, M\"{o}sner, Neubert, Wohlfarth, Alard, Basrak, Bastid, Belayev, Blaich
  et~al.}}]{Kampfer1993Velocity}
\bibinfo{author}{\bibfnamefont{B.}~\bibnamefont{K\"{a}mpfer}},
  \bibinfo{author}{\bibfnamefont{R.}~\bibnamefont{Kotte}},
  \bibinfo{author}{\bibfnamefont{J.}~\bibnamefont{M\"{o}sner}},
  \bibinfo{author}{\bibfnamefont{W.}~\bibnamefont{Neubert}},
  \bibinfo{author}{\bibfnamefont{D.}~\bibnamefont{Wohlfarth}},
  \bibinfo{author}{\bibfnamefont{J.~P.} \bibnamefont{Alard}},
  \bibinfo{author}{\bibfnamefont{Z.}~\bibnamefont{Basrak}},
  \bibinfo{author}{\bibfnamefont{N.}~\bibnamefont{Bastid}},
  \bibinfo{author}{\bibfnamefont{I.~M.} \bibnamefont{Belayev}},
  \bibinfo{author}{\bibfnamefont{T.}~\bibnamefont{Blaich}},
  \bibnamefont{et~al.}, \bibinfo{journal}{Physical Review C}
  \textbf{\bibinfo{volume}{48}}, \bibinfo{pages}{R955} (\bibinfo{year}{1993}).

\bibitem[{\citenamefont{Verde et~al.}(2007)\citenamefont{Verde, Danielewicz,
  Lynch, Chan, Gelbke, Kwong, Liu, Liu, Seymour, Shomin
  et~al.}}]{Verde2007Correlation}
\bibinfo{author}{\bibfnamefont{G.}~\bibnamefont{Verde}},
  \bibinfo{author}{\bibfnamefont{P.}~\bibnamefont{Danielewicz}},
  \bibinfo{author}{\bibfnamefont{W.~G.} \bibnamefont{Lynch}},
  \bibinfo{author}{\bibfnamefont{C.~F.} \bibnamefont{Chan}},
  \bibinfo{author}{\bibfnamefont{C.~K.} \bibnamefont{Gelbke}},
  \bibinfo{author}{\bibfnamefont{L.~K.} \bibnamefont{Kwong}},
  \bibinfo{author}{\bibfnamefont{T.~X.} \bibnamefont{Liu}},
  \bibinfo{author}{\bibfnamefont{X.~D.} \bibnamefont{Liu}},
  \bibinfo{author}{\bibfnamefont{D.}~\bibnamefont{Seymour}},
  \bibinfo{author}{\bibfnamefont{R.}~\bibnamefont{Shomin}},
  \bibnamefont{et~al.}, \bibinfo{journal}{Physics Letters B}
  \textbf{\bibinfo{volume}{653}}, \bibinfo{pages}{12} (\bibinfo{year}{2007}).

\bibitem[{\citenamefont{Harrach et~al.}(1982)\citenamefont{Harrach,
  Gl\"{a}ssel, Grodzins, Kapoor, and Specht}}]{Harrach1982Direct}
\bibinfo{author}{\bibfnamefont{D.}~\bibnamefont{Harrach}},
  \bibinfo{author}{\bibfnamefont{P.}~\bibnamefont{Gl\"{a}ssel}},
  \bibinfo{author}{\bibfnamefont{L.}~\bibnamefont{Grodzins}},
  \bibinfo{author}{\bibfnamefont{S.~S.} \bibnamefont{Kapoor}},
  \bibnamefont{and} \bibinfo{author}{\bibfnamefont{H.~J.}
  \bibnamefont{Specht}}, \bibinfo{journal}{Physical Review Letters}
  \textbf{\bibinfo{volume}{48}}, \bibinfo{pages}{1093} (\bibinfo{year}{1982}).

\bibitem[{\citenamefont{Northcliffe and
  Schilling}(1970)}]{NorthcliffeSchilling}
\bibinfo{author}{\bibfnamefont{L.}~\bibnamefont{Northcliffe}} \bibnamefont{and}
  \bibinfo{author}{\bibfnamefont{R.}~\bibnamefont{Schilling}},
  \bibinfo{journal}{Atomic Data and Nuclear Data Tables}
  \textbf{\bibinfo{volume}{7}}, \bibinfo{pages}{233} (\bibinfo{year}{1970}).

\bibitem[{\citenamefont{Hubert et~al.}(1990)\citenamefont{Hubert, Bimbot, and
  Gauvin}}]{HubertBimbotGauvin}
\bibinfo{author}{\bibfnamefont{F.}~\bibnamefont{Hubert}},
  \bibinfo{author}{\bibfnamefont{R.}~\bibnamefont{Bimbot}}, \bibnamefont{and}
  \bibinfo{author}{\bibfnamefont{H.}~\bibnamefont{Gauvin}},
  \bibinfo{journal}{Atomic Data and Nuclear Data Tables}
  \textbf{\bibinfo{volume}{46}}, \bibinfo{pages}{1} (\bibinfo{year}{1990}).

\bibitem[{\citenamefont{Frankland et~al.}(2001)\citenamefont{Frankland, Bacri,
  Borderie, Rivet, Squalli, Auger, Bellaize, Bocage, Bougault, Brou
  et~al.}}]{Frankland2001SingleSource}
\bibinfo{author}{\bibfnamefont{J.~D.} \bibnamefont{Frankland}},
  \bibinfo{author}{\bibfnamefont{C.}~\bibnamefont{Bacri}},
  \bibinfo{author}{\bibfnamefont{B.}~\bibnamefont{Borderie}},
  \bibinfo{author}{\bibfnamefont{M.}~\bibnamefont{Rivet}},
  \bibinfo{author}{\bibfnamefont{M.}~\bibnamefont{Squalli}},
  \bibinfo{author}{\bibfnamefont{G.}~\bibnamefont{Auger}},
  \bibinfo{author}{\bibfnamefont{N.}~\bibnamefont{Bellaize}},
  \bibinfo{author}{\bibfnamefont{F.}~\bibnamefont{Bocage}},
  \bibinfo{author}{\bibfnamefont{R.}~\bibnamefont{Bougault}},
  \bibinfo{author}{\bibfnamefont{R.}~\bibnamefont{Brou}}, \bibnamefont{et~al.},
  \bibinfo{journal}{Nuclear Physics A} \textbf{\bibinfo{volume}{689}},
  \bibinfo{pages}{905} (\bibinfo{year}{2001}).

\bibitem[{\citenamefont{Cugnon and L'Hote}(1983)}]{Cugnon1983Global}
\bibinfo{author}{\bibfnamefont{J.}~\bibnamefont{Cugnon}} \bibnamefont{and}
  \bibinfo{author}{\bibfnamefont{D.}~\bibnamefont{L'Hote}},
  \bibinfo{journal}{Nuclear Physics A} \textbf{\bibinfo{volume}{397}},
  \bibinfo{pages}{519} (\bibinfo{year}{1983}), ISSN \bibinfo{issn}{03759474},
  \urlprefix\url{http://dx.doi.org/10.1016/0375-9474(83)90614-0}.

\bibitem[{\citenamefont{Bizard et~al.}(1992)\citenamefont{Bizard, Durand,
  Genoux-Lubain, Louvel, Bougault, Brou, Doubre, El-Masri, Fugiwara, Hagel
  et~al.}}]{Bizard1992413}
\bibinfo{author}{\bibfnamefont{G.}~\bibnamefont{Bizard}},
  \bibinfo{author}{\bibfnamefont{D.}~\bibnamefont{Durand}},
  \bibinfo{author}{\bibfnamefont{A.}~\bibnamefont{Genoux-Lubain}},
  \bibinfo{author}{\bibfnamefont{M.}~\bibnamefont{Louvel}},
  \bibinfo{author}{\bibfnamefont{R.}~\bibnamefont{Bougault}},
  \bibinfo{author}{\bibfnamefont{R.}~\bibnamefont{Brou}},
  \bibinfo{author}{\bibfnamefont{H.}~\bibnamefont{Doubre}},
  \bibinfo{author}{\bibfnamefont{Y.}~\bibnamefont{El-Masri}},
  \bibinfo{author}{\bibfnamefont{H.}~\bibnamefont{Fugiwara}},
  \bibinfo{author}{\bibfnamefont{K.}~\bibnamefont{Hagel}},
  \bibnamefont{et~al.}, \bibinfo{journal}{Physics Letters B}
  \textbf{\bibinfo{volume}{276}}, \bibinfo{pages}{413 } (\bibinfo{year}{1992}).

\bibitem[{\citenamefont{Viola et~al.}(1985)\citenamefont{Viola, Kwiatkowski,
  and Walker}}]{PhysRevC.31.1550}
\bibinfo{author}{\bibfnamefont{V.~E.} \bibnamefont{Viola}},
  \bibinfo{author}{\bibfnamefont{K.}~\bibnamefont{Kwiatkowski}},
  \bibnamefont{and} \bibinfo{author}{\bibfnamefont{M.}~\bibnamefont{Walker}},
  \bibinfo{journal}{Physical Review C} \textbf{\bibinfo{volume}{31}},
  \bibinfo{pages}{1550} (\bibinfo{year}{1985}).

\bibitem[{\citenamefont{Hinde et~al.}(1987)\citenamefont{Hinde, Leigh,
  Bokhorst, Newton, Walsh, and Boldeman}}]{Hinde1987318}
\bibinfo{author}{\bibfnamefont{D.}~\bibnamefont{Hinde}},
  \bibinfo{author}{\bibfnamefont{J.}~\bibnamefont{Leigh}},
  \bibinfo{author}{\bibfnamefont{J.}~\bibnamefont{Bokhorst}},
  \bibinfo{author}{\bibfnamefont{J.}~\bibnamefont{Newton}},
  \bibinfo{author}{\bibfnamefont{R.}~\bibnamefont{Walsh}}, \bibnamefont{and}
  \bibinfo{author}{\bibfnamefont{J.}~\bibnamefont{Boldeman}},
  \bibinfo{journal}{Nuclear Physics A} \textbf{\bibinfo{volume}{472}},
  \bibinfo{pages}{318 } (\bibinfo{year}{1987}).

\bibitem[{\citenamefont{Tassan-Got and St\'{e}phan}(1991)}]{TassanGot1991Deep}
\bibinfo{author}{\bibfnamefont{L.}~\bibnamefont{Tassan-Got}} \bibnamefont{and}
  \bibinfo{author}{\bibfnamefont{C.}~\bibnamefont{St\'{e}phan}},
  \bibinfo{journal}{Nuclear Physics A} \textbf{\bibinfo{volume}{524}},
  \bibinfo{pages}{121} (\bibinfo{year}{1991}).

\bibitem[{\citenamefont{Gl\"{a}ssel et~al.}(1982)\citenamefont{Gl\"{a}ssel,
  Harrach, Grodzins, and Specht}}]{Glassel1982Direct}
\bibinfo{author}{\bibfnamefont{P.}~\bibnamefont{Gl\"{a}ssel}},
  \bibinfo{author}{\bibfnamefont{D.}~\bibnamefont{Harrach}},
  \bibinfo{author}{\bibfnamefont{L.}~\bibnamefont{Grodzins}}, \bibnamefont{and}
  \bibinfo{author}{\bibfnamefont{H.~J.} \bibnamefont{Specht}},
  \bibinfo{journal}{Physical Review Letters} \textbf{\bibinfo{volume}{48}},
  \bibinfo{pages}{1089} (\bibinfo{year}{1982}).

\bibitem[{\citenamefont{R\"{o}sch et~al.}(1989)\citenamefont{R\"{o}sch,
  Cassing, Gemmeke, Gentner, Keller, Lassen, Lucking, Richter, Schreck, and
  Schrieder}}]{Rosch1989Preequilibrium}
\bibinfo{author}{\bibfnamefont{W.}~\bibnamefont{R\"{o}sch}},
  \bibinfo{author}{\bibfnamefont{W.}~\bibnamefont{Cassing}},
  \bibinfo{author}{\bibfnamefont{H.}~\bibnamefont{Gemmeke}},
  \bibinfo{author}{\bibfnamefont{R.}~\bibnamefont{Gentner}},
  \bibinfo{author}{\bibfnamefont{K.}~\bibnamefont{Keller}},
  \bibinfo{author}{\bibfnamefont{L.}~\bibnamefont{Lassen}},
  \bibinfo{author}{\bibfnamefont{W.}~\bibnamefont{Lucking}},
  \bibinfo{author}{\bibfnamefont{A.}~\bibnamefont{Richter}},
  \bibinfo{author}{\bibfnamefont{R.}~\bibnamefont{Schreck}}, \bibnamefont{and}
  \bibinfo{author}{\bibfnamefont{G.}~\bibnamefont{Schrieder}},
  \bibinfo{journal}{Nuclear Physics A} \textbf{\bibinfo{volume}{496}},
  \bibinfo{pages}{141} (\bibinfo{year}{1989}).

\bibitem[{\citenamefont{Vigdor et~al.}(1980)\citenamefont{Vigdor, Karwowski,
  Jacobs, Kailas, Singh, Soga, and Yip}}]{Vigdor1980Anisotropy}
\bibinfo{author}{\bibfnamefont{S.~E.} \bibnamefont{Vigdor}},
  \bibinfo{author}{\bibfnamefont{H.~J.} \bibnamefont{Karwowski}},
  \bibinfo{author}{\bibfnamefont{W.~W.} \bibnamefont{Jacobs}},
  \bibinfo{author}{\bibfnamefont{S.}~\bibnamefont{Kailas}},
  \bibinfo{author}{\bibfnamefont{P.~P.} \bibnamefont{Singh}},
  \bibinfo{author}{\bibfnamefont{F.}~\bibnamefont{Soga}}, \bibnamefont{and}
  \bibinfo{author}{\bibfnamefont{P.}~\bibnamefont{Yip}},
  \bibinfo{journal}{Physics Letters B} \textbf{\bibinfo{volume}{90}},
  \bibinfo{pages}{384} (\bibinfo{year}{1980}).

\bibitem[{\citenamefont{Cussol et~al.}(1993)\citenamefont{Cussol, Bizard, Brou,
  Durand, Louvel, Patry, Peter, Regimbart, Steckmeyer, and
  Sullivan}}]{Cussol1993Chargedparticle}
\bibinfo{author}{\bibfnamefont{D.}~\bibnamefont{Cussol}},
  \bibinfo{author}{\bibfnamefont{G.}~\bibnamefont{Bizard}},
  \bibinfo{author}{\bibfnamefont{R.}~\bibnamefont{Brou}},
  \bibinfo{author}{\bibfnamefont{D.}~\bibnamefont{Durand}},
  \bibinfo{author}{\bibfnamefont{M.}~\bibnamefont{Louvel}},
  \bibinfo{author}{\bibfnamefont{J.}~\bibnamefont{Patry}},
  \bibinfo{author}{\bibfnamefont{J.}~\bibnamefont{Peter}},
  \bibinfo{author}{\bibfnamefont{R.}~\bibnamefont{Regimbart}},
  \bibinfo{author}{\bibfnamefont{J.~C.} \bibnamefont{Steckmeyer}},
  \bibnamefont{and} \bibinfo{author}{\bibfnamefont{J.}~\bibnamefont{Sullivan}},
  \bibinfo{journal}{Nuclear Physics A} \textbf{\bibinfo{volume}{561}},
  \bibinfo{pages}{298} (\bibinfo{year}{1993}).

\bibitem[{\citenamefont{Marie et~al.}(1997)\citenamefont{Marie, Laforest,
  Bougault, Wieleczko, Durand, Bacri, Lecolley, Saint-Laurent, Auger, Benlliure
  et~al.}}]{Marie199715}
\bibinfo{author}{\bibfnamefont{N.}~\bibnamefont{Marie}},
  \bibinfo{author}{\bibfnamefont{R.}~\bibnamefont{Laforest}},
  \bibinfo{author}{\bibfnamefont{R.}~\bibnamefont{Bougault}},
  \bibinfo{author}{\bibfnamefont{J.}~\bibnamefont{Wieleczko}},
  \bibinfo{author}{\bibfnamefont{D.}~\bibnamefont{Durand}},
  \bibinfo{author}{\bibfnamefont{C.}~\bibnamefont{Bacri}},
  \bibinfo{author}{\bibfnamefont{J.}~\bibnamefont{Lecolley}},
  \bibinfo{author}{\bibfnamefont{F.}~\bibnamefont{Saint-Laurent}},
  \bibinfo{author}{\bibfnamefont{G.}~\bibnamefont{Auger}},
  \bibinfo{author}{\bibfnamefont{J.}~\bibnamefont{Benlliure}},
  \bibnamefont{et~al.}, \bibinfo{journal}{Physics Letters B}
  \textbf{\bibinfo{volume}{391}}, \bibinfo{pages}{15 } (\bibinfo{year}{1997}).

\bibitem[{\citenamefont{Bonnet et~al.}(2009)\citenamefont{Bonnet, Borderie,
  Neindre, Rivet, Bougault, Chbihi, Dayras, Frankland, Galichet, Gagnon-Moisan
  et~al.}}]{Bonnet20091}
\bibinfo{author}{\bibfnamefont{E.}~\bibnamefont{Bonnet}},
  \bibinfo{author}{\bibfnamefont{B.}~\bibnamefont{Borderie}},
  \bibinfo{author}{\bibfnamefont{N.~L.} \bibnamefont{Neindre}},
  \bibinfo{author}{\bibfnamefont{M.-F.} \bibnamefont{Rivet}},
  \bibinfo{author}{\bibfnamefont{R.}~\bibnamefont{Bougault}},
  \bibinfo{author}{\bibfnamefont{A.}~\bibnamefont{Chbihi}},
  \bibinfo{author}{\bibfnamefont{R.}~\bibnamefont{Dayras}},
  \bibinfo{author}{\bibfnamefont{J.}~\bibnamefont{Frankland}},
  \bibinfo{author}{\bibfnamefont{E.}~\bibnamefont{Galichet}},
  \bibinfo{author}{\bibfnamefont{F.}~\bibnamefont{Gagnon-Moisan}},
  \bibnamefont{et~al.}, \bibinfo{journal}{Nuclear Physics A}
  \textbf{\bibinfo{volume}{816}}, \bibinfo{pages}{1 } (\bibinfo{year}{2009}).

\bibitem[{\citenamefont{Bougault et~al.}(1989)\citenamefont{Bougault, Colin,
  Delaunay, Genoux-Lubain, Hajfani, Le~Brun, Lecolley, Louvel, and
  Steckmeyer}}]{Bougault1989Timescale}
\bibinfo{author}{\bibfnamefont{R.}~\bibnamefont{Bougault}},
  \bibinfo{author}{\bibfnamefont{J.}~\bibnamefont{Colin}},
  \bibinfo{author}{\bibfnamefont{F.}~\bibnamefont{Delaunay}},
  \bibinfo{author}{\bibfnamefont{A.}~\bibnamefont{Genoux-Lubain}},
  \bibinfo{author}{\bibfnamefont{A.}~\bibnamefont{Hajfani}},
  \bibinfo{author}{\bibfnamefont{C.}~\bibnamefont{Le~Brun}},
  \bibinfo{author}{\bibfnamefont{J.~F.} \bibnamefont{Lecolley}},
  \bibinfo{author}{\bibfnamefont{M.}~\bibnamefont{Louvel}}, \bibnamefont{and}
  \bibinfo{author}{\bibfnamefont{J.~C.} \bibnamefont{Steckmeyer}},
  \bibinfo{journal}{Physics Letters B} \textbf{\bibinfo{volume}{232}},
  \bibinfo{pages}{291} (\bibinfo{year}{1989}).

\bibitem[{\citenamefont{Louvel et~al.}(1994)\citenamefont{Louvel,
  Genoux-Lubain, Bizard, Bougault, Brou, Buta, Doubre, Durand, El~Masri,
  Fugiwara et~al.}}]{Louvel1994Rapid}
\bibinfo{author}{\bibfnamefont{M.}~\bibnamefont{Louvel}},
  \bibinfo{author}{\bibfnamefont{A.}~\bibnamefont{Genoux-Lubain}},
  \bibinfo{author}{\bibfnamefont{G.}~\bibnamefont{Bizard}},
  \bibinfo{author}{\bibfnamefont{R.}~\bibnamefont{Bougault}},
  \bibinfo{author}{\bibfnamefont{R.}~\bibnamefont{Brou}},
  \bibinfo{author}{\bibfnamefont{A.}~\bibnamefont{Buta}},
  \bibinfo{author}{\bibfnamefont{H.}~\bibnamefont{Doubre}},
  \bibinfo{author}{\bibfnamefont{D.}~\bibnamefont{Durand}},
  \bibinfo{author}{\bibfnamefont{Y.}~\bibnamefont{El~Masri}},
  \bibinfo{author}{\bibfnamefont{H.}~\bibnamefont{Fugiwara}},
  \bibnamefont{et~al.}, \bibinfo{journal}{Physics Letters B}
  \textbf{\bibinfo{volume}{320}}, \bibinfo{pages}{221} (\bibinfo{year}{1994}).

\bibitem[{\citenamefont{Durand et~al.}(1995)\citenamefont{Durand, Colin,
  Lecolley, Meslin, Aboufirassi, Bilwes, Bougault, Brou, Cosmo, Galin
  et~al.}}]{Durand1995Nuclear}
\bibinfo{author}{\bibfnamefont{D.}~\bibnamefont{Durand}},
  \bibinfo{author}{\bibfnamefont{J.}~\bibnamefont{Colin}},
  \bibinfo{author}{\bibfnamefont{J.~F.} \bibnamefont{Lecolley}},
  \bibinfo{author}{\bibfnamefont{C.}~\bibnamefont{Meslin}},
  \bibinfo{author}{\bibfnamefont{M.}~\bibnamefont{Aboufirassi}},
  \bibinfo{author}{\bibfnamefont{B.}~\bibnamefont{Bilwes}},
  \bibinfo{author}{\bibfnamefont{R.}~\bibnamefont{Bougault}},
  \bibinfo{author}{\bibfnamefont{R.}~\bibnamefont{Brou}},
  \bibinfo{author}{\bibfnamefont{F.}~\bibnamefont{Cosmo}},
  \bibinfo{author}{\bibfnamefont{J.}~\bibnamefont{Galin}},
  \bibnamefont{et~al.}, \bibinfo{journal}{Physics Letters B}
  \textbf{\bibinfo{volume}{345}}, \bibinfo{pages}{397} (\bibinfo{year}{1995}).

\bibitem[{\citenamefont{Bauge et~al.}(1993)\citenamefont{Bauge, Elmaani, Lacey,
  Lauret, Ajitanand, Craig, Cronqvist, Gualtieri, Hannuschke, Li
  et~al.}}]{Bauge1993Observation}
\bibinfo{author}{\bibfnamefont{E.}~\bibnamefont{Bauge}},
  \bibinfo{author}{\bibfnamefont{A.}~\bibnamefont{Elmaani}},
  \bibinfo{author}{\bibfnamefont{R.~A.} \bibnamefont{Lacey}},
  \bibinfo{author}{\bibfnamefont{J.}~\bibnamefont{Lauret}},
  \bibinfo{author}{\bibfnamefont{N.~N.} \bibnamefont{Ajitanand}},
  \bibinfo{author}{\bibfnamefont{D.}~\bibnamefont{Craig}},
  \bibinfo{author}{\bibfnamefont{M.}~\bibnamefont{Cronqvist}},
  \bibinfo{author}{\bibfnamefont{E.}~\bibnamefont{Gualtieri}},
  \bibinfo{author}{\bibfnamefont{S.}~\bibnamefont{Hannuschke}},
  \bibinfo{author}{\bibfnamefont{T.}~\bibnamefont{Li}}, \bibnamefont{et~al.},
  \bibinfo{journal}{Physical Review Letters} \textbf{\bibinfo{volume}{70}},
  \bibinfo{pages}{3705} (\bibinfo{year}{1993}).

\bibitem[{\citenamefont{Bocage et~al.}(2000)\citenamefont{Bocage, Colin,
  Louvel, Auger, Bacri, Bellaize, Borderie, Bougault, Brou, Buchet
  et~al.}}]{Bocage2000Dynamical}
\bibinfo{author}{\bibfnamefont{F.}~\bibnamefont{Bocage}},
  \bibinfo{author}{\bibfnamefont{J.}~\bibnamefont{Colin}},
  \bibinfo{author}{\bibfnamefont{M.}~\bibnamefont{Louvel}},
  \bibinfo{author}{\bibfnamefont{G.}~\bibnamefont{Auger}},
  \bibinfo{author}{\bibfnamefont{C.}~\bibnamefont{Bacri}},
  \bibinfo{author}{\bibfnamefont{N.}~\bibnamefont{Bellaize}},
  \bibinfo{author}{\bibfnamefont{B.}~\bibnamefont{Borderie}},
  \bibinfo{author}{\bibfnamefont{R.}~\bibnamefont{Bougault}},
  \bibinfo{author}{\bibfnamefont{R.}~\bibnamefont{Brou}},
  \bibinfo{author}{\bibfnamefont{P.}~\bibnamefont{Buchet}},
  \bibnamefont{et~al.}, \bibinfo{journal}{Nuclear Physics A}
  \textbf{\bibinfo{volume}{676}}, \bibinfo{pages}{391} (\bibinfo{year}{2000}).

\bibitem[{\citenamefont{Colin et~al.}(2003)\citenamefont{Colin, Cussol,
  Normand, Bellaize, Bougault, Buta, Durand, Lopez, Manduci, Marie
  et~al.}}]{Colin2003Dynamical}
\bibinfo{author}{\bibfnamefont{J.}~\bibnamefont{Colin}},
  \bibinfo{author}{\bibfnamefont{D.}~\bibnamefont{Cussol}},
  \bibinfo{author}{\bibfnamefont{J.}~\bibnamefont{Normand}},
  \bibinfo{author}{\bibfnamefont{N.}~\bibnamefont{Bellaize}},
  \bibinfo{author}{\bibfnamefont{R.}~\bibnamefont{Bougault}},
  \bibinfo{author}{\bibfnamefont{A.}~\bibnamefont{Buta}},
  \bibinfo{author}{\bibfnamefont{D.}~\bibnamefont{Durand}},
  \bibinfo{author}{\bibfnamefont{O.}~\bibnamefont{Lopez}},
  \bibinfo{author}{\bibfnamefont{L.}~\bibnamefont{Manduci}},
  \bibinfo{author}{\bibfnamefont{J.}~\bibnamefont{Marie}},
  \bibnamefont{et~al.}, \bibinfo{journal}{Physical Review C}
  \textbf{\bibinfo{volume}{67}}, \bibinfo{pages}{064603}
  (\bibinfo{year}{2003}).

\bibitem[{\citenamefont{De~Filippo
  et~al.}(2005{\natexlab{b}})\citenamefont{De~Filippo, Pagano, Piasecki,
  Amorini, Anzalone, Auditore, Baran, Berceanu, Blicharska, Brzychczyk
  et~al.}}]{Filippo2005Dynamical}
\bibinfo{author}{\bibfnamefont{E.}~\bibnamefont{De~Filippo}},
  \bibinfo{author}{\bibfnamefont{A.}~\bibnamefont{Pagano}},
  \bibinfo{author}{\bibfnamefont{E.}~\bibnamefont{Piasecki}},
  \bibinfo{author}{\bibfnamefont{F.}~\bibnamefont{Amorini}},
  \bibinfo{author}{\bibfnamefont{A.}~\bibnamefont{Anzalone}},
  \bibinfo{author}{\bibfnamefont{L.}~\bibnamefont{Auditore}},
  \bibinfo{author}{\bibfnamefont{V.}~\bibnamefont{Baran}},
  \bibinfo{author}{\bibfnamefont{I.}~\bibnamefont{Berceanu}},
  \bibinfo{author}{\bibfnamefont{J.}~\bibnamefont{Blicharska}},
  \bibinfo{author}{\bibfnamefont{J.}~\bibnamefont{Brzychczyk}},
  \bibnamefont{et~al.}, \bibinfo{journal}{Physical Review C}
  \textbf{\bibinfo{volume}{71}}, \bibinfo{pages}{064604}
  (\bibinfo{year}{2005}{\natexlab{b}}).

\bibitem[{\citenamefont{Wilcke et~al.}(1980)\citenamefont{Wilcke, Birkelund,
  Wollersheim, Hoover, Huizenga, Schr\"{o}der, and Tubbs}}]{Wilcke1980Reaction}
\bibinfo{author}{\bibfnamefont{W.~W.} \bibnamefont{Wilcke}},
  \bibinfo{author}{\bibfnamefont{J.~R.} \bibnamefont{Birkelund}},
  \bibinfo{author}{\bibfnamefont{H.~J.} \bibnamefont{Wollersheim}},
  \bibinfo{author}{\bibfnamefont{A.~D.} \bibnamefont{Hoover}},
  \bibinfo{author}{\bibfnamefont{J.~R.} \bibnamefont{Huizenga}},
  \bibinfo{author}{\bibfnamefont{W.~U.} \bibnamefont{Schr\"{o}der}},
  \bibnamefont{and} \bibinfo{author}{\bibfnamefont{L.~E.} \bibnamefont{Tubbs}},
  \bibinfo{journal}{Atomic Data and Nuclear Data Tables}
  \textbf{\bibinfo{volume}{25}}, \bibinfo{pages}{389} (\bibinfo{year}{1980}).

\end{thebibliography}

\end{document}